\documentclass[preprint,11pt]{elsarticle}
 \usepackage{float}
\usepackage{amsmath,amssymb}
\usepackage{graphicx}
\usepackage{hyperref}
\usepackage{bm}
\usepackage{color}
\pdfoutput=1
\usepackage{tikz}
\usepackage{tabularx, booktabs}
\RequirePackage[T1]{fontenc}

\RequirePackage{graphicx}
\RequirePackage{flushend}
\usepackage{arydshln}
\usepackage{url} 
\usepackage{pdflscape} 
\usepackage{amsmath}
\usepackage{amssymb} 
\usepackage{dcolumn}
\usepackage{bm} 
\usepackage{color}
\usepackage{epsfig} 
\usepackage{amsfonts}
\usepackage{ulem}
\usepackage{graphicx}
\usepackage{dcolumn,pifont}
\usepackage{epsfig}
\usepackage{graphicx}
\usepackage{pgfplots}
\usepackage{tikz}
\usepackage{graphicx}
\usepackage{subcaption}
\usepackage{amsmath}
\usepackage{epsfig}
\usepackage{bm}
\usepackage{times,fancyhdr}
\usepackage{threeparttable}
\usepackage{rotating}

\def\case#1/#2{\textstyle\frac{#1}{#2}}

\newcommand{\be}{\begin{equation}}
\newcommand{\ee}{\end{equation}}
\newcommand{\ben}{\begin{eqnarray}}
\newcommand{\een}{\end{eqnarray}}
\newtheorem{thm}{Theorem}

\def\case#1/#2{\textstyle\frac{#1}{#2}}

\def\bea{\begin{eqnarray}}
\def\eea{\end{eqnarray}}

\PassOptionsToPackage{linktocpage}{hyperref}
\def\case#1/#2{\textstyle\frac{#1}{#2}}

\newcommand{\om}{{\omega}}

\newcommand{\bear}{\begin{eqnarray}}
\newcommand{\ear}{\end{eqnarray}}

\providecommand{\U}[1]{\protect\rule{.1in}{.1in}}

\newcommand{\mincir}{\raise
-3.truept\hbox{\rlap{\hbox{$\sim$}}\raise4.truept\hbox{$<$}\ }}
\newcommand{\magcir}{\raise
-3.truept\hbox{\rlap{\hbox{$\sim$}}\raise4.truept\hbox{$>$}\ }}

	\DeclareMathAlphabet{\mathpzc}{OT1}{pzc}{m}{it}
	
\begin{document}

\begin{frontmatter}

\title{Slow--Fast Evolution of Scalar Fields in Higher-Order Cosmological Gravity: Dynamics Inspired by the Pais--Uhlenbeck Oscillator}

\author[addr1]{Manuel Gonzalez-Espinoza}
\ead{manuel.gonzalez@pucv.cl}

\author[addr2,addr3]{Genly Leon\corref{cor1}}
\ead{genly.leon@ucn.cl}

\author[addr4]{Yoelsy Leyva}
\ead{yoelsy.leyva@academicos.uta.cl}

\author[addr4]{Giovanni Otalora}
\ead{giovanni.otalora@academicos.uta.cl}
\author[addr2,addr3,addr6]{Andronikos Paliathanasis}
\ead{anpaliat@phys.uoa.gr}

\author[addr2]{Aleksander Kozak}
\ead{aleksander.kozak2@uwr.edu.pl}

\cortext[cor1]{Corresponding author}

\address[addr1]{Instituto de Física, Pontificia Universidad Católica de Valparaíso, Casilla 4950, Valparaíso, Chile}
\address[addr2]{Departamento de Matemáticas, Universidad Católica del Norte, Avda. Angamos 0610, Casilla 1280, Antofagasta, Chile}
\address[addr3]{Department of Mathematics, Faculty of Applied Sciences, Durban University of Technology, Durban 4000, South Africa}
\address[addr4]{Departamento de Física, Facultad de Ciencias, Universidad de Tarapacá, Casilla 7-D, Arica, Chile}
\address[addr6]{National Institute for Theoretical and Computational Sciences (NITheCS), South Africa}

\begin{abstract}
We investigate the cosmological dynamics of scalar fields governed by higher-order gravity, with particular emphasis on models inspired by the Pais-Uhlenbeck oscillator—a prototypical fourth-order system known for its connection to ghost-free formulations. By recasting the field equations into a slow-fast dynamical system, we analyze phase space evolution across exponential and power-law coupling regimes. Our approach integrates numerical simulations and geometric methods to visualize trajectories, stream flows, and asymptotic behavior under varying potential parameters. The underlying system
admits singular surfaces and non-smooth transitions, revealing intricate dynamical structures. We examine the stability of de Sitter solutions, the crossing of the phantom divide, and the emergence of cyclic behavior through multiple-scale analysis. The inclusion of radiation and dust fluids enables the creation of realistic cosmological scenarios, including a transient matter-dominated era and a late-time accelerated expansion. Our results highlight the viability of Pais--Uhlenbeck scalar models in accounting for inflationary dynamics and dark energy, offering diagnostic tools for characterizing attractors and bifurcation phenomena in higher-derivative cosmology.
\end{abstract}

\end{frontmatter}

\section{Introduction}
In General Relativity and Cosmology, the Cosmological Principle is represented by the homogeneous and isotropic Friedmann-Lemaître-Robertson-Walker (FLRW) metrics. That is, the universe appears uniform to all observers on a sufficiently large scale. Although cold dark matter is widely accepted, its particles remain undetected; evidence of its existence comes solely from its gravitational effects on galaxies and larger structures \citep{Primack:1983xj, Peebles:1984zz, Bond:1984fp, Trimble:1987ee, Turner:1991id}.
Dark energy was introduced to explain the Universe's accelerated expansion \citep{Carroll:1991mt}, supported by extensive observations \citep{SupernovaCosmologyProject:1997zqe, SupernovaSearchTeam:1998fmf, SupernovaCosmologyProject:1998vns}.

The $\Lambda$CDM paradigm remains the most successful cosmological model, providing a framework that explains observations with the least number of parameters \citep{Planck:2018vyg}. However, the fundamental physics underlying dark matter and dark energy remains uncertain, prompting investigations into modified gravity theories \citep{Leon:2009rc, Nojiri:2010wj, DeFelice:2010aj, Clifton:2011jh, DeFelice:2011bh, Xu:2012jf, Bamba:2012cp, Leon:2012mt, Kofinas:2014aka, Bahamonde:2015zma, Momeni:2015uwx, Cai:2015emx, Krssak:2018ywd, CANTATA:2021ktz, Dehghani:2023yph, Paliathanasis:2023nkb, Anagnostopoulos:2021ydo, Odintsov:2022bpg, Nojiri:2022ski, Oikonomou:2020qah, Nojiri:2018ouv}.

Scalar fields are central to cosmology, playing a critical role in explaining both early-time phenomena, such as inflation \cite{Guth:1980zm}, and the late-time acceleration of the Universe. They are used in various models, including quintessence \cite{Ratra:1987rm, Parsons:1995kt, Rubano:2001xi, Saridakis:2008fy, Cai:2009zp, WaliHossain:2014usl, Barrow:2016qkh}, phantom fields \cite{Elizalde:2004mq, Nojiri:2005pu, Urena-Lopez:2005pzi}, quintom fields \cite{Guo:2004fq, Feng:2004ff, Zhang:2005eg, Zhang:2005kj, Lazkoz:2006pa, Lazkoz:2007mx, Setare:2008pz, Setare:2008pc, Leon:2012vt, Leon:2018lnd, Mishra:2018dzq, Marciu:2020vve, Dimakis:2020tzc}, chiral cosmology \cite{Dimakis:2020tzc, Chervon:2013btx, Paliathanasis:2020wjl}, and multi-scalar field models describing various cosmological epochs \cite{Rainer:1996gw, Copeland:1999cs, Coley:1999mj, Green:1999vv, Tsujikawa:2000wc, Elizalde:2008yf, Paliathanasis:2018vru, Kim:2006ys, Ashoorioon:2008qr, Langlois:2008mn, Achucarro:2010jv, White:2013ufa, Deffayet:2013lga, Gergely:2014rna, Christodoulidis:2019jsx, Christodoulidis:2019mkj, Akrami:2020zfz, Vazquez:2023kyx,Carloni:2023egi,Carloni:2024rrk,Luongo:2023aaq,DAgostino:2021vvv}.

Analytical approaches, including asymptotic methods and averaging theory \cite{verhulst2006method, sanders2010averaging, kevorkian2013perturbation}, have provided insights into scalar field cosmologies with generalized harmonic potentials, both in vacuum and in the presence of matter \cite{Fajman:2020yjb, Fajman:2021cli, Leon:2021lct, Leon:2021rcx, Leon:2020pfy, Leon:2020ovw, Leon:2021hxc, Leon:2020pvt}. Single and multi-field cosmologies have been extensively studied in previous works \cite{Alho:2015cza, Alho:2019pku, Chakraborty:2021vcr}. 

Slow-fast dynamics were applied to analyze theories based on the Generalized Uncertainty Principle (GUP) \cite{Paliathanasis:2015cza, Paliathanasis:2021egx}. For instance, the modified equations in the flat FLRW metric were transformed into an autonomous dynamical system for specific interaction functions \cite{Paliathanasis:2021egx}. Notably, the exponential potential described different asymptotic solutions with physical relevance, highlighting how GUP components enhance scalar perturbation growth in fast manifolds. At the same time, allowing for diverse behaviors in slow manifolds, including decay, growth, or oscillatory solutions.
GUP-based theories have broad implications in astrophysics and cosmology, including the explanation of the origin of cosmic magnetic fields \cite{Ashoorioon:2004rs}, black hole thermodynamics \cite{Adler:2001vs}, and cosmic acceleration \cite{Paliathanasis:2025dcr, Aghababaei:2021gxe}. Effects of GUP on both early and late Universe physics are critical for understanding the complete dynamical picture. In \cite{Giacomini:2020zmv}, the GUP-modified quintessence scalar field exhibits unique features, including critical points that describe the de Sitter Universe independently of the potential choice. Modifications to the Heisenberg uncertainty principle introduced a quadratic GUP, with diverse implications for quantum \cite{Jizba:2009qf, Buoninfante:2019fwr, Scardigli:2016pjs, Luciano:2019mrz} and gravitational systems \cite{Buoninfante:2020cqz, Giardino:2020myz}. GUP alters the Klein-Gordon equation, transforming it into a singular fourth-order partial differential equation.

Recent work in \cite{Paliathanasis:2024ozx} explored the equivalence between GUP and modified $f(R, \Box{R})$-gravity \cite{Gottlober:1989ww, Berkin:1990nu, RomeroCastellanos:2018inv, Carloni:2018yoz}, which introduces quantum corrections through higher-order terms. These corrections, derived from minimum length-scale effects, significantly modify cosmological dynamics. For a specific functional form, $f(R, \Box{R})=R+F(\Box{R})$, the gravitational model reduces from sixth-order to fourth-order, with $F(\Box{R})$ representing quantum-induced modifications to the General Relativity action \cite{Paliathanasis:2024ozx}.

The Pais-Uhlenbeck (PU) Oscillator \cite{Pais:1950za} is a foundational example of higher-order derivative theories. It consists of a fourth-order scalar field theory, and by suppressing spatial field dependence, the action simplifies \cite{Hawking:2001yt}:
 \begin{align} I &= \int d t \left [\frac{1}{2} \left [ {\ddot \phi}^2 - (\om_1^2 + \om_2^2) \dot\phi^2 + \om_1^2 \om_2^2 \phi^2 \right ] -\frac{\Lambda \phi^4}{4} \right ] + {\rm a ~ boundary ~term}. \label{2.6} \end{align}

Using the Ostrogradski formalism, a new variable is introduced to write the Hamiltonian, ensuring that the classical Hamilton equations align with the fourth-order equations of motion. Defining $ x = \dot\phi $, the Hamiltonian $ H(\phi, x; p_\phi, p_x) $ becomes:
\begin{align}
H(\phi, x; p_\phi, p_x) &= p_\phi x + \frac{1}{2} \left[ p_x^2 + (\omega_1^2 + \omega_2^2)x^2 - \omega_1^2 \omega_2^2 \phi^2 \right] + \frac{\Lambda \phi^4}{4}. \label{2.9}
\end{align}

After eliminating variables $ x, p_x, p_\phi $, the motion corresponds to:
\begin{align} &\phi^{(4)}+ ( \om_1^2 + \om_2^2)\ddot \phi + \om_1^2 \om_2^2 \phi - \Lambda \phi^3=0. \label{2.10} \end{align}

However, the Hamiltonian is not positive definite, leading to Ostrogradsky instability \cite{Ostrogradsky:1850fid}, often manifesting as ghost perturbations in scalar or vector theories. Following Smilga's approach \cite{Smilga:2004cy}, as in \cite{Pulgar:2014cba}, we analyze the phase space dynamics.

Efforts to address Ostrogradski's instability have focused on transforming the Hamiltonian into a form with positive energies. While this approach is effective for free PU oscillators \cite{Bolonek, Mostafazadeh, Nucci:2008ya, Stephen, Pavsic:2013noa, Banerjee:2013upa, Bolonek:2006ir, Bagarello, Masterov:2015ija}, it fails in the case of self-interacting systems with quartic interaction terms \cite{Pavsic:2013noa}. Smilga \cite{Smilga:2005gb, Smilga:2004cy, Smilga:2008pr} demonstrated that stability can be achieved for certain initial conditions, though quantization introduces instability due to tunneling effects. Other proposed solutions include bounded interaction terms \cite{Pavsic:2013noa}, varying oscillator masses with coupling \cite{Pavsic:2013noa, Pavsic:2012pw}, and incorporating both damping and third-order derivative terms for stability \cite{Pavsic:2013mja}.
An unconditionally stable system has also been identified \cite{Robert:2006nj}. For instance, \cite{Pavsic:2013noa} investigated a PU oscillator with a bounded interaction term (the fourth power of sine), demonstrating stability for all initial conditions. As a fourth-order system in time derivatives, the PU oscillator can be reformulated as a system of two coupled oscillators. Stability persists even when the masses differ, provided a coupling term is present \cite{Pavsic:2013noa, Pavsic:2012pw}. Additionally, \cite{Nesterenko:2006tu} found that damping can destabilize the PU oscillator. However, \cite{Pavsic:2013mja} showed that combining damping with a third-order derivative term prevents instability, ensuring the system's stability.

Having established a rigorous theoretical foundation and critically reviewed prevailing strategies to address Ostrogradski instability, we now turn to the framework and structure of our analysis.
The paper is organized as follows. In Section~\ref{sect:II}, we introduce the gravitational model under consideration: a spatially flat FLRW geometry with a scalar field governed by the Pais–Uhlenbeck oscillator. We present the associated cosmological field equations and identify the components of dark energy. Section~\ref{f-s-devisers} revisits the f-devisers method~\cite{delCampo:2013vka}, which allows for an autonomous reformulation of scalar field cosmologies using normalized variables and potential-adapted parameters, thereby enabling a broader potential analysis.
Section~\ref{sect:III} contains the core results of this work, including a phase space investigation and a reconstruction of asymptotic behaviors, emphasizing the role of higher-order derivatives in shaping cosmological evolution. In Section~\ref{sect:two_field_model}, we extend the model to a two-field formulation. Section~\ref{sect:V} focuses on the reconstruction of exact inflationary solutions.
In Section~\ref{sect:VI}, we apply singular perturbation techniques to reduce the system, offering analytical insights into the structure, stability, and cosmological implications of the slow–fast dynamics.
Finally, Section~\ref{sect:VII} presents our concluding remarks.

\section{The model}
\label{sect:II}

We consider the Pais-Uhlenbeck (PU) model as a prototype of higher-order derivative theory. Following Smilga's approach \cite{Smilga:2004cy}, we investigate the cosmological dynamics in phase space. This approach provides conditions for the stability of de Sitter solutions or the crossing of the phantom divide ($w_{DE} = -1$). Additionally, it enables an analysis of the parameter space where cyclic behavior is possible and examines regions where the ghost, according to Smilga's classification, exhibits benign or malicious behavior.

In \cite{Pulgar:2014cba}, the Pais-Uhlenbeck modification in a radiation background was investigated, yielding two solutions where the higher derivatives modification mimics a dust fluid, both being saddle points. These solutions are candidates for the transient matter-dominated era preceding the accelerated expansion phase. However, a more complete scenario should include a radiation field and a dust fluid. Therefore, we investigate the Lagrangian density: \begin{equation} \mathcal{L}=\frac{1}{2}\sqrt{-g}\left(R+\nabla_{\mu}\phi\nabla^{\mu}\phi+\alpha\nabla_{\mu}\nabla^{\mu}\phi\nabla_{\nu}\nabla^{\nu}\phi-2 V(\phi)\right), \end{equation} where $\alpha$ is the coupling parameter. Additionally, we consider a radiation source with energy density $\rho_r=\rho_{r,0} a^{-4}$ and a dust matter fluid with energy density $\rho_m=\rho_{m,0} a^{-3}$ as the background.

For a homogeneous, isotropic, and spatially flat 
universe, the line element is described by
\begin{equation}
 ds^2=-dt^2+a(t)^2d\mathbf{x}^2.
\end{equation}
The background equations are 
\begin{eqnarray}
3 H^2 - V - \dfrac{1}{2}\dot{\phi}^2-\dfrac{9}{2} \alpha H^2 \dot{\phi}^2 + 3 \alpha \dot{H} \dot{\phi}^2 - \dfrac{1}{2} \alpha\ddot{\phi}^2 + \alpha \dot{\phi}\dddot{\phi} - \rho &=& 0, \label{eq3}\\
3 H^2 + 2 \dot{H} - V + \dfrac{1}{2}\dot{\phi}^2-\dfrac{9}{2} \alpha H^2 \dot{\phi}^2 - 3 \alpha \dot{H} \dot{\phi}^2- 6 \alpha H \dot{\phi}\ddot{\phi} - \dfrac{1}{2} \alpha\ddot{\phi}^2 - \alpha \dot{\phi}\dddot{\phi} + p &=& 0, \label{eq4}\\
V_{,\phi}+3 H \dot{\phi} + \ddot{\phi}-9\alpha H \dot{H} \dot{\phi}-3 \alpha \dot{\phi} \ddot{H}- 9 \alpha H^2 \ddot{\phi}- 6\alpha \dot{H}\ddot{\phi}- 6 \alpha H \dddot{\phi}-\alpha \ddddot{\phi}&=& 0.
\end{eqnarray}
The functions $\rho$ and $p$ are the density and isotropic pressure of the total background fluid, which will be considered a mixture of dust and radiation.

Additionally, we can define an effective Dark Energy (DE) source with energy density and pressure given by 
\begin{eqnarray}
\rho_{DE} &=& \dfrac{1}{2}\dot{\phi}^2 + V+\dfrac{1}{2} \alpha \left(\ddot{\phi} + 3 H \dot{\phi} \right)^2 - 3 \alpha \dot{H} \dot{\phi}^2  - \alpha \dot{\phi}\dddot{\phi} - 3 \alpha H \dot{\phi}\ddot{\phi},\\
p_{DE} &=& \dfrac{1}{2}\dot{\phi}^2 - V-\dfrac{1}{2} \alpha \left(\ddot{\phi} + 3 H \dot{\phi} \right)^2 - 3 \alpha \dot{H} \dot{\phi}^2 - \alpha \dot{\phi}\dddot{\phi} - 3 \alpha H \dot{\phi}\ddot{\phi}.
\end{eqnarray}

Therefore, we can combine the Friedmann equations
 \begin{eqnarray}
\label{Fr1b}
3H^2& =& \rho_r +\rho_m + \rho_{DE}, \\
\label{Fr2b}
2\dot{H}& =&-\left(\frac{4}{3}\rho_r+\rho_m+\rho_{DE}+p_{DE}\right).
\end{eqnarray}
The conservation equation for radiation is 
\be\label{Cons1} \dot\rho_r=-4 H \rho_r, 
\ee and the conservation equation for matter is
\be\label{Cons1b} \dot\rho_m=-3 H \rho_m. 
\ee
The dark energy density and pressure satisfy
the usual evolution equation
\begin{eqnarray}\label{Cons2}
\dot{\rho}_{DE} +3H(\rho_{DE}+p_{DE})=0,
\end{eqnarray}
and we can  define the dark energy equation-of-state parameter as
usual
\begin{eqnarray}
w_{DE}\equiv \frac{p_{DE}}{\rho_{DE}}.
\end{eqnarray}
Finally, we defined the effective (total) equation of state parameter given by 
\be
w_{\text{eff}}\equiv \frac{p_{DE}+\frac{1}{3}\rho_r}{\rho_{DE}+\rho_r+\rho_m},
\ee which satisfies $w_{\text{eff}}:=\frac{1}{3}(2 q-1)$.

\subsection{General equations}

We assume $\alpha \neq 0$. The case $\alpha=0$ is Quintessence, and the case  $\alpha<0$ is related to quintom models. Using  \eqref{eq3} to remove $3H^2$ from \eqref{eq4}, we obtain 
\begin{align}
    2 \dot{H}(1-3 \alpha {\dot{\phi}}^2)= - \left({\dot{\phi}}^2 + p + \rho \right) + 2 \alpha \dot{\phi}\left(\dddot{\phi}+ 3 \alpha H \ddot{\phi}\right)
\end{align}
Consider 
\begin{equation}
    (1-3 \alpha {\dot{\phi}}^2) \neq 0.
\end{equation}
In this case, we have the modified Raychaudhuri equation: 
\begin{equation}
  \dot{H}= - \frac{\left({\dot{\phi}}^2 + p + \rho \right)}{2   (1-3 \alpha {\dot{\phi}}^2)} + \frac{\alpha \dot{\phi}\left(\dddot{\phi}+ 3 \alpha H \ddot{\phi}\right)}{(1-3 \alpha {\dot{\phi}}^2)}
\end{equation}
It can be studied the regime 
${\dot{\phi}}^2 \ll \frac{1}{3 \alpha}, \alpha>0$, where the factor $ (1-3 \alpha {\dot{\phi}}^2)>0$. 

The modified Friedmann equation and the Klein-Gordon equation are
\begin{align}
   & \frac{9 \alpha ^2 H \dot{\phi}^3 \ddot{\phi}}{1-3 \alpha  \dot{\phi}^2}+H^2 \left(3-\frac{9}{2} \alpha  \dot{\phi}^2\right)+\frac{1}{2} \left(\frac{p+\rho }{3 \alpha  \dot{\phi}^2-1}+p-\rho \right) \nonumber \\
    & -\frac{1}{2}
   \alpha  \ddot{\phi}^2+\frac{\dot{\phi}^2}{6 \alpha  \dot{\phi}^2-2}+\frac{\alpha  \dddot{\phi} \dot{\phi}}{1-3 \alpha  \dot{\phi}^2}-V(\phi )=0, \label{newFried}
\\
 & H \left(\frac{9 \alpha (\rho+p) \dot{\phi}}{2-6 \alpha  \dot{\phi}^2}-\frac{27 \alpha
   ^2 \dot{\phi} \ddot{\phi}^2}{\left(1-3 \alpha  \dot{\phi}^2\right)^2}+\frac{6 \dot{\phi}-9 \alpha  \dot{\phi}^3}{2-6 \alpha  \dot{\phi}^2}+\frac{6 \alpha  \dddot{\phi}}{3 \alpha  \dot{\phi}^2-1}\right)+\frac{9 \alpha  H^2 \ddot{\phi}}{3 \alpha  \dot{\phi}^2-1} \nonumber \\
   & +\frac{3 \alpha  \dot{\phi} \left(\dot{p}+ \dot{\rho}\right)}{2-6 \alpha  \dot{\phi}^2}+\ddot{\phi} \left(\frac{3 \alpha 
   (p+\rho ) \left(3 \alpha  \dot{\phi}^2+2\right)}{2 \left(1-3 \alpha  \dot{\phi}^2\right)^2}+\frac{9 \alpha ^2 \dot{\phi}^4+2}{2 \left(1-3 \alpha  \dot{\phi}^2\right)^2}\right) \nonumber \\
   & -\frac{9 \alpha ^2 \dddot{\phi} \dot{\phi} 
   \ddot{\phi}}{\left(1-3 \alpha  \dot{\phi}^2\right)^2}+\frac{\alpha  \ddddot{\phi}}{3 \alpha  \dot{\phi}^2-1}+V'(\phi )=0.\label{newKG}
\end{align}
Now we substitute 
\begin{equation}
    p= \frac{1}{3} \rho_r, \quad \rho= \rho_m + \rho_r,  \quad \dot\rho_r=-4 H \rho_r, \quad \dot\rho_m=-3 H \rho_m. 
\end{equation}
Then, the  effective Dark Energy (DE) source with energy density and pressure becomes 
\begin{eqnarray}
\rho_{DE} &=&\frac{9 \alpha ^2 H \dot{\phi}^3 \ddot{\phi}}{3 \alpha  \dot{\phi}^2-1}+\frac{\dot{\phi}^2 \left(9
   \alpha  H^2 \left(3 \alpha  \dot{\phi}^2-1\right)-1\right)}{6 \alpha  \dot{\phi}^2-2}+\frac{3 \alpha 
   \rho_m \dot{\phi}^2}{2-6 \alpha  \dot{\phi}^2}+\frac{2 \alpha  \rho_r \dot{\phi}^2}{1-3 \alpha  \dot{\phi}^2} \nonumber \\
   && +\frac{1}{2} \alpha  \ddot{\phi}^2+\frac{\alpha  \dddot{\phi} \dot{\phi}}{3 \alpha  \dot{\phi}^2-1}+V(\phi ),\\
p_{DE} &=& \frac{\dot{\phi}^2 \left(9 \alpha  H^2 \left(1-3 \alpha 
   \dot{\phi}^2\right)-1\right)}{6 \alpha  \dot{\phi}^2-2}+3 \alpha  H \dot{\phi} \ddot{\phi}
   \left(\frac{1}{3 \alpha  \dot{\phi}^2-1}-1\right)+\frac{3 \alpha  \rho_m \dot{\phi}^2}{2-6
   \alpha  \dot{\phi}^2}+\frac{2 \alpha  \rho_r \dot{\phi}^2}{1-3 \alpha  \dot{\phi}^2} \nonumber \\
   && -\frac{1}{2} \alpha  \ddot{\phi}^2+\frac{\alpha  \dddot{\phi} \dot{\phi}}{3 \alpha  \dot{\phi}^2-1}-V(\phi ),
\end{eqnarray}
Equations \eqref{newFried}, \eqref{newKG}, can be used as rules for the derivatives $\dddot{\phi}$ and $\ddddot{\phi}$. 

\section{The \texorpdfstring{$f$}{f}-Devisers Method: Framework, Reconstruction, and Dynamical Classification}
\label{f-s-devisers}
To extend the dynamical systems approach to a broader class of scalar field potentials in cosmology, one may employ the \textit{$f$-devisers method} (recently used in \cite{Leon:2022ipc}; see references therein). This technique facilitates the reformulation of the field equations into a closed autonomous system using normalized variables and a potential-adapted parameter $\lambda$.

\subsection{General Formulation and Reconstruction}

The method introduces two auxiliary variables:
\begin{align}
\lambda &\equiv -\frac{V'(\phi)}{V(\phi)}, \label{eq:lambda_def} \\
f &\equiv \frac{V''(\phi)}{V(\phi)} - \left( \frac{V'(\phi)}{V(\phi)} \right)^2, \label{eq:f_def}
\end{align}
which implies:
\begin{align}
V'(\phi) &= -\lambda V(\phi), \label{eq:dV_dphi} \\
V''(\phi) &= \left(f + \lambda^2\right) V(\phi). \label{eq:d2V_dphi2}
\end{align}

Provided that $ f = f(\lambda) $ is a well-defined function, one obtains:
\begin{align}
\frac{d\lambda}{d\phi} &= -f(\lambda), \label{eq:dlambda_dphi} \\
\frac{dV}{d\phi} &= -\lambda V, \label{eq:dv_dphi_again}
\end{align}
which leads to the parametric reconstruction:
\begin{align}
\phi(\lambda) &= -\int \frac{1}{f(\lambda)}\, d\lambda, \label{eq:phi_lambda} \\
V(\lambda) &= V_0 \exp\left( \int \frac{\lambda}{f(\lambda)}\, d\lambda \right). \label{eq:V_lambda}
\end{align}

\subsection{Representative Examples}

\subsubsection{Linear Case: $ f(\lambda) = \alpha\lambda + \beta $}

\[
\frac{d\lambda}{d\phi} = -\alpha\lambda - \beta
\quad \implies \quad
\lambda(\phi) = C e^{-\alpha\phi} + \frac{\beta}{\alpha}, \quad (\alpha \neq 0).
\]
For $\alpha = 0$, $\lambda(\phi) = -\beta\phi + C$.

\textit{Reconstructed potentials:}
\begin{itemize}
    \item If $\alpha \neq 0$:
    \[
    V(\phi) = V_0 \exp\left( \frac{C}{\alpha} e^{-\alpha\phi} - \frac{\beta}{\alpha} \phi \right).
    \]
    \item If $\alpha = 0$:
    \[
    V(\phi) = V_0 \exp\left( \frac{\beta}{2}\phi^2 - C\phi \right).
    \]
\end{itemize}

\subsubsection{Quadratic Case: $ f(\lambda) = \alpha\lambda^2 + \beta\lambda + \gamma $}

Define the discriminant:
\[
\Delta = \beta^2 - 4\alpha\gamma.
\]

\textit{Closed-form solutions:}
\begin{itemize}
\item $\Delta < 0$: Define $A = \frac{|\Delta|}{4\alpha^2}$,
    \[
    \lambda(\phi) = \sqrt{A} \tan(-\alpha \sqrt{A} \phi + C) - \frac{\beta}{2\alpha},
    \]
    \[
    V(\phi) = V_0 \cos^{-1}(-\alpha \sqrt{A} \phi + C) \cdot e^{\frac{\beta}{2\alpha} \phi}.
    \]
    \item $\Delta = 0$: $\lambda_0 = -\frac{\beta}{2\alpha}$,
    \[
    \lambda(\phi) = \lambda_0 + \frac{1}{\alpha \phi + C}, \quad
    V(\phi) = V_0 (\alpha\phi + C)^{-1/\alpha} e^{-\lambda_0 \phi}.
    \]

    \item $\Delta > 0$: Real roots $\lambda_{1,2} = \frac{-\beta \pm \sqrt{\Delta}}{2\alpha}$,
    \[
    \lambda(\phi) = \frac{\lambda_1 - D\lambda_2 e^{-\delta\phi}}{1 - D e^{-\delta\phi}},
    \quad \delta = \alpha(\lambda_2 - \lambda_1), \quad D = e^{-\delta C}.
    \]
    Potential reconstructed via:
    \[
    V(\phi) = V_0 \exp\left( - \int \lambda(\phi) \, d\phi \right).
    \]
\end{itemize}
To compute the integral, let us change variables:
\[
u = e^{-\delta \phi} \quad \Rightarrow \quad du = -\delta u \, d\phi \quad \Rightarrow \quad d\phi = \frac{-1}{\delta u} du
\]

Substituting into the integral:
\begin{align}
\int \lambda(\phi) \, d\phi 
&= -\frac{1}{\delta} \int \frac{\lambda_1 - D\lambda_2 u}{u(1 - D u)} \, du
\end{align}

We use partial fractions:
\[
\frac{\lambda_1 - D\lambda_2 u}{u(1 - D u)} = \frac{A}{u} + \frac{B}{1 - D u}
\]

Solving for $ A $ and $ B $:
\[
A = \lambda_1, \quad B = D(\lambda_1 - \lambda_2)
\]

Then:
\begin{align}
\int \lambda(\phi) \, d\phi 
&= -\frac{1}{\delta} \left[ \lambda_1 \int \frac{1}{u} du + (\lambda_1 - \lambda_2) \int \frac{D}{1 - D u} du \right] \\
&= -\frac{1}{\delta} \left[ \lambda_1 \ln u - (\lambda_2 - \lambda_1) \ln(1 - D u) \right]
\end{align}

Substituting back $ u = e^{-\delta \phi} $:
\[
\int \lambda(\phi) \, d\phi = \lambda_1 \phi + \frac{\lambda_2 - \lambda_1}{\delta} \ln(1 - D e^{-\delta \phi})
\]

Thus, the potential becomes:
\begin{equation}
V(\phi) = V_0 \exp\left(- \lambda_1 \phi - \frac{\lambda_2 - \lambda_1}{\delta} \ln(1 - D e^{-\delta \phi}) \right)
\end{equation}

Using exponent rules:
\begin{equation}
V(\phi) = V_0 \, e^{-\lambda_1 \phi} \left(1 - D e^{-\delta \phi} \right)^{-\frac{\lambda_2 - \lambda_1}{\delta}}
\end{equation}
If $\lambda_1 = 0$, $D = 1$, and $\delta = -\lambda_2/2$ we obtain the Starobinsky potential
\[
V_{\text{Star}}(\phi) = V_0 \left(1 - e^{-\delta \phi} \right)^2.
\]

\begin{table}[!ht]
\centering
\caption{Explicit forms of $ V(\phi) $ for representative linear and quadratic $ f(\lambda) $.}
\label{table:linear_quadratic_f}
\begin{tabularx}{\linewidth}{@{}lX@{}}
\hline
\textbf{Form of $ f(\lambda) $} & \textbf{Reconstructed Potential $ V(\phi) $} \\
\hline
$ \alpha \lambda + \beta $, $ \alpha \neq 0 $ & $ V_0 \exp\left( \frac{C}{\alpha} e^{-\alpha\phi} - \frac{\beta}{\alpha} \phi \right) $ \\
$ \beta $, $ \alpha = 0 $ & $ V_0 \exp\left( \frac{\beta}{2} \phi^2 - C\phi \right) $ \\
$ \alpha \lambda^2 + \beta \lambda + \gamma $, $ \Delta = 0 $ & $ V_0 (\alpha\phi + C)^{-1/\alpha} e^{-\lambda_0 \phi} $ \\
$ \Delta > 0 $ & $ V(\phi) = V_0 \, e^{-\lambda_1 \phi} \left(1 - D e^{-\delta \phi} \right)^{-\frac{\lambda_2 - \lambda_1}{\delta}} $ \\
$ \Delta < 0 $ & $ V(\phi) = V_0 \cos^{-1}(-\alpha \sqrt{A} \phi + C) \cdot e^{\frac{\beta}{2\alpha} \phi} $ \\
\hline
\end{tabularx}
\end{table}

\subsection{Analytical Validation of Representative \texorpdfstring{$f(\lambda)$}{f(lambda)} Expressions}

Table~\ref{fsform} summarizes closed-form expressions for the function $ f(\lambda) $ corresponding to several widely studied quintessence potentials. Each entry is derived using the $f$-devisers definitions \eqref{eq:lambda_def}-\eqref{eq:f_def}.

\begin{table}[!ht]
\centering
\caption{Closed-form expressions for $ f(\lambda) $ derived from representative quintessence potentials~\cite{Escobar:2013js}.}
\label{fsform}
\begin{tabularx}{\linewidth}{@{}Xl@{}}
\hline
\textbf{Potential $ V(\phi) $} & \textbf{$ f(\lambda) $} \\
\hline
$ \left|\dfrac{\mu}{n}\right| \phi^n $~\cite{Ratra:1987rm, Peebles:1987ek, Abramo:2003cp, Aguirregabiria:2004xd, Copeland:2004hq, Saridakis:2009pj, Saridakis:2009ej, Leon:2009dt, Chang:2013cba, Skugoreva:2013ooa, Pavlov:2013nra} & $ -\dfrac{\lambda^2}{n} $ \\
$ V_0 e^{-\alpha \phi} + V_1 $~\cite{Yearsley:1996yg, Pavluchenko:2003ge, Cardenas:2002np} & $ -\lambda(\lambda - \alpha) $ \\
$ V_1 e^{\alpha \phi} + V_2 e^{\beta \phi} $~\cite{Barreiro:1999zs, Gonzalez:2007hw, Gonzalez:2006cj} & $ -(\lambda + \alpha)(\lambda + \beta) $ \\
$ \dfrac{V_0}{\sinh^\alpha(\beta \phi)} $~\cite{Ratra:1987rm, Wetterich:1987fm, Copeland:2009be, Leyva:2009zz, Pavluchenko:2003ge, Sahni:1999gb, Urena-Lopez:2000ewq} & $ \dfrac{\lambda^2}{\alpha} - \alpha\beta^2 $ \\
$ V_0 [\cosh(\xi \phi) - 1] $~\cite{Ratra:1987rm, Wetterich:1987fm, Matos:2009hf, Copeland:2009be, Leyva:2009zz, Pavluchenko:2003ge, delCampo:2013vka, Sahni:1999qe, Sahni:1999gb, Lidsey:2001nj, Matos:2000ng} & $ -\dfrac{1}{2}(\lambda^2 - \xi^2) $ \\
\hline
\end{tabularx}
\end{table}

\subsubsection{Power-Law Potential: $ V(\phi) = \left| \dfrac{\mu}{n} \right| \phi^n $}

\begin{align}
&V' = n V / \phi, \quad V'' = n(n-1) V / \phi^2 \implies \\
&\lambda = -\frac{n}{\phi}, \quad f = \frac{n(n-1)}{\phi^2} - \lambda^2 = -\frac{n}{\phi^2} = -\frac{\lambda^2}{n}.
\end{align}

\subsubsection{Exponential Plus Constant: $ V(\phi) = V_0 e^{-\alpha \phi} + V_1 $}

\begin{align}
&V' = -\alpha V_0 e^{-\alpha \phi} = -\alpha(V - V_1), \quad V'' = \alpha^2(V - V_1) \implies \\
&\lambda = \alpha \left(1 - \frac{V_1}{V} \right), \quad f = \alpha^2 \left(1 - \frac{V_1}{V} \right) - \lambda^2 = -\lambda(\lambda - \alpha).
\end{align}

\subsubsection{Double Exponential: $ V(\phi) = V_1 e^{\alpha \phi} + V_2 e^{\beta \phi} $}

Let $ x = V_1 e^{\alpha \phi},\ y = V_2 e^{\beta \phi} $:

\begin{align}
&\lambda = -\frac{\alpha x + \beta y}{x + y}, \quad V'' = \alpha^2 x + \beta^2 y \implies \\
&f = \frac{V''}{V} - \lambda^2 = -(\lambda + \alpha)(\lambda + \beta).
\end{align}

\subsubsection{Hyperbolic Potential: $ V(\phi) = \dfrac{V_0}{\sinh^\alpha(\beta \phi)} $}

Let $ S = \sinh(\beta \phi) $:

\begin{align}
&V' = -\alpha V_0 \beta \cosh(\beta \phi) S^{-\alpha - 1}, \quad \lambda = \alpha \beta \coth(\beta \phi),\\
&f = \frac{V''}{V} - \lambda^2 = \frac{\lambda^2}{\alpha} - \alpha\beta^2.
\end{align}

\subsubsection{Cosh-Based Potential: $ V(\phi) = V_0[\cosh(\xi \phi) - 1] $ }

\begin{align}
&V' = V_0 \xi \sinh(\xi \phi), \quad V'' = V_0 \xi^2 \cosh(\xi \phi), \\
&\lambda = -\xi \dfrac{\sinh(\xi \phi)}{\cosh(\xi \phi) - 1}, \quad f = -\dfrac{1}{2}(\lambda^2 - \xi^2).
\end{align}
These derivations confirm the analytic forms listed in Table~\ref{fsform}. They demonstrate how standard potentials map cleanly into the $ f $-devisers framework, enabling systematic phase space analysis and attractor classification within autonomous dynamical models.

\subsection{Dynamical Implications and Attractor Classification}

The structure of $ f(\lambda) $ directly determines critical points, their stability, and the nature of attractor behavior:
\begin{itemize}
    \item Fixed points arise at $ f(\tilde{\lambda}) = 0 $, often corresponding to scaling, matter-dominated, or inflationary regimes.
    \item Linear forms yield scaling solutions with fixed energy ratios between matter and scalar field.
    \item Quadratic functions permit bifurcations, multiple isolated attractors, and saddle-node transitions.
    \item A  smooth, steady decrease in $f(\lambda)$ supports tracker behavior, allowing the scalar field to evolve alongside the energy content of the universe—such as matter or radiation—before eventually dominating and driving the overall dynamics.
    \item Asymptotic de Sitter attractors typically appear when $ \lambda \to 0 $ and $ f(\lambda) \to 0 $.
\end{itemize}

Center manifold theory may be required to analyze non-hyperbolic fixed points, particularly near extrema or inflection points. Moreover, the geometric correspondence between neighboring phase space trajectories and smoothly varying potentials enables studies of attractor tuning and parameter deformation.

\subsection{Scope and Limitations}

The $f$-devisers method based on \eqref{eq:lambda_def}--\eqref{eq:f_def} provides an elegant framework for classifying scalar field cosmologies through the autonomous system \eqref{eq:dlambda_dphi}--\eqref{eq:dv_dphi_again}, 
assuming the function $ f(\lambda) $ is single-valued and analytically expressible. This assumption ensures dynamical closure via normalized variables and $ \lambda $.
However, this structure excludes a wide class of physically motivated models where the inversion $ \phi = \phi(\lambda) $ is not globally feasible. These include transcendental, piecewise-defined, and multi-branched potentials, such as
\begin{equation}
V(\phi) \propto \phi^p \ln^q(\phi), \quad
V(\phi) \propto \phi^n e^{-q\phi^m},
\end{equation}
as discussed in~\cite{Barrow:1995xb, Parsons:1995ew}.

A representative example is the Peebles–Vilenkin potential~\cite{Peebles:1998qn}, designed to interpolate between inflation and quintessence:
\begin{equation} \label{eq:PVpotential}
V(\phi) = 
\begin{cases}
V_0 (\phi^4 + M^4), & \phi < 0 \\
\displaystyle \dfrac{V_0 M^8}{\phi^4 + M^4}, & \phi \geq 0
\end{cases}.
\end{equation}

From this, we compute
\begin{equation}
  \lambda(\phi) =
\begin{cases}
\displaystyle -\frac{4\phi^3}{\phi^4 + M^4}, & \phi < 0 \\[0.8em]
\displaystyle +\frac{4\phi^3}{\phi^4 + M^4}, & \phi \geq 0
\end{cases},  \label{eq:46}
\end{equation}

and 

\begin{equation}
f(\phi) =
\begin{cases}
\displaystyle +\frac{4\phi^2(-3M^4 +\phi^4)}{(M^4 + \phi^4)^2}, & \phi < 0 \\[0.8em]
\displaystyle -\frac{4\phi^2(-3M^4 +\phi^4)}{(M^4 + \phi^4)^2}, & \phi \geq 0
\end{cases}.
\label{eq:fphi_PV_factorized}
\end{equation}
Both functions are constructed from the piecewise-defined Peebles–Vilenkin potential and exhibit the following differentiability properties at $ \phi = 0 $:
\begin{itemize}
    \item \textbf{Continuity:} Both $ \lambda(\phi) $ and $ f(\phi) $ are continuous at $ \phi = 0 $.
    \item \textbf{First Derivative:} Both functions are continuously differentiable at $ \phi = 0 $; that is, they belong to $ \mathcal{C}^1 $.
    \item \textbf{Second and Higher Derivatives:} $\lambda(\phi)$ belongs to $ \mathcal{C}^2 $ at $ \phi = 0 $, while $f(\phi)$ displays discontinuity in its derivatives of order $n = 4k +2$, for $k\in\mathbb{N}$. Similarly, derivatives $\lambda^{(n)}(\phi)$ become discontinuous at $\phi = 0$ for $n = 4k + 3$, $k\in\mathbb{N}$.
\end{itemize}

Despite being rational in $ \phi $, neither $ \lambda(\phi) $ nor $ f(\phi) $ admits a closed-form inversion that eliminates $ \phi $. For example, expressing $ f = f(\lambda) $ would require inverting \eqref{eq:46}
which yields transcendental dependencies. This prevents closure of the autonomous system in terms of normalized variables and $ \lambda $ alone.

As a result, the dynamical evolution remains explicitly dependent on $\phi$, and the geometry of the phase space captures only functional analogies between potentials rather than a unified classification of trajectories. The lack of higher-order differentiability near $\phi = 0$ further restricts the applicability of analytical techniques such as center manifold theory and matched asymptotics. Nevertheless, alternative methods have been developed to analyze such non-closed systems—see~\cite{Leon:2008de, Leon:2018skk, Leon:2020ovw, Leon:2020pfy} and references therein—broadening the analytical scope and clarifying the formal limitations of the $f$-devisers method.
In summary, although the $f$-devisers framework provides an elegant tool for classifying cosmological dynamics via functions $f(\lambda)$, its applicability is confined to potentials that admit explicit analytical representations. For many physically motivated models—including piecewise-defined and transcendental forms—the failure to express $f$ exclusively in terms of $\lambda$ obstructs system closure. In such cases, the scalar field $\phi$ continues to influence the dynamics explicitly, and phase space proximity reflects potential similarity rather than shared evolution. This marks a well-defined structural boundary of the method’s domain of validity; however, this limitation does not apply to the dynamical systems analyzed in the present study, which remain formally closed and amenable to rigorous trajectory classification.

\section{Qualitative behavior on the Phase space}
\label{sect:III}
In this section, we perform the stability analysis of the cosmological scenario. To do that, we first transform it into its autonomous form \cite{Perko,Ellis,tavakol_1997,Copeland:1997et,Ferreira:1997au,Coley:2003mj,Chen:2008ft,Cotsakis:2013zha,Giambo:2009byn}
\be \label{eomscol}
\textbf{X}'=\textbf{f(X)}, \ee 
where $\bf{X}$ is a column vector of auxiliary variables,
and primes denote derivatives
with respect to $N=\ln a$. Then, one extracts the  critical points
$\bf{X_c}$  which satisfy  $\bf{X}'=0$. To determine
their stability properties, one takes the Taylor expansion around them  up to the first
order as \be\label{perturbation} \textbf{U}'={\bf{Q}}\cdot
\textbf{U},\ee with $\textbf{U}$ the
column vector of the perturbations of the variables and ${\bf {Q}}$ the matrix  containing the coefficients of the
perturbation equations. The eigenvalues of ${\bf {Q}}$ evaluated at the specific critical point determine their type and
stability.

In this concrete model, we define the dimensionless variables 
\begin{subequations}
\label{vars}
\begin{align}
\mathbf{u}:= (u_1, u_2, u_3, u_4, u_5, v)=\left(\frac{\dot\phi}{\sqrt{6} H}, \frac{\sqrt{V(\phi)}}{\sqrt{3} H}, -\frac{\ddot \phi}{\sqrt{6} H^2}, \Omega_r, \frac{1}{H},  -\frac{\sqrt{\frac{2}{3}}\left(3 H \ddot\phi+\dddot{\phi}\right)}{H}\right)
\end{align}
where
\begin{align}
&	\Omega_m= \frac{\rho_{m}}{3 H^2}, \quad  \Omega_r= \frac{\rho_{r}}{3 H^2}
	\end{align}
\end{subequations}	
which are related through 
\begin{equation}
\Omega_m=  \frac{\alpha  u_5^2 \left(3 u_1^2 \left(6 u_2^2+2 \Omega_r-9\right)+6 u_1 u_3-u_3^2\right)+u_5^4 \left(-\left(u_1^2+\alpha  u_1
   v+u_2^2+\Omega_r-1\right)\right)+18 \alpha ^2 u_1^2 (u_3-3
   u_1)^2}{u_5^2 \left(u_5^2-9 \alpha  u_1^2\right)}, 
\end{equation}	
and use the $f$-devisers definitions \eqref{eq:lambda_def}-\eqref{eq:f_def}.

Taking the derivatives of \eqref{vars} with respect to $N=\ln a$ we obtain the dynamical system: 
\begin{subequations}
\label{syst1}
\begin{align}
 &\frac{d u_1}{d N} = \frac{3 u_1^3 \left(u_5^2-9 \alpha \right)+3 \alpha  u_1^2 \left(12 u_3+v u_5^2\right)+u_1 \left(u_5^2 \left(-3 u_2^2+\Omega_r+3\right)-3 \alpha  u_3^2\right)-2 u_3
   u_5^2}{2 \left(u_5^2-9 \alpha  u_1^2\right)},\\
&\frac{d u_2}{d N} =  -\frac{u_2 \left(3 \alpha  \left(-3 \sqrt{6} \lambda  u_1^3+9 u_1^2-6 u_1 u_3+u_3^2\right)-u_5^2 \left(u_1 \left(-\sqrt{6} \lambda +3
   u_1+3 \alpha  v\right)-3 u_2^2+\Omega_r+3\right)\right)}{2 \left(u_5^2-9 \alpha  u_1^2\right)},\\
    &\frac{d u_3}{d N} = \frac{3 u_1^2 u_5^2 (2 u_3-3 \alpha  v)+6 \alpha  u_1 u_3 \left(6
   u_3+v u_5^2\right)-6 u_2^2 u_3 u_5^2-6 \alpha  u_3^3+2 u_3 \Omega_r u_5^2+v u_5^4}{2 \left(u_5^2-9 \alpha  u_1^2\right)},\\
&\frac{d v}{d N} = \frac{1}{2 \alpha  u_5^4 \left(u_5^2-9 \alpha  u_1^2\right)} \Bigg(-324 \alpha ^3 u_1^2 u_3
   (u_3-3 u_1)^2 \nonumber \\
   & +\alpha  u_5^4 \left(v u_5^2 \left(3 u_1^2-3 u_2^2+\Omega_r-3\right)+6 \left(36 u_1^3-3 u_1^2 \left(3 \sqrt{6} \lambda  u_2^2+u_3\right)+4 u_1
   \Omega_r+u_3 \left(-3 u_2^2+\Omega_r+3\right)\right)\right) \nonumber \\
   & +3 \alpha ^2 u_5^2 \Big(v u_5^2 \left(9 u_1^2+24 u_1 u_3-u_3^2\right) +u_1 v^2
   u_5^4 \nonumber \\
   & +6 \left(-54 u_1^5+18 u_1^4
   \left(\sqrt{6} \lambda  u_2^2+u_3\right)-12 u_1^3 \Omega_r+3 u_1^2 u_3 \left(-6 u_2^2+2 \Omega_r+3\right)+6 u_1 u_3^2-u_3^3\right)\Big) \nonumber \\
   & +2 u_5^6 \left(-6 u_1+\sqrt{6} \lambda  u_2^2+2 u_3\right)\Bigg),\\
& \frac{d u_5}{d N}= \frac{u_5^3 \left(3
   u_1 (u_1+\alpha  v)-3 u_2^2+\Omega_r+3\right)-3 \alpha  u_5 (u_3-3 u_1)^2}{2 \left(u_5^2-9 \alpha  u_1^2\right)},\\
& \frac{d \Omega_r}{d N}= \frac{\Omega_r \left(3 \alpha  \left(3 u_1^2+6 u_1
   u_3-u_3^2\right)+u_5^2 \left(3 u_1 (u_1+\alpha  v)-3 u_2^2+\Omega_r-1\right)\right)}{u_5^2-9 \alpha  u_1^2},\\
& \frac{d \lambda}{d N}= -\sqrt{6} f(\lambda) u_1,
\end{align}
\end{subequations}
where $f(\lambda)$ is defined by \eqref{eq:f_def} and can be explicitly expressed as a function of \eqref{eq:lambda_def}. 

The system given by equation \eqref{syst1} can be symbolically represented as:
\begin{align}
&\frac{d \mathbf{u}}{d N} = \frac{\mathbf{G} (\mathbf{u}, \lambda; \alpha)}{s(\mathbf{u}; \alpha)} , \quad
 \frac{d \lambda}{d N}= -\sqrt{6} f(\lambda) u_1, \label{syst-symbolic}
\end{align}
where $\mathbf{G}(\mathbf{u}, \lambda; \alpha)$ is a smooth vector field, and $s(\mathbf{u}; \alpha)$ is a smooth scalar function. The function $f(\lambda)$ is an arbitrary smooth function.

The dynamical system \eqref{syst-symbolic} possesses a singular line defined by $s(\mathbf{u}; \alpha)=0$, indicating the presence of non-smooth dynamical systems, which may be non-differentiable, discontinuous, or piecewise smooth. Inequality constraints further restrict solutions. Classical dynamical systems rely on smoothness, so generalizing these concepts to non-smooth systems is essential but often challenging. Techniques for analyzing non-smooth dynamical systems, such as differential inclusions (multi-valued differential equations), have been developed in the literature \cite{Saavedra:2000wk, Markus_Kunze-2000, kunze2001non, LaureaMariodiBernardo2008PDST, AwrejcewiczJan2014ANDS}.

The system describes a slow–fast dynamical model in which most variables evolve gradually while one—specifically, $v$—changes rapidly. It’s governed by a small parameter, $\alpha$, whose presence amplifies the fast dynamics relative to the slow ones. When this parameter becomes negligible, the model reduces to a simplified form constrained to a lower-dimensional surface, known as the slow manifold, where the fast variable remains constant. This reduction allows the system to be analyzed using a smaller set of variables. To study its behavior in this regime, both analytical tools from dynamical systems theory and numerical techniques are employed. The equilibrium states of the reduced system can be examined using Tikhonov’s theorem \cite{Berglund, tikhonov1952systems}.

The second problem is investigating the dynamics of the slow manifold using analytical dynamical systems tools and numerical methods. Conditions for the stability of de Sitter solutions and the crossing of the phantom divide, $w_{DE} = -1$ (quintom behavior), can be established, along with an analysis of cyclic behavior and the classification of the ghost as either benign or malicious, as per Smilga's framework. This study involves examining perturbative solutions of the dynamical systems associated with modified gravity theories \cite{Quiros:2016oln, Herrera:2016fdo, Alvarez:2016qky, Quiros:2017gsu, Otalora:2018bso, Gonzalez-Espinoza:2018gyl, Gonzalez-Espinoza:2019ajd, Quiros:2020bcg, Cataldo:2020cxm, Gonzalez-Espinoza:2021mwr, Gonzalez-Espinoza:2021qnv, Leyva:2021fuo}.

Finally, we investigate whether the Pais-Uhlenbeck model can provide insights into inflation and dark energy issues for $\alpha>0$. Using the assumption $ \phi(t) = \pm \sqrt{\frac{1}{3 \alpha}} t + \phi_0 $, the background equations simplify, and exact solutions can be obtained by imposing a solution for $ a(t) $ and deducing $ V(t) $.

\subsection{Dynamical systems analysis}

Consider $\alpha>0$. 
The system \eqref{syst1} admits the (lines of) equilibrium points
\begin{enumerate}
\item $\left(u_1, u_2, u_3, v, \Omega_r,\lambda, u_5\right)=\left( 0,  -1,  0,  0,  0 , 0, {u_{5c}}\right)$, with eigenvalues \newline $\Big\{0,-4,-3,-\frac{3 \alpha +\sqrt{\alpha  \left(9 \alpha +2 {u_{5c}} \left({u_{5c}}-\sqrt{12 \alpha 
   f(0)+{u_{5c}}^2}\right)\right)}}{2 \alpha },\frac{\sqrt{\alpha  \left(9 \alpha +2 {u_{5c}}
   \left({u_{5c}}-\sqrt{12 \alpha  f(0)+{u_{5c}}^2}\right)\right)}-3 \alpha }{2 \alpha }$, \newline $-\frac{3 \alpha
   +\sqrt{\alpha  \left(9 \alpha +2 {u_{5c}} \left(\sqrt{12 \alpha  f(0)+{u_{5c}}^2}+{u_{5c}}\right)\right)}}{2
   \alpha },\frac{\sqrt{\alpha  \left(9 \alpha +2 {u_{5c}} \left(\sqrt{12 \alpha 
   f(0)+{u_{5c}}^2}+{u_{5c}}\right)\right)}-3 \alpha }{2 \alpha }\Big\}$. It is normally hyperbolic. 

 \item$\left(u_1, u_2, u_3, v, \Omega_r,\lambda, u_5\right)=\left( 0,  1,  0,  0,  0 , 0, {u_{5c}}\right)$, with eigenvalues \newline 
 $\Big\{0,-4,-3,-\frac{3 \alpha +\sqrt{\alpha  \left(9 \alpha +2 {u_{5c}} \left({u_{5c}}-\sqrt{12 \alpha 
   f(0)+{u_{5c}}^2}\right)\right)}}{2 \alpha },\frac{\sqrt{\alpha  \left(9 \alpha +2 {u_{5c}}
   \left({u_{5c}}-\sqrt{12 \alpha  f(0)+{u_{5c}}^2}\right)\right)}-3 \alpha }{2 \alpha }$,\newline $-\frac{3 \alpha
   +\sqrt{\alpha  \left(9 \alpha +2 {u_{5c}} \left(\sqrt{12 \alpha  f(0)+{u_{5c}}^2}+{u_{5c}}\right)\right)}}{2
   \alpha },\frac{\sqrt{\alpha  \left(9 \alpha +2 {u_{5c}} \left(\sqrt{12 \alpha 
   f(0)+{u_{5c}}^2}+{u_{5c}}\right)\right)}-3 \alpha }{2 \alpha }\Big\}$.  It is normally hyperbolic. 
   
   \end{enumerate}
Additionally, when simultaneously $u_5^2-9 \alpha u_1^2=0$ and the numerators of \eqref{syst1} are zero, we have the singular lines will be examined numerically. 

\subsection{Unperturbed system ($\alpha=0$; quintessence)}
In the case $\alpha=0$, we recover quintessence as follows. 
\begin{subequations}
\begin{align}
 & \frac{d u_1}{dN}=  \frac{1}{2} \left(3 u_1^3+u_1 \left(-3 u_2^2+\Omega_r+3\right)-2 u_3\right),\\
 & \frac{d u_2}{dN}= \frac{1}{2} u_2 \left(3 u_1^2-\sqrt{6} \lambda  u_1-3 u_2^2+\Omega_r+3\right),\\
   & \frac{d u_3}{dN}= u_3 \left(3 u_1^2-3 u_2^2+\Omega_r\right)+\frac{v u_5^2}{2}, \label{eq51c}\\
   &\frac{d \Omega_r}{dN}= \Omega_r \left(3 u_1^2-3 u_2^2+\Omega_r-1\right),\\
   & \frac{d \lambda}{d N}=-\sqrt{6} u_1 f(\lambda ),\\
   &  \frac{d u_5}{d N}= \frac{1}{2} u_5 \left(3 u_1^2-3 u_2^2+\Omega_r+3\right), \label{eq51f}\\
   & 0= -6 u_1+\sqrt{6} \lambda  u_2^2+2 u_3.
\end{align}
\end{subequations}
 The last gives the slow manifold $$\sqrt{6} \lambda  u_2^2-6 u_1+2 u_3=0, \quad v= \text{constant}.$$
Then, $$u_3= 3 u_1-\sqrt{\frac{3}{2}} \lambda  u_2^2.$$ Therefore the equation \eqref{eq51c} for $u_3$ decouples. Because the dependence of the vector field with respect to $u_5$ only appears in equations \eqref{eq51c} and \eqref{eq51f} , then equation for \eqref{eq51f}  decouples too, obtaining the  the dynamical system for the ``reduced'' vector of states $(u_1,u_2, \Omega_r, \lambda)$: 
\begin{subequations}
\label{reduced}
\begin{align}
   &\frac{d u_1}{dN}= \frac{1}{2} \left(3 u_1^3+u_1 \left(-3 u_2^2+\Omega_r-3\right)+\sqrt{6} \lambda  u_2^2\right), \\
     & \frac{d u_2}{dN}= \frac{1}{2} u_2 \left(3 u_1^2-\sqrt{6} \lambda  u_1-3 u_2^2+\Omega_r+3\right),\\
      &\frac{d \Omega_r}{dN}= \Omega_r \left(3 u_1^2-3 u_2^2+\Omega_r-1\right),\\
       & \frac{d \lambda}{d N}=-\sqrt{6} u_1 f(\lambda ).
\end{align}
\end{subequations}
The phase space of \eqref{reduced} is given by \begin{align}
\left\{\left(u_1,u_2,  \Omega_r, \lambda\right)\in\mathbb{R}^4:  u_1^2+u_2^2+\Omega_r\leq 1, \lambda \in \mathbb{R}, u_2\geq 0\right\}. 
\end{align}
We will investigate this system further in section \ref{slow-dynamics}.

\subsection{Singularly perturbed model as $0<\alpha\ll 1$}

Assume that $0<\alpha\ll 1$. Then, the autonomous system \eqref{eomscol}
can be written as 
\begin{small}
\begin{subequations}
\label{singular_ds}
\begin{align}
 & \frac{d u_1}{dN}=   \frac{1}{2} \left(3 u_1^3+\left(-3 u_2^2+\Omega_r+3\right) u_1-2 u_3\right)+\frac{3 u_1 \left(9 u_1^4+3 \left(\Omega_r-3 u_2^2\right) u_1^2+\left(v
   u_5^2+6 u_3\right) u_1-u_3^2\right) \alpha }{2 u_5^2}+O\left(\alpha ^2\right),\label{dinsysialphap}\\
 & \frac{d u_2}{dN}= \frac{1}{2} u_2 \left(3 u_1^2-\sqrt{6} \lambda  u_1-3 u_2^2+\Omega_r+3\right)+\frac{3 u_2
   \left(9 u_1^4+3 \left(\Omega_r-3 u_2^2\right) u_1^2+\left(v u_5^2+6 u_3\right) u_1-u_3^2\right) \alpha }{2 u_5^2}+O\left(\alpha ^2\right),\\
 & \frac{d u_3}{dN}=  \left(\frac{v u_5^2}{2}+u_3
   \left(3 u_1^2-3 u_2^2+\Omega_r\right)\right)+\frac{3 u_3 \left(9 u_1^4+3 \left(\Omega_r-3 u_2^2\right) u_1^2+\left(v u_5^2+6 u_3\right)
   u_1-u_3^2\right) \alpha }{u_5^2}+O\left(\alpha ^2\right),\\
& \frac{d \Omega_r}{dN}= \Omega_r \left(3 u_1^2-3 u_2^2+\Omega_r-1\right)+\frac{3 \Omega_r \left(9 u_1^4+3 \left(\Omega_r-3 u_2^2\right)
   u_1^2+\left(v u_5^2+6 u_3\right) u_1-u_3^2\right) \alpha }{u_5^2}+O\left(\alpha ^2\right),\\
 & \frac{d \lambda}{dN}= -\sqrt{6} f(\lambda) u_1,\\
    & \frac{d u_5}{dN}= \frac{1}{2} u_5 \left(3 u_1^2-3 u_2^2+\Omega_{r}+3\right)+\frac{3 \left(9 u_1^4+3 \left(\Omega_r-3 u_2^2\right) u_1^2+\left(v u_5^2+6 u_3\right) u_1-u_3^2\right) \alpha }{2 u_5}+O\left(\alpha ^2\right),\\
 & \alpha \frac{d v}{dN}= \left(\sqrt{6} \lambda  u_2^2-6 u_1+2 u_3\right) \nonumber \\
& + \frac{\left(108 u_1^3+3 v u_5^2 u_1^2+18 u_3 u_1^2-36 \sqrt{6}
   u_2^2 \lambda  u_1^2+24 \Omega_r u_1-3 u_2^2 v u_5^2-3 v u_5^2-18 u_2^2 u_3+18 u_3+v u_5^2 \Omega_r+6 u_3 \Omega_r\right) \alpha }{2
   u_5^2}+O\left(\alpha ^2\right). \label{dinsysfalphap}
\end{align}
\end{subequations}
\end{small}
The system \eqref{singular_ds} is a singular perturbation system; specifically, it is a slow-fast system of first-order ordinary differential equations, where $v$ is the fast variable \cite{slow-fast1, slow-fast2, slow-fast3, slow-fast4, Paliathanasis:2015cza}.

The system can be written symbolically as 
\begin{align}
\frac{d \mathbf{U}}{d N}=\frac{\mathbf{F}(\mathbf{U})}{\alpha}+\mathbf{S}(\mathbf{U})+ \mathcal{O}(\alpha), \label{slow-fast-symbolic}
\end{align}
where we have defined the state vector $\mathbf{U}=\left(u_1,u_2, u_3, \Omega_r, \lambda, u_5, v\right)^T$, the fast part of the system
\begin{equation}
\mathbf{F}(\mathbf{U})=\left(
\begin{array}{c}
0\\
0\\
0\\
0\\
0\\
0\\
\sqrt{6} \lambda  u_2^2-6 u_1+2 u_3
 \end{array}\right),
\end{equation}
and the slow part of the system
\begin{equation}
\mathbf{S}(\mathbf{U})=\left(\begin{array}{c}
 \frac{1}{2} \left(3 u_1^3+\left(-3 u_2^2+\Omega_r+3\right) u_1-2 u_3\right)\\
\frac{1}{2} u_2 \left(3 u_1^2-\sqrt{6} \lambda  u_1-3 u_2^2+\Omega_r+3\right)\\
 \frac{v u_5^2}{2}+u_3
   \left(3 u_1^2-3 u_2^2+\Omega_r\right)\\
 \Omega_r \left(3 u_1^2-3 u_2^2+\Omega_r-1\right)\\
-\sqrt{6} f(\lambda) u_1\\
\frac{1}{2} u_5 \left(3 u_1^2-3 u_2^2+\Omega_{r}+3\right)\\
0\end{array}\right).
\end{equation}

Under the time rescaling $\tilde{N}=\alpha N$, the dynamical system \eqref{singular_ds} becomes
\begin{equation}
\frac{d \mathbf{U}}{d \tilde{N}}=\mathbf{F}{(\mathbf{U})}.
\end{equation}
Tikhonov's theorem \cite{slow-fast1, slow-fast4} can be used to investigate the fixed points of the system  \eqref{singular_ds} in the slow manifold defined by
\begin{equation}
    \sqrt{6} \lambda  u_2^2-6 u_1+2 u_3=0, \quad v=y_{2 c}= \text{constant}.
\end{equation}
In terms of the scalar field and its derivatives, a solution on the slow manifold satisfies:
\begin{align}
    & \sqrt{6} \lambda u_2^2 - 6 u_1 + 2 u_3 = 0 \quad \implies \quad \ddot{\phi} + 3 H \dot{\phi} + V'(\phi) = 0 \quad \text{(Klein--Gordon equation)}, \label{EQ.(43)} \\
    & v = y_{2c} \quad \implies \quad 
    \dddot{\phi} + 3 H \ddot{\phi} = -\sqrt{\frac{3}{2}} y_{2c} H \quad \implies \quad
    \phi(t) = \int_0^t \int_0^{\xi} \left( \frac{c_1}{a(\zeta)^3} - \frac{y_{2c}}{\sqrt{6}} \right) d\zeta \, d\xi + c_3 t + c_2.
\end{align}

This solution is referred to as the \emph{outer solution} of the singularly perturbed system.

Let
\[
g(\zeta) = \frac{c_1}{a(\zeta)^3} - \frac{y_{2c}}{\sqrt{6}}, \quad \text{and} \quad r(t - \zeta) = (t - \zeta) \cdot \theta(t - \zeta),
\]
where $ \theta(t) $ denotes the Heaviside step function.

Then, the expression
\[
\phi(t) = \int_0^t \int_0^{\xi} g(\zeta) \, d\zeta \, d\xi + c_3 t + c_2
\]
can be rewritten as a convolution product:
\[
\phi(t) = (g * r)(t) + c_3 t + c_2 = \int_0^t (t - \zeta) \, g(\zeta) \, d\zeta + c_3 t + c_2.
\]

In the limit $\alpha=0$, the dynamical system is equivalent to the algebraic dynamical system 
\begin{align}
\frac{d \mathbf{U}}{d N}=\mathbf{S}(\mathbf{U}),\\
\mathbf{F}(\mathbf{U})=0,
\end{align}
i.e., 
\begin{subequations}
\label{outer-solution}
\begin{align}
& \left(\begin{array}{c}
\frac{d u_1}{d N}  \\
\frac{d u_2}{d N}  \\
\frac{d u_3}{d N} \\
\frac{d \Omega_r}{d N}\\ 
\frac{d \lambda}{d N} \\
\frac{d u_5}{d N} \\
\frac{d v}{d N}
\end{array}\right)=\left(\begin{array}{c}
 \frac{1}{2} \left(3 u_1^3+\left(-3 u_2^2+\Omega_r+3\right) u_1-2 u_3\right)\\
\frac{1}{2} u_2 \left(3 u_1^2-\sqrt{6} \lambda  u_1-3 u_2^2+\Omega_r+3\right)\\
 \frac{v u_5^2}{2}+u_3
   \left(3 u_1^2-3 u_2^2+\Omega_r\right)\\
 \Omega_r \left(3 u_1^2-3 u_2^2+\Omega_r-1\right)\\
-\sqrt{6} f(\lambda) u_1\\
\frac{1}{2} u_5 \left(3 u_1^2-3 u_2^2+\Omega_{r}+3\right)\\
0\end{array}\right),\\
& \sqrt{6} \lambda  u_2^2-6 u_1+2 u_3=0.
\end{align}
\end{subequations}
 The last gives the slow manifold $$\sqrt{6} \lambda  u_2^2-6 u_1+2 u_3=0, \quad v= \text{constant}.$$

The phase space of \eqref{outer-solution} is given by \begin{align}
\left\{\left(u_1,u_2, u_3, \Omega_r, \lambda, u_5, v\right)\in\mathbb{R}^7: \sqrt{6} \lambda  u_2^2-6 u_1+2 u_3=0, \quad v= \text{constant}, \quad u_1^2+u_2^2+\Omega_r\leq 1\right\},
\end{align}
and physical values are given by
\begin{equation}
    \Omega_{DE} =u_1^2+u_2^2, \ \ \ w_{DE} = \dfrac{u_1^2-u_2^2}{u_1^2+u_2^2}, \ \ \ w_{eff} = u_1^2-u_2^2 + \dfrac{\Omega_r}{3}.
\end{equation}

\subsection{Equilibrium points on the slow manifold}
\label{slow-dynamics}
The dynamics on the slow manifold are given by
\begin{equation}
  \left(\begin{array}{c}
\frac{d u_1}{d N}  \\
\frac{d u_2}{d N}  \\
\frac{d \Omega_r}{d N}\\ 
\frac{d \lambda}{d N}
\end{array}\right)=  \left(
\begin{array}{c}
 \frac{1}{2} \left(3 u_1^3+u_1 \left(-3 u_2^2+\Omega_r-3\right)+\sqrt{6} \lambda  u_2^2-6 u_1\right) \\
 \frac{1}{2} u_2 \left(3 u_1^2-\sqrt{6} \lambda  u_1-3 u_2^2+\Omega_r+3\right) \\
 \Omega_r \left(3 u_1^2-3 u_2^2+\Omega_r-1\right) \\
- \sqrt{6} u_1 f(\lambda )
\end{array}
\right), \label{dinsysalpha0}
\end{equation}
defined on the phase space
\begin{align}
\left\{\left(u_1,u_2,  \Omega_r, \lambda\right)\in\mathbb{R}^4:  u_1^2+u_2^2+\Omega_r\leq 1, \lambda \in \mathbb{R}, u_2\geq 0\right\}. 
\end{align}
The system \eqref{dinsysalpha0} admits the (lines of) equilibrium points 
\begin{enumerate}
    \item $(u_1, u_2, \Omega_r, \lambda)=(0, 0, 0, \lambda_c)$, with eigenvalues $\left\{-\frac{3}{2},\frac{3}{2},-1,0\right\}$. This is a saddle-point solution that is dominated by matter. 
    
    \item $(u_1, u_2, \Omega_r, \lambda)=(0,  0,  1, \lambda_c)$, with eigenvalues $ \{2,-1,1,0\}$. This is a saddle-point solution that is dominated by radiation. 
    
    \item $(u_1, u_2, \Omega_r, \lambda)=\left(-1,  0,  0, \tilde{\lambda }\right)$, with eigenvalues $\left\{3,2,\sqrt{\frac{3}{2}} \tilde{\lambda }+3,\sqrt{6} f'\left(\tilde{\lambda }\right)\right\}$. This point represents a kinetic-dominated solution. That is a source provided $\lambda > -\sqrt{6}$ and $f'\left(\tilde{\lambda }\right) > 0$. It is nohyperbolic when $\lambda=-\sqrt{6}$ or when $f'\left(\tilde{\lambda }\right)=0$. It is a saddle otherwise. 
    
    \item $(u_1, u_2, \Omega_r, \lambda)=\left(1, 0,  0,  \tilde{\lambda }\right)$, with eigenvalues $\left\{3,2,3-\sqrt{\frac{3}{2}} \tilde{\lambda },-\sqrt{6} f'\left(\tilde{\lambda }\right)\right\}$. This point represents a kinetic-dominated solution. That is a source provided $\lambda<\sqrt{6}$ and $f'\left(\tilde{\lambda }\right)<0$. It is nohyperbolic when $\lambda=\sqrt{6}$ or when $f'\left(\tilde{\lambda }\right)=0$. It is a saddle otherwise. 
    
   \item $(u_1, u_2, \Omega_r, \lambda)=\left(\frac{\sqrt{\frac{3}{2}}}{\tilde{\lambda }}, \frac{\sqrt{\frac{3}{2}}}{|\tilde{\lambda }|}, 0, \tilde{\lambda }\right)$, with eigenvalues $ \left\{-1,-\frac{3 \sqrt{24-7 \tilde{\lambda
   }^2}}{4 \tilde{\lambda }}-\frac{3}{4},\frac{3 \sqrt{24-7 \tilde{\lambda }^2}}{4 \tilde{\lambda }}-\frac{3}{4}, -\frac{3 f'\left(\tilde{\lambda }\right)}{\tilde{\lambda }}\right\}$. It is a scalar field-matter scaling solution that it is sink for $f'(\tilde{\lambda })<0, \tilde{\lambda } <-\sqrt{3}$ or $f'(\tilde{\lambda })>0, \tilde{\lambda } >\sqrt{3}$. Is is nonhyperbolic provided $f'(\tilde{\lambda })=0$, or  $\tilde{\lambda }=-\sqrt{3}$ or $\tilde{\lambda } =\sqrt{3}$. 
  
   \item $(u_1, u_2, \Omega_r, \lambda)=\left(\frac{\tilde{\lambda }}{\sqrt{6}},  \sqrt{1-\frac{\tilde{\lambda }^2}{6}},  0,  \tilde{\lambda }\right)$, with eigenvalues $ \left\{\frac{1}{2} \left(\tilde{\lambda
   }^2-6\right),\tilde{\lambda }^2-4,\tilde{\lambda }^2-3,-\tilde{\lambda } f'\left(\tilde{\lambda }\right)\right\}$.

   \item $(u_1, u_2, \Omega_r, \lambda)=(0,  1,  0,  0)$, with eigenvalues $\left\{-4,-3,-\frac{1}{2} \sqrt{9-12 f(0)}-\frac{3}{2},\frac{1}{2} \sqrt{9-12 f(0)}-\frac{3}{2}\right\}$. If $f(0)>0$, it is a sink. If $f(0)<0$ is a saddle and when $f(0)=0$ is nonhyperbolic.

   \item $(u_1, u_2, \Omega_r, \lambda)=\left(\frac{2 \sqrt{\frac{2}{3}}}{\tilde{\lambda }},  \frac{2}{\sqrt{3} |\tilde{\lambda }|},  1-\frac{4}{\tilde{\lambda }^2},  \tilde{\lambda }\right)$, with eigenvalues $
   \left\{1,-\frac{\sqrt{64-15 \tilde{\lambda }^2}}{2 \tilde{\lambda }}-\frac{1}{2},\frac{\sqrt{64-15 \tilde{\lambda }^2}}{2 \tilde{\lambda }}-\frac{1}{2},-\frac{4 f'\left(\tilde{\lambda }\right)}{\tilde{\lambda }}\right\}$. It is nonhyperbolic for $f'(\tilde{\lambda })<0, \tilde{\lambda }\in\{-2, 2\}$, or $f'(\lambda )=0,  \lambda \neq 0$, or $f'(\tilde{\lambda })>0, \tilde{\lambda }\in\{-2, 2\}$. They are saddle otherwise. 
\end{enumerate}
where $\tilde{\lambda }$ denotes the zeroes of $f(\lambda)$ (see the discussion of $f$-devisers in section \ref{f-s-devisers}).
\subsection{Stability of de Sitter solution through center manifold}

The de Sitter solution is important in cosmology, especially in models that describe the universe’s accelerated expansion. Studying its stability helps clarify how different gravitational theories behave at late times. When standard linear methods are not enough—due to non-hyperbolic fixed points—the center manifold approach becomes useful. This work applies that method to analyze the stability of the de Sitter solution and its implications for cosmological dynamics.

\subsubsection{Theorem (Reduced Dynamics and Stability via Center Manifold)}

Consider again the nonlinear autonomous system near a fixed point:
\begin{align}
    \frac{dy}{dN} &= Ay + f(y,z), \tag{2.34a} \\
    \frac{dz}{dN} &= Bz + g(y,z), \tag{2.34b}
\end{align}
where:
\begin{itemize}
    \item $ y \in \mathbb{R}^c $, $ z \in \mathbb{R}^s $,
    \item $ A $ has eigenvalues with zero real part (center directions),
    \item $ B $ has eigenvalues with negative real part (stable directions),
    \item $ f $ and $ g $ are $ \mathcal{C}^r $ with $ f(0,0) = g(0,0) = 0 $, $ Df(0,0) = Dg(0,0) = 0 $.
\end{itemize}

Then the following results hold:

\begin{thm}[Existence]\label{thm:existence_CM}
There exists a $ \mathcal{C}^r $ center manifold $ W^c $ locally represented as $ z = h(y) $, with:
\[
h(0) = 0, \quad Dh(0) = 0,
\]
such that the dynamics restricted to $ W^c $ are governed by the reduced system:
\begin{equation}
\frac{dy}{dN} = Ay + f(y,h(y)). \tag{2.36}
\end{equation}
\end{thm}

\begin{thm}[Stability]\label{thm:stability_CM}
Suppose the origin $ y = 0 $ is stable (respectively asymptotically stable or unstable) for the reduced system~\textup{(2.36)}. Then the origin $ (y,z) = (0,0) $ of the full system~\textup{(2.34a)--(2.34b)} is also stable (respectively asymptotically stable or unstable).

Furthermore, if $ (y(N), z(N)) $ is a solution of the full system with sufficiently small initial data, then there exists a solution $ y(N) $ of~\textup{(2.36)} such that, as $ N \to \infty $,
\[
y(N) = y_{\text{cm}}(N) + \mathcal{O}(e^{-rN}), \quad
z(N) = h(y_{\text{cm}}(N)) + \mathcal{O}(e^{-rN}),
\]
where $ y_{\text{cm}}(N) $ solves~\textup{(2.36)} and $ r > 0 $ is a constant.
\end{thm}

\subsubsection{Application to Gradient-Like Evolution}
Analysing the case of the de Sitter solution $dS$ (with $f(0)=0$) in detail, note that the eigenvalues are $-4, -3, -3, 0$, i.e., non-hyperbolic. 
We first introduced the linear transformation 
\begin{equation}
x_1=\Omega_{r}, \quad x_2= u_2+\frac{\Omega_{r}}{2}-1, \quad x_3= u_1-\frac{\lambda
   }{\sqrt{6}}, \quad y=\lambda 
\end{equation}
that translates the equilibrium point to the origin and produces the canonical Jordan decomposition of the linear part. 
The full system is expressed as: 

\begin{equation}
\left(\begin{array}{c}
\frac{d x_1}{d N} \\ \\ \\ \\  \hdashline
\frac{d x_2}{d N}  \\\\ \\ \\ \\  \hdashline
\frac{d x_3}{d N}\\ \\ \\ \\   
\\ \hdashline
\frac{d y}{d N} \\ 
\end{array}\right)=\left(
\begin{array}{c}
 \frac{1}{4} x_1 \Bigg(-3 x_1^2+4
   x_1 (3 x_2+4)\\+2 \left(-6 x_2
   (x_2+2)+6 x_3^2+2 \sqrt{6}
   x_3 y+y^2-8\right)\Bigg) \\ \hdashline 
 \frac{1}{16} \Bigg(-3 x_1^3+2 x_1^2
   (3 x_2+5)+2 x_1 \left(2 x_2
   (3 x_2+8)+6 x_3^2+4 \sqrt{6}
   x_3 y+3 y^2\right)\\ -4 (x_2+1)
   \left(6 x_2 (x_2+2)-6
   x_3^2+y^2\right)\Bigg) \\ \hdashline
 \frac{1}{48} \Bigg(6 x_3 \left(8 f(y)-3
   x_1^2+4 x_1 (3 x_2+4)+6
   \left(-2 x_2
   (x_2+2)+y^2-4\right)\right)\\ +\sqrt{6} y
   \left(8 f(y)+3 x_1^2-4 x_1 (3
   x_2+2)+2 \left(6 x_2
   (x_2+2)+y^2\right)\right)+72
   x_3^3+36 \sqrt{6} x_3^2 y\Bigg)
   \\  \hdashline
 -f(y) \left(\sqrt{6}
   x_3+y\right) \\\\
\end{array}
\right).
\end{equation}

Using the Centre Manifold theorem, we obtain that the graph locally gives the centre manifold of the origin
\begin{align}
& \Big\{(x_1, x_2, x_3, y)\in [0,1] \times [-1,5/2] \times \mathbb{R} \times \mathbb{R}: x_1=h_1(y), x_2=h_2(y), x_3=h_3(y), \\
& h_1(0)=0, h_2(0)=0, h_3(0)=0,  h_1'(0)=0, h_2'(0)=0, h_3'(0)=0, |y|<\delta\Big\},
\end{align}
for a small enough $\delta$, where the functions $h_1$, $h_2$ and $h_3$ satisfy the differential equations:

\begin{align}
& 0=f(y) \left(\sqrt{6} h_3(y)+y\right)
   h_1'(y) \nonumber\\
   & +\frac{1}{2} h_1(y) \left(-6
   h_2(y) (h_2(y)+2)+2 \sqrt{6} y
   h_3(y)+6
   h_3(y)^2+y^2-8\right) \nonumber\\
   & +h_1(y)^2 (3
   h_2(y)+4)-\frac{3}{4} h_1(y)^3,\\
& 0= f(y)\left(\sqrt{6} h_3(y)+y\right)
   h_2'(y) \nonumber\\ 
   &+\frac{1}{16} \Bigg(2 h_1(y)
   \left(2 h_2(y) (3 h_2(y)+8)+4
   \sqrt{6} y h_3(y)+6 h_3(y)^2+3
   y^2\right)\nonumber\\
   & +2 h_1(y)^2 (3
   h_2(y)+5)-3 h_1(y)^3-4
   (h_2(y)+1) \left(6 h_2(y)
   (h_2(y)+2)-6
   h_3(y)^2+y^2\right)\Bigg),\\
   &0=f(y) \left(\sqrt{6}
   h_3(y)+y\right) h_3'(y) \nonumber\\
   & +\frac{1}{48}\Bigg(6 h_3(y) \left(8 f(y)+4
   h_1(y) (3 h_2(y)+4)-3
   h_1(y)^2+6 \left(-2 h_2(y)
   (h_2(y)+2)+y^2-4\right)\right)\nonumber\\
   & +\sqrt{6} y
   \left(8 f(y)-4 h_1(y) (3
   h_2(y)+2)+3 h_1(y)^2+2 \left(6
   h_2(y)
   (h_2(y)+2)+y^2\right)\right) \nonumber\\
   & +72
   h_3(y)^3+36 \sqrt{6} y
   h_3(y)^2\Bigg).
\end{align}

We propose the center manifold coordinates in the form
\begin{align}
    h_1(y) &= a_0 y^2 + a_1 y^3 + a_2 y^4 + a_3 y^5 + \mathcal{O}(y^6), \\
    h_2(y) &= b_0 y^2 + b_1 y^3 + b_2 y^4 + b_3 y^5 + \mathcal{O}(y^6), \\
    h_3(y) &= c_0 y^2 + c_1 y^3 + c_2 y^4 + c_3 y^5 + \mathcal{O}(y^6).
\end{align}
Assuming $ f(0) = 0 $ and imposing the required vanishing of nonlinear terms, the coefficients are:
\begin{align}
    & a_0 = a_1 = a_2 = a_3 = 0, \\
    & b_0 = -\frac{1}{12}, \quad b_1 = -\frac{1}{18} f'(0), \\
    & b_2 = \frac{1}{864} \left(-24 f''(0) - 56 f'(0)^2 - 3\right), \\
    & b_3 = \frac{1}{648} \left(-6 f^{(3)}(0) - f'(0)(48 f''(0) + 68 f'(0)^2 + 9)\right), \\
    & c_0 = \frac{f'(0)}{3\sqrt{6}}, \quad c_1 = \frac{f''(0) + 2f'(0)^2}{6\sqrt{6}}, \\
    & c_2 = \frac{3f^{(3)}(0) + 28f'(0)^3 + 3f'(0)(7f''(0) + 1)}{54\sqrt{6}}, \\
    & c_3 = \frac{680f'(0)^4 + 9(f^{(4)}(0) + 8f''(0)^2 + 2f''(0)) + 96f^{(3)}(0)f'(0) + 24f'(0)^2(27f''(0) + 5)}{648\sqrt{6}}.
\end{align}

The evolution equation on the center manifold takes the gradient-like form:
\begin{equation}
    \frac{dy}{dN} = -\nabla U(y), \label{gradient}
\end{equation}
with potential
\begin{align}
    U(y) &= \frac{1}{3} f'(0) y^3 + \frac{1}{24} \left(2f'(0)^2 + 3f''(0)\right) y^4 \nonumber \\
         &\quad + \frac{1}{30} \left(2f'(0)\left(f'(0)^2 + f''(0)\right) + f^{(3)}(0)\right) y^5 \nonumber \\
         &\quad + \frac{1}{1296} \left[2\left(56f'(0)^4 + (60f''(0)+6)f'(0)^2 + 12f^{(3)}(0)f'(0) + 9f''(0)^2\right) + 9f^{(4)}(0)\right] y^6 + \mathcal{O}(y^7).\label{expandV}
\end{align}

\subsubsection{Interpretation as Energy Descent}

The system evolves toward minima of $U(y)$. The critical points $ y_c $ satisfying $ \nabla U(y_c) = 0 $ correspond to fixed points, and their nature determines stability:

\begin{itemize}
    \item \textbf{Local minimum}: stable, with positive-definite Hessian $ \nabla^2 U $.
    \item \textbf{Local maximum}: unstable, repelling nearby trajectories.
    \item \textbf{Saddle point}: partially stable; depends on direction of perturbations.
\end{itemize}
Table \ref{lowest-nonzero-derivative} presents the stability classification of the origin based on the lowest nonzero derivative of the potential $ U(y) $.

\begin{table}[h!]
\centering
\caption{Stability classification of the origin based on the lowest nonzero derivative of the potential $ U(y) $.}
\label{lowest-nonzero-derivative}
\renewcommand{\arraystretch}{1.5}
\begin{tabularx}{\textwidth}{@{}lXX@{}}
\toprule
\textbf{Conditions on $ f^{(n)}(0) $} & \textbf{Leading Term in $ U(y) $} & \textbf{Stability at $ y=0 $} \\
\midrule
$ f'(0) \neq 0 $ & $ y^3 $ &  Unstable \\
$ f'(0) = 0,\ f''(0) > 0 $ & $ y^4 $ &  Stable \\
$ f'(0) = 0,\ f''(0) < 0 $ & $ y^4 $ &  Unstable \\
$ f^{(n)}(0)=0\ \text{for}\ n<3,\ f^{(3)}(0) \neq 0 $ & $ y^5 $ & Depends on sign;  \\ && likely unstable due to absence of true minimum \\
$ f^{(n)}(0)=0\ \text{up to even } n $ & $ y^6 $ or higher &  Stable if leading coefficient is positive \\
\bottomrule
\end{tabularx}
\end{table}

\subsubsection{Local Stability at the Origin}

A case-by-case classification of stability at the origin $ y = 0 $ based on the lowest non-zero derivative of $ U(y) $. These cases reflect the classical theory of degenerate extrema:
\begin{itemize}
    \item Odd-order nonzero derivative: origin cannot be a local extremum; typically a saddle.
    \item Even-order positive coefficient: origin is a flat minimum; stability persists, but convergence may be slow.
    \item Even-order negative: origin behaves as a flat maximum; repulsive and unstable.
\end{itemize}

This framework enables precise classification of critical behavior using only truncated expansions and sign analysis.

\subsection{Application 1: Exponential Potential with a Cosmological Constant and the de Sitter Solution}

We consider a scalar field model with an exponential potential and additive cosmological constant:
\begin{equation}
    V(\phi) = V_0 e^{-\beta \phi} + \Lambda. \label{exp}
\end{equation}
The corresponding auxiliary function is
\begin{equation}
    f(\lambda) = -\lambda(\lambda - \beta),
\end{equation}
which leads to nontrivial structure in the dynamics of the slow manifold.

The resulting equations are 
\begin{equation}
\frac{d}{dN}
\begin{pmatrix}
u_1 \\
u_2 \\
\Omega_r \\
\lambda
\end{pmatrix}
=
\begin{pmatrix}
\frac{1}{2}\left(3u_1^3 + u_1(\Omega_r - 3u_2^2 - 3) + \sqrt{6}\lambda u_2^2 - 6u_1\right) \\
\frac{1}{2}u_2\left(3u_1^2 - \sqrt{6}\lambda u_1 - 3u_2^2 + \Omega_r + 3\right) \\
\Omega_r\left(3u_1^2 - 3u_2^2 + \Omega_r - 1\right) \\
\sqrt{6}u_1 \cdot \lambda(\lambda - \beta)
\end{pmatrix}.
\label{systemEXP}
\end{equation}

\subsubsection{Critical Points and Cosmological Parameters: Discussion}

\begin{itemize}
    \item\textbf{$ a_M $}: This point corresponds to a matter-dominated solution characterized by $ \Omega_m = 1 $ and vanishing curvature ($ \Omega_{rc} = 0 $). Despite all dynamical variables being zero, the energy budget is entirely matter-driven. The parameter $ \lambda $ remains arbitrary, indicating degeneracy in the scalar field sector at this configuration.

    \item \textbf{$ b_R $}: Describes a regime with pure curvature dominance ($ \Omega_{rc} = 1 $) and no scalar field dynamics. As with $ a_M $, $ \lambda $ is arbitrary, indicating its role as a passive parameter in this configuration.

    \item \textbf{$ c $}: Characterized by non-zero $ u_2 $ and zero curvature and scalar field contributions. The fixed value $ \lambda_c = 0 $ suggests a potential extrema or stationary configuration, with an unconstrained $ u_{5c} $.

    \item \textbf{$ d^\pm $}: These represent kinetic-dominated solutions with $ u_{1c} = \pm 1 $ and all other variables zero. Such configurations typically correspond to stiff fluid behavior and are often unstable unless fine-tuned.

    \item \textbf{$ e^\pm $}: Similar to $ d^\pm $, but here the slope $ \lambda $ is fixed at $ \beta $, coupling the kinetic structure with the potential’s steepness. These may correspond to scaling solutions under appropriate conditions.

    \item \textbf{$ f_M $}: This point enforces a specific relationship between $ u_1 $ and $ u_2 $ through $ \beta $, consistent with matter-like behavior. The root expressions suggest a cosmologically relevant attractor when $ \beta $ is real and nonzero.

    \item \textbf{$ g $}: Describes a canonical scalar field configuration with $ u_1 \propto \beta $ and $ u_2 $ constrained by a unit-norm condition. Viability demands $ \beta^2 \leq 6 $ to keep $ u_2 $ real, delimiting the parameter space.

    \item \textbf{$ h_R $}: Corresponds to a radiation-like solution modified by the potential slope $ \beta $. Positive curvature ($ \Omega_{rc} > 0 $) appears when $ \beta^2 > 4 $, with stability conditions depending on the full Jacobian spectrum.
\end{itemize}

\begin{table}[h!]
\centering
\caption{Critical points on the slow manifold for $ f(\lambda) = -\lambda(\lambda - \beta) $.}
\begin{tabular}{c c c c c c}
\toprule
Name & $ u_{1c} $ & $ u_{2c} $ & $ \Omega_{rc} $ & $ \lambda_c $ & $ u_{5c} $ \\
\midrule
$ a_M $ & $0 $& $0$ & $0$ & arbitrary & $0$ \\
$ b_R $ & $0$ & $0$ & $1$ & arbitrary & $0$ \\
$ c $ & 0 & $1$ & $0$ & $0$ & arbitrary \\
$ d^\pm $ & $ \pm 1 $ & $0$ & $0$ &$ 0$ & $0$ \\
$ e^\pm $ & $ \pm 1 $ & $0$ &$ 0$ & $ \beta $ & $0 $\\
$ f_M $ & $ \sqrt{\frac{3}{2}}\frac{1}{\beta} $ & $ \sqrt{\frac{3}{2}}\frac{1}{|\beta|} $ & 0 & $ \beta $ & 0 \\
$ g $ & $ \frac{\beta}{\sqrt{6}} $ & $ \sqrt{1 - \frac{\beta^2}{6}} $ & 0 & $ \beta $ & 0 \\
$ h_R $ & $ \sqrt{\frac{8}{3}} \frac{1}{\beta} $ & $ \frac{2}{\sqrt{3}|\beta|} $ & $ 1 - \frac{4}{\beta^2} $ & $ \beta $ & 0 \\
\bottomrule
\end{tabular}
\label{table-exp-critical}
\end{table}

\subsubsection{Linear Stability Analysis}

The stability characteristics of the critical points presented in Table~\ref{table-exp-critical} are:

\begin{itemize}
    \item \textbf{$ a_M $}: eigenvalues: $\left\{-\frac{3}{2},\frac{3}{2},-1,0\right\}$, non-hyperbolic for all values of $\beta$, behaves as a saddle point;
    \item \textbf{$ b_R $}: eigenvalues: $\left\{2,-1,2,0\right\}$, non-hyperbolic for all values of $\beta$, behaves as a saddle point;
    \item \textbf{$ c $}: eigenvalues: $\left\{-4,-3,-3,0\right\}$, non-hyperbolic for all values of $\beta$;
    \item \textbf{$ d^{+} $}: eigenvalues: $\left\{3,2,3,-\sqrt{6}\beta\right\}$, a source for $\beta<0$, non-hyperbolic for $\beta=0$, otherwise a saddle;
    \item \textbf{$ e^{+} $}: eigenvalues: $\left\{3,2,3-\sqrt{\frac{3}{2}}\beta,\sqrt{6}\beta\right\}$, a source for $0<\beta<\sqrt{6}$, non-hyperbolic for $\beta\in\{0,\sqrt{6}\}$, otherwise a saddle;
    \item \textbf{$ f_M $}: eigenvalues: $\left\{ -1,-\frac{3}{4}-\frac{3\sqrt{24-7\beta^2}}{4\beta}, -\frac{3}{4}+\frac{3\sqrt{24-7\beta^2}}{4\beta},3\right\}$, non-hyperbolic for $\beta\in\{-\sqrt{3},\sqrt{3}\}$, otherwise a saddle;
    \item \textbf{$ g $}: eigenvalues: $\left\{ \frac{1}{2}(\beta^2-6),\beta^2-4,\beta^2-3,\beta^2 \right\}$, a source for $\beta<-\sqrt{6}$ or $\beta>\sqrt{6}$, non-hyperbolic for $\beta\in\{-\sqrt{6},-2,-\sqrt{3},0,\sqrt{3},2,\sqrt{6}\}$, otherwise a saddle;
    \item \textbf{$ h_R $}: eigenvalues: $\left\{1,-\frac{1}{2}-\frac{\sqrt{64-15\beta^2}}{2\beta},-\frac{1}{2}+\frac{\sqrt{64-15\beta^2}}{2\beta}, 4\right\}$, non-hyperbolic for $\beta\in\{-2,2\}$, otherwise a saddle.
\end{itemize}

 Summarizing, the eigenvalue structure of the critical points shows the following general features:

\begin{itemize}
    \item Points $ a_M $, $ b_R $, and $ c $ are always non-hyperbolic with one vanishing eigenvalue. They act generically as saddle-type solutions, and $c$ requires center manifold analysis in at least one direction.
    
    \item Points $ d^\pm $ and $ e^\pm $ exhibit bifurcation behavior: their stability transitions between source and saddle depending on the magnitude and sign of $ \beta $. The transition thresholds correspond to algebraic expressions involving $ \beta $, such as $ \beta = 0 $, $ \sqrt{6} $, or $ \sqrt{\tfrac{3}{2}} $.
    
    \item Point $ f_M $ undergoes bifurcation at $ \beta = \pm \sqrt{3} $, dictated by the sign of the square-root terms in its eigenvalues.
    
    \item Point $ g $ depends polynomially on $ \beta^2 $, leading to symmetric bifurcation domains around critical values like $ \beta^2 = 6 $; it becomes a source only outside those thresholds.
    
    \item Point $ h_R $ features eigenvalues with rational dependencies on $ \beta $ and square-root bifurcations at $ \beta = \pm 2 $. Outside these bifurcation zones, it behaves as a saddle.
    
    \item All critical points involve at least one eigenvalue that vanishes or crosses zero for specific values of $ \beta $, indicating structural instability or the need for nonlinear analysis via center manifold theory.
\end{itemize}

\subsubsection{Center Manifold Analysis of the de Sitter Point $ c $}

Point $ c $ corresponds to $ (u_1, u_2, \Omega_r, \lambda) = (0,1,0,0) $. To assess nonlinear stability, we consider the gradient-like evolution:
\begin{equation}
    \frac{dy}{dN} = -\nabla U(y), \quad \text{where} \quad f(\lambda) = -\lambda(\lambda - \beta).
\end{equation}

Evaluating derivatives:
\[
f(0) = 0, \quad f'(0) = \beta, \quad f''(0) = -2, \quad f^{(n)}(0) = 0 \quad \text{for } n \geq 3.
\]

Expanding the potential given by \eqref{expandV}:
\begin{align}
U(y) &= \frac{1}{3} \beta y^3 + \left( \frac{1}{12} \beta^2 - \frac{1}{4} \right) y^4 + \left( \frac{1}{15} \beta^3 - \frac{2}{15} \beta \right) y^5 \nonumber \\
&\quad + \left( \frac{7}{81} \beta^4 - \frac{1}{12} \beta^2 + \frac{1}{36} \right) y^6 + \mathcal{O}(y^7).
\end{align}

Table \ref{table-cm-exp} summarizes the Center manifold classification of the de Sitter point $ c $ for the exponential model.

\begin{table}[h!]
\centering
\caption{Center manifold classification of the de Sitter point $ c $ for the exponential model.}
\label{table-cm-exp}
\renewcommand{\arraystretch}{1.3}
\begin{tabularx}{\textwidth}{@{}Xll@{}}
\toprule
\textbf{Condition on $ \beta $} & \textbf{Leading Term in $ U(y) $} & \textbf{Stability at $ y=0 $} \\
\midrule
$ \beta \neq 0 $ & $ y^3 $ (inflection point) &  Unstable \\
$ \beta = 0 $ & $ y^4 $ (degenerate maximum) & Unstable \\
\bottomrule
\end{tabularx}
\end{table}

The de Sitter solution $ c $ is non-hyperbolic for all values of $ \beta $. The Center Manifold Theorem ensures that its stability is determined by the potential $ U(y) $. For any $ \beta $, the origin is either an inflection point (for $ \beta \neq 0 $) or a degenerate maximum (for $ \beta = 0 $), leading to divergence of nearby trajectories. Therefore, the exponential potential with a cosmological constant does not admit the de Sitter point $ c $ as a late-time attractor.

\subsection{Cosmological Interpretation of Critical Points}

Each fixed point in Table~\ref{table-exp-critical} corresponds to a cosmological regime governed by distinct scalar field dynamics. Based on the parameters listed in Table~\ref{table-cosmo-interpretation}, we summarize their physical interpretations below:

\begin{table}[h!]
\centering
\caption{Cosmological classification of critical points under an exponential potential with cosmological constant.}
\label{table-cosmo-interpretation}
\renewcommand{\arraystretch}{1.3}
\begin{tabularx}{\textwidth}{@{}ccccX@{}}
\toprule
\textbf{Point} & $ \Omega_{de} $ & $ \Omega_m $ & $ \omega_{eff} $ & \textbf{Interpretation} \\
\midrule
$ a_M $ & $0$ & $1$ & $0$ & Pure matter-dominated epoch. No scalar field activity; consistent with early or mid-universe behavior. Dynamically unstable. \\
$ b_R $ & $0$ & $0$ & $ \tfrac{1}{3} $ & Radiation-dominated expansion. Dark energy is latent (zero density), despite a vacuum-like equation of state. \\
$ c $ & $1$ & $0$ & $-1$ & True de Sitter geometry: characterized by full scalar field dominance and accelerated expansion. In this model, the condition $ f(0) = 0 $ is satisfied regardless of $ \beta $, since $ f(\lambda) = -\lambda(\lambda - \beta) $ implies $ f(0) = 0 $. Despite this, center manifold analysis reveals that the origin is always structurally unstable. For $ \beta \neq 0 $, the potential exhibits a cubic inflection; for $ \beta = 0 $, it reduces to a quartic degenerate maximum. In both cases, the de Sitter solution is not a viable late-time attractor.\\
$ d^\pm $, $ e^\pm $ & $1$ & $0$ & $1$ & Kinetic energy–dominated solutions. Correspond to stiff-fluid regimes unsuitable for cosmic acceleration. Typically, repellers or transient saddles. \\
$ f_M $ & $ \frac{3}{\beta^2} $ & $ 1 - \frac{3}{\beta^2} $ & 0 & Scaling solution involving both scalar field and matter. Expansion behaves matter-like, making it viable for intermediate epochs. Stability varies with $ \beta $. \\
$ g $ & 1 & 0 & $ \frac{\beta^2 - 3}{3} $ & Scalar field–dominated with adjustable pressure. Accelerated expansion possible for small $ \beta^2 $. Source or saddle, depending on the parameter range. \\
$ h_R $ & $ \frac{4}{\beta^2} $ & 0 & $ \tfrac{1}{3} $ & Scalar field radiative scaling. Models radiation-like behavior at early times, characterized by a nonzero scalar energy. Typically unstable. \\
\bottomrule
\end{tabularx}
\end{table}

\begin{itemize}
    \item \textbf{Exact de Sitter solution:} The only fixed point that realizes exact de Sitter expansion is point $ c $, characterized by $ \Omega_{de} = 1 $ and $ \omega_{\text{eff}} = -1 $. However, it is intrinsically unstable on the center manifold for all $ \beta $, as confirmed through gradient analysis.

    \item \textbf{Scalar field–driven acceleration:} Points $ f_M $ and $ g $ represent regimes with either mixed or pure scalar field dominance. Depending on the value of $ \beta $, they can approximate late-time acceleration. Nonetheless, they lack the necessary nonlinear stability to function as genuine cosmological attractors.

    \item \textbf{Transitional and early-time behavior:} Points $ a_M $, $ b_R $, $ d^\pm $, $ e^\pm $, and $ h_R $ correspond to matter, radiation, kinetic, or scaling-dominated phases. These are typically associated with early or intermediate cosmic evolution, but none of them support a viable long-term endpoint for the universe.
\end{itemize}

\noindent
In summary, despite the presence of a cosmological constant in the potential, the system does not support a dynamically stable de Sitter endpoint. This reinforces the importance of nonlinear stability criteria beyond equation-of-state diagnostics when evaluating scalar-field cosmologies.

\subsection{Exponential Potential Dynamics}
\label{sec:exp}

We consider the coupling function $ f(\lambda) = \beta \lambda - \lambda^2 $, where $ \beta $ is the exponential slope parameter. The dynamical behavior across a range of $ \beta $ values reveals how steepness influences trajectory confinement, vector field geometry, and scalar field evolution.

\begin{figure}[H]
\centering
\includegraphics[width=0.49\textwidth]{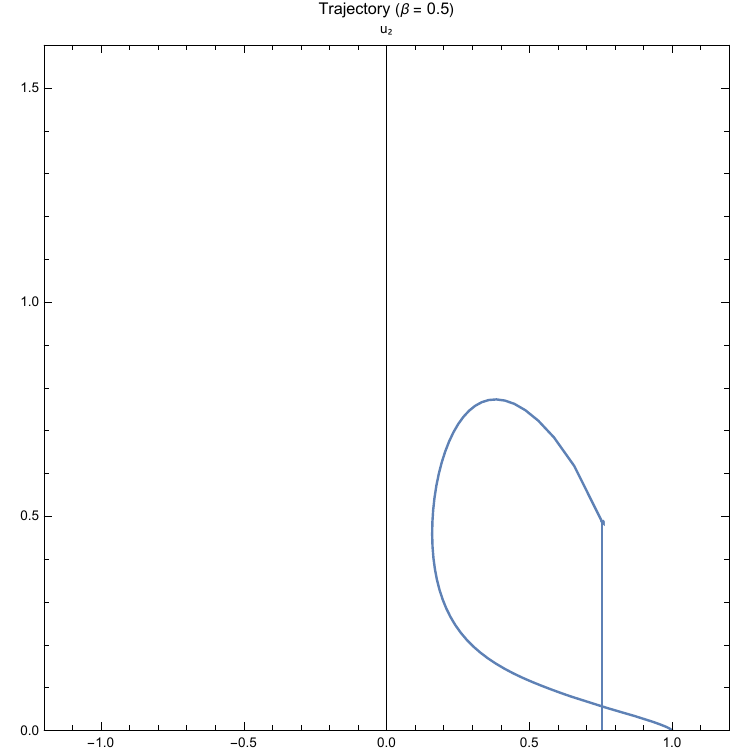}
\includegraphics[width=0.49\textwidth]{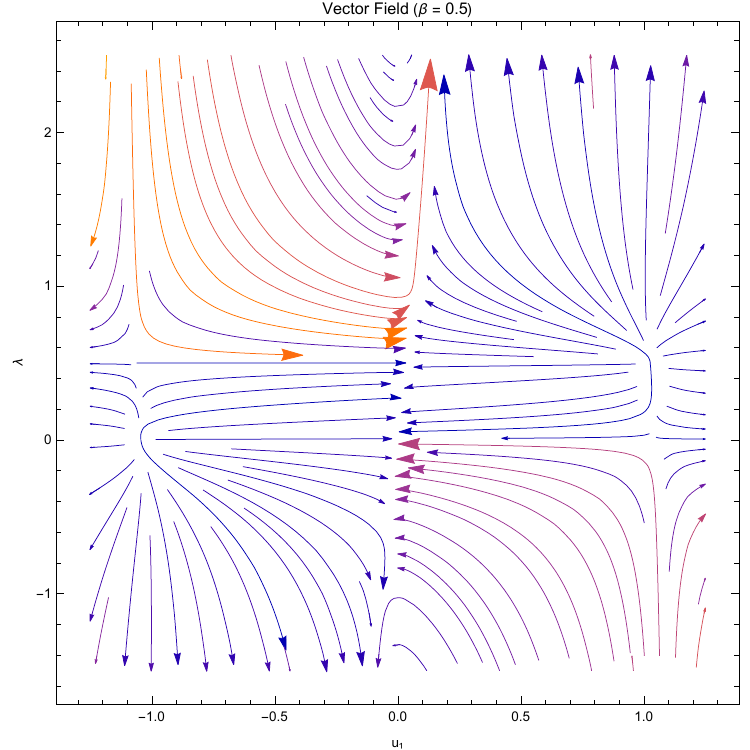}
\includegraphics[width=0.49\textwidth]{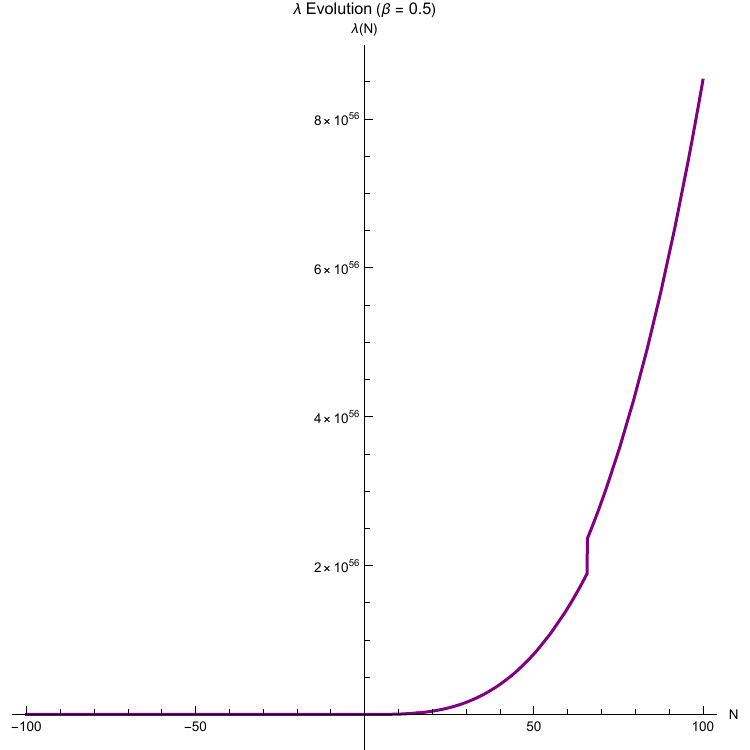}
\includegraphics[width=0.49\textwidth]{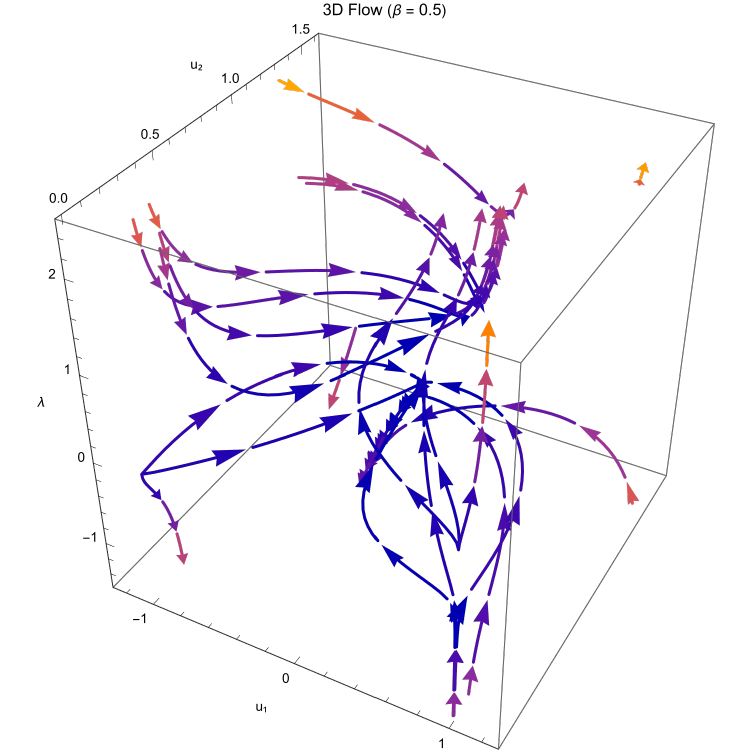}
\caption{
Dynamical behavior for exponential slope parameter $ \beta = 0.5 $, with fixed initial conditions $ u_1(0) = 0.2 $, $ u_2(0) = 0.3 $, $ \Omega_r(0) = 10^{-5} $, and $ \lambda(0) = 1.0 $. Panels show the trajectory in $(u_1, u_2)$, stream plot in $(u_1, \lambda)$, evolution of $\lambda(n)$, 3D stream plot in $(u_1, u_2, \lambda)$, and vector field in $(u_1, u_2, \Omega_r)$.
}
\label{fig:beta05}
\end{figure}

Figure~\ref{fig:beta05} shows dynamics for $ \beta = 0.5 $, under initial conditions $ u_1(0) = 0.2 $, $ u_2(0) = 0.3 $, $ \Omega_r(0) = 10^{-5} $, and $ \lambda(0) = 1.0 $. Panels illustrate inward trajectories in the $(u_1, u_2)$ phase space, a shallow vector field in $(u_1, \lambda)$, early stabilization of $\lambda(N)$, and convergent flow structures in the full 3D stream plot.

\begin{figure}[H]
\centering
\includegraphics[width=0.49\textwidth]{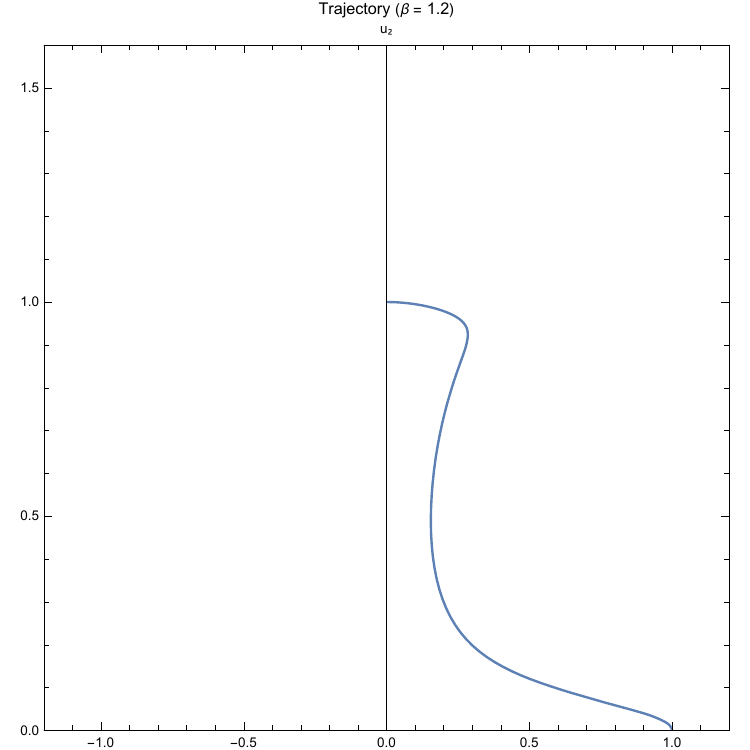}
\includegraphics[width=0.49\textwidth]{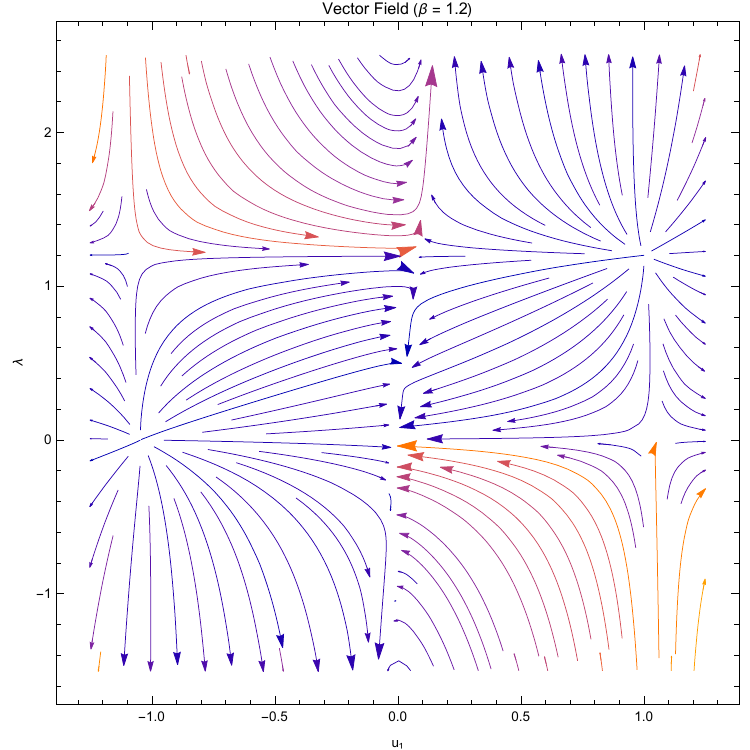}
\includegraphics[width=0.49\textwidth]{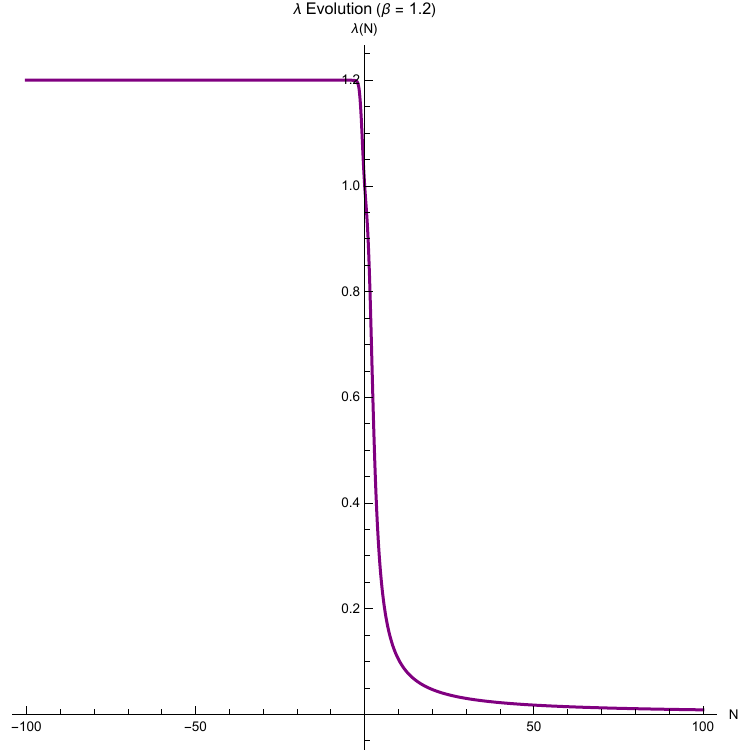}
\includegraphics[width=0.49\textwidth]{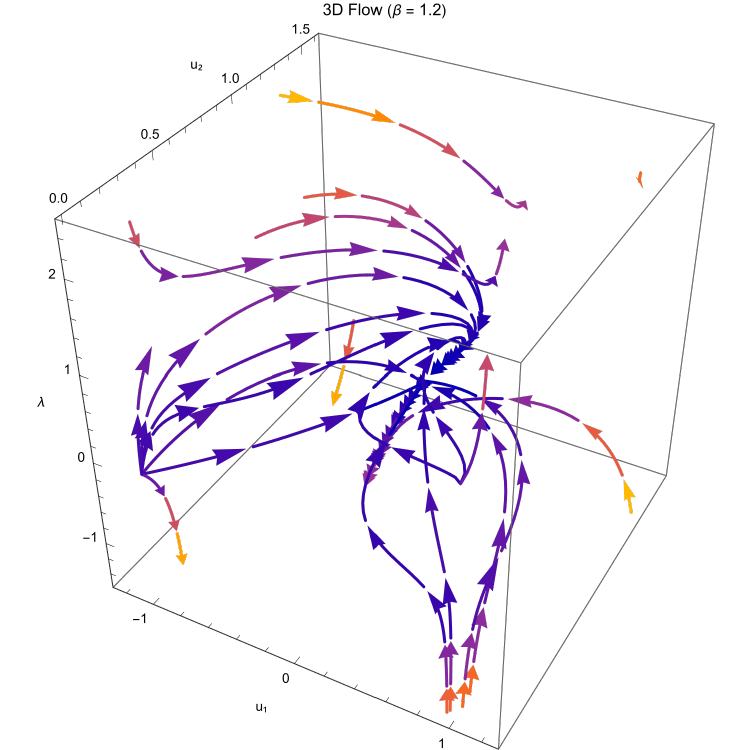}
\caption{
Dynamical behavior for exponential slope parameter $ \beta = 1.2 $, with fixed initial conditions $ u_1(0) = 0.2 $, $ u_2(0) = 0.3 $, $ \Omega_r(0) = 10^{-5} $, and $ \lambda(0) = 1.0 $. Panels as in Figure~\ref{fig:beta05}.
}
\label{fig:beta12}
\end{figure}

For $ \beta = 1.2 $, shown in Figure~\ref{fig:beta12}, the system develops stronger nonlinearities and delayed convergence. Vector arrows in $(u_1, \lambda)$ show steep gradients, and $\lambda(N)$ exhibits transient excursions before settling.

\begin{figure}[H]
\centering
\includegraphics[width=0.49\textwidth]{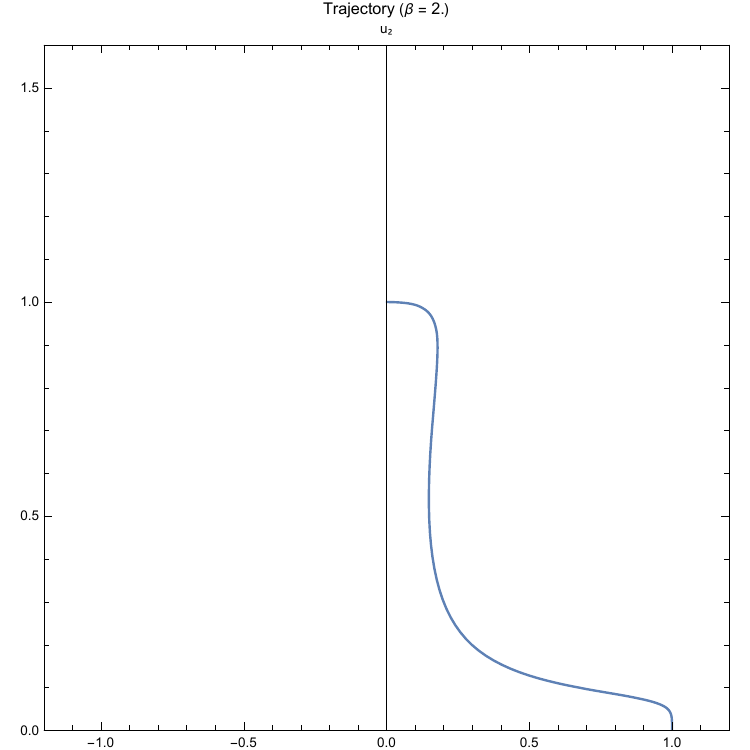}
\includegraphics[width=0.49\textwidth]{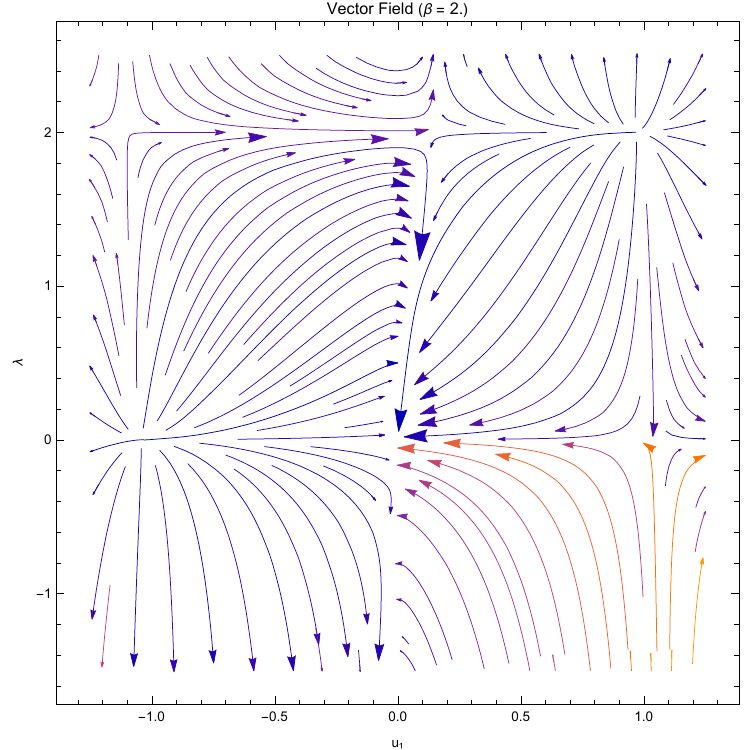}
\includegraphics[width=0.49\textwidth]{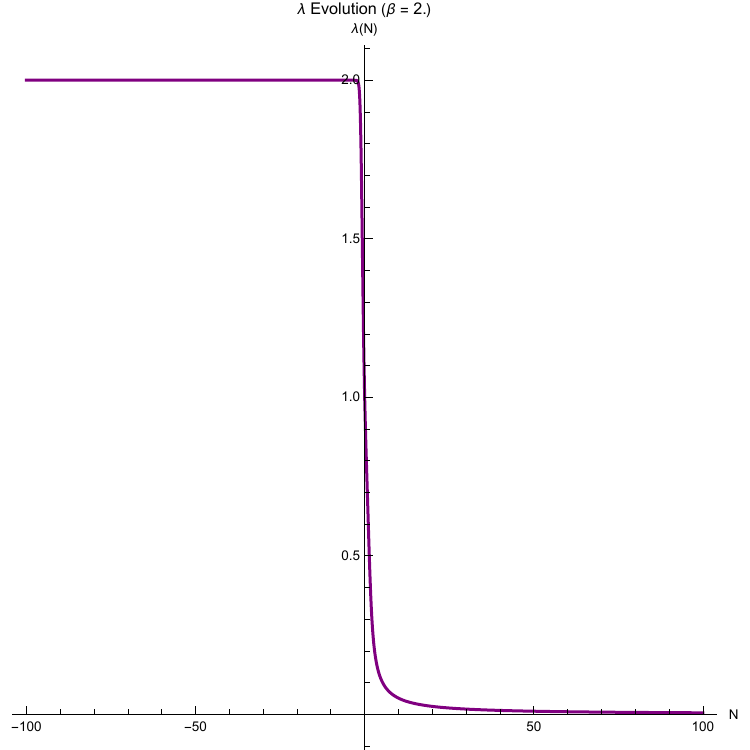}
\includegraphics[width=0.49\textwidth]{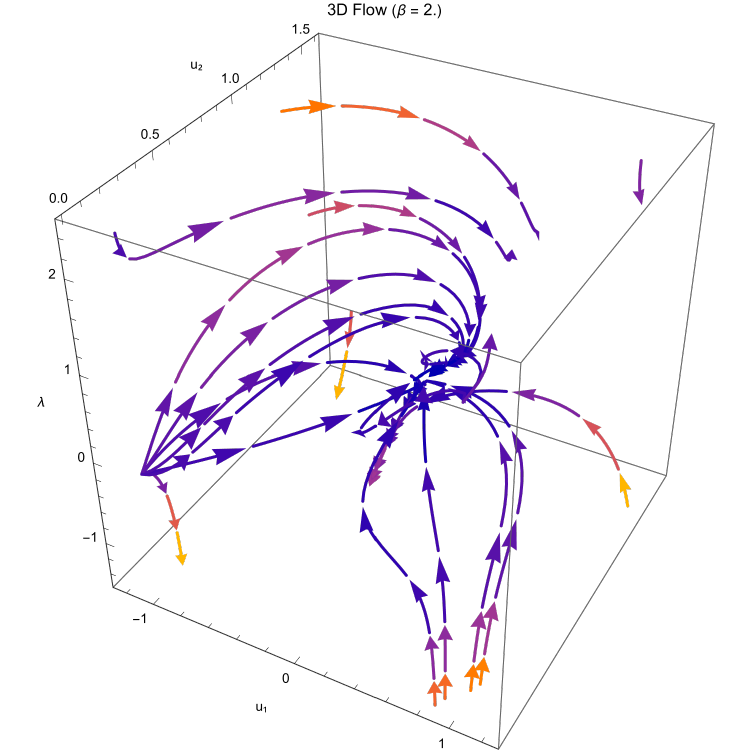}
\caption{
Dynamical behavior for exponential slope parameter $ \beta = 2.0 $, with fixed initial conditions $ u_1(0) = 0.2 $, $ u_2(0) = 0.3 $, $ \Omega_r(0) = 10^{-5} $, and $ \lambda(0) = 1.0 $. Panels as in Figure~\ref{fig:beta05}.
}
\label{fig:beta20}
\end{figure}

\begin{figure}[H]
\centering
\includegraphics[width=0.49\textwidth]{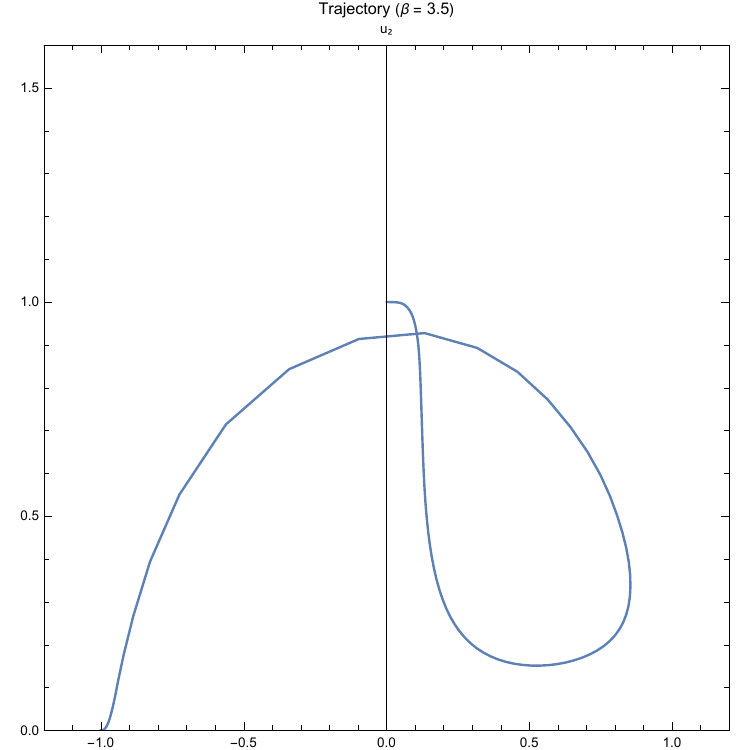}
\includegraphics[width=0.49\textwidth]{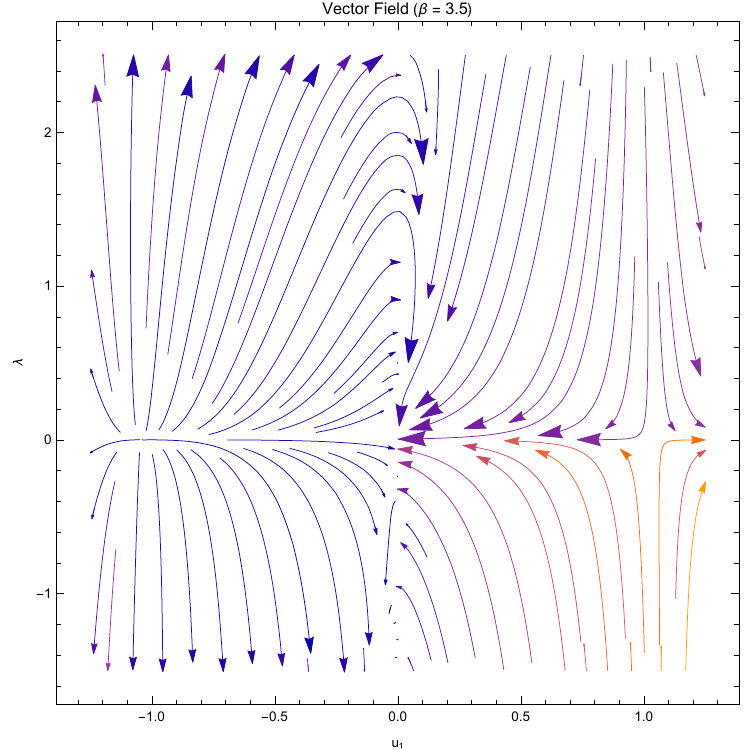}
\includegraphics[width=0.49\textwidth]{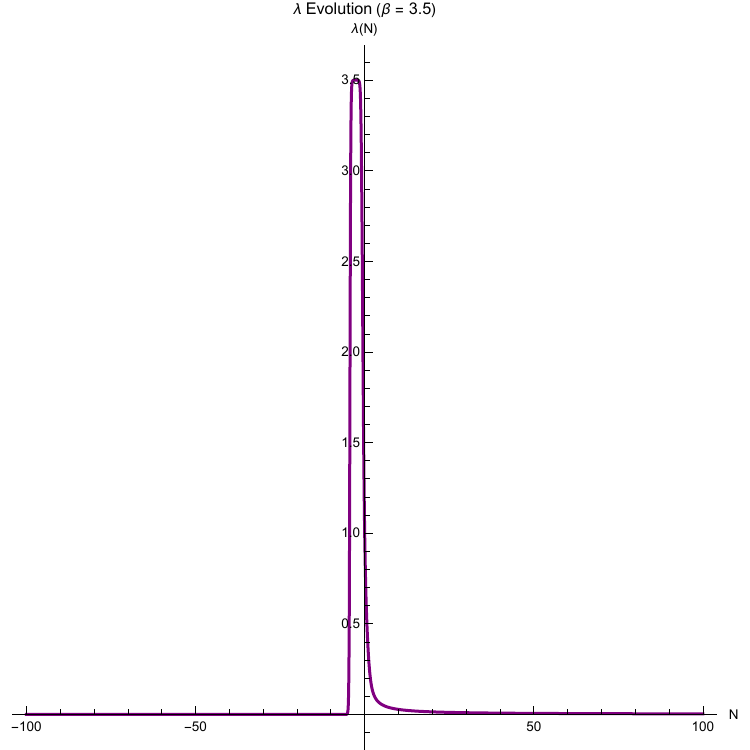}
\includegraphics[width=0.49\textwidth]{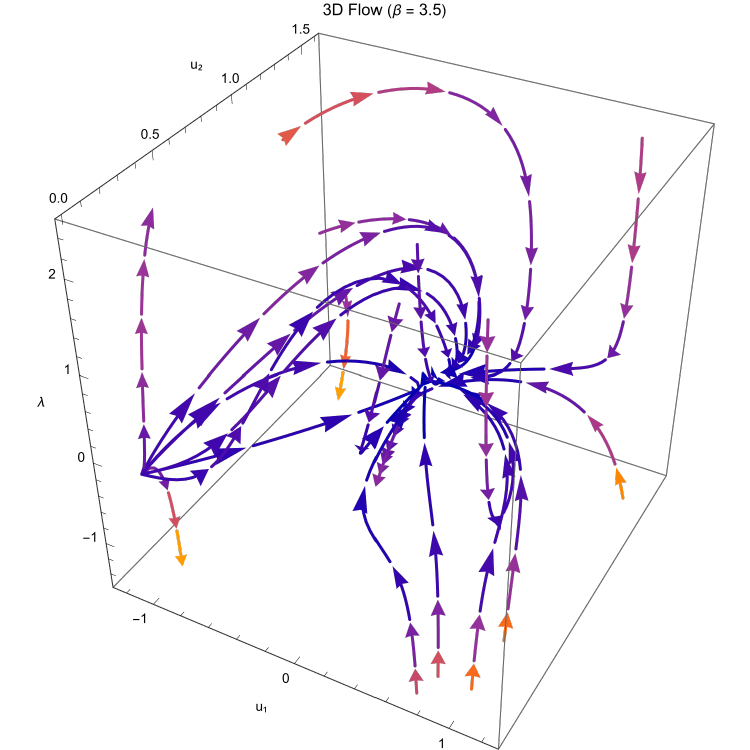}
\caption{
Dynamical behavior for exponential slope parameter $ \beta = 3.5 $, with fixed initial conditions $ u_1(0) = 0.2 $, $ u_2(0) = 0.3 $, $ \Omega_r(0) = 10^{-5} $, and $ \lambda(0) = 1.0 $. Panels as in Figure~\ref{fig:beta05}.
}
\label{fig:beta35}
\end{figure}

Figures~\ref{fig:beta20} and~\ref{fig:beta35} ($ \beta = 2.0,\ 3.5 $) illustrate increasingly stiff dynamics: trajectories widen, stream plots reveal curvature-dominated zones, and $\lambda(N)$ declines monotonically. These regimes suggest decoupled scaling behavior where the field evolves but lacks late-time stabilization.

\subsection{Evaluation of Cosmological Parameters for Small \texorpdfstring{$\alpha$}{α}}

\begin{figure}[H]
	\centering
		\includegraphics[width=0.8\textwidth]{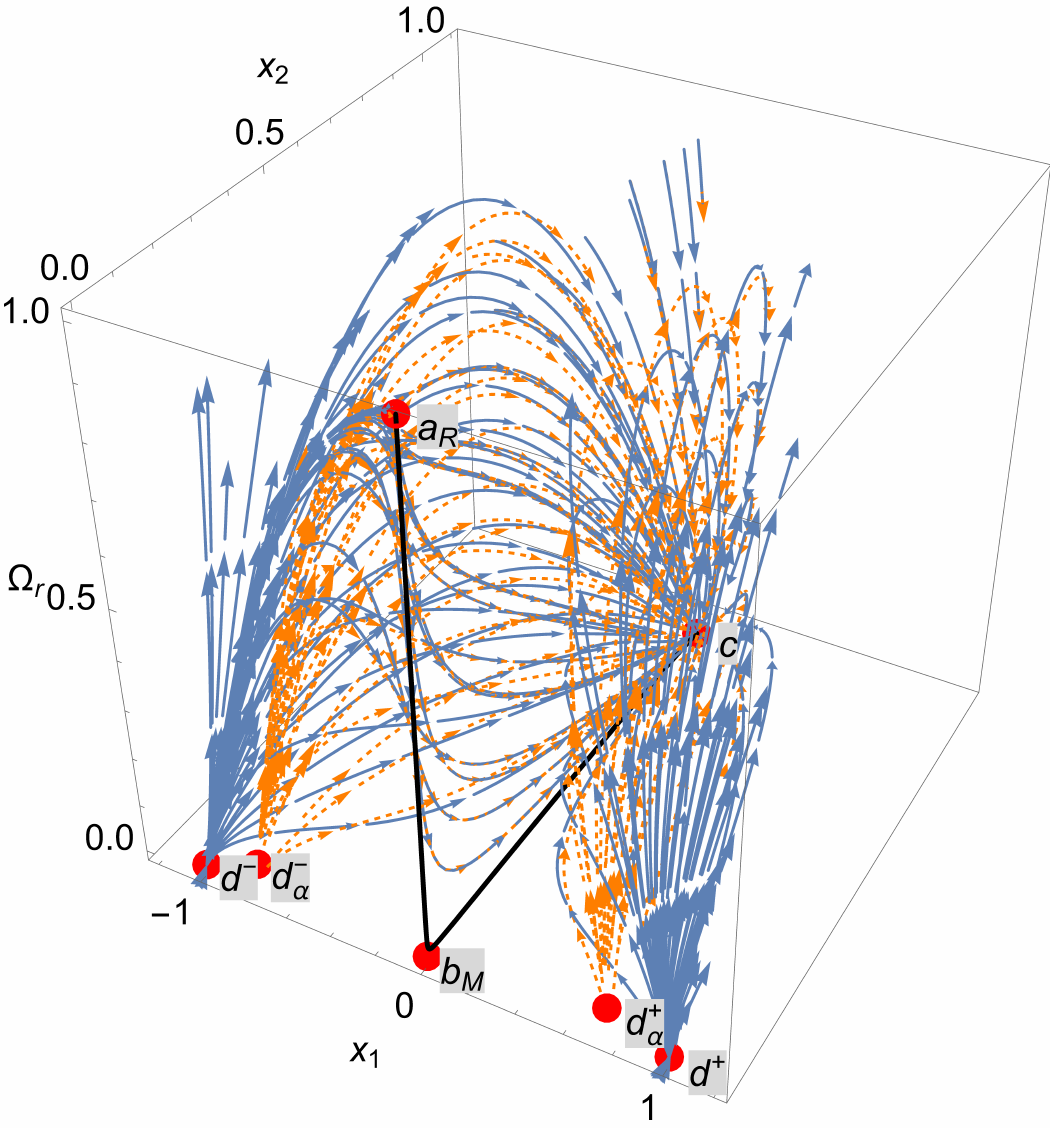}
	\caption{This figure illustrates the phase space trajectories for the dynamical system defined by Eqs. \eqref{dinsysialphap} and \eqref{dinsysfalphap}, with parameter values set to $\lambda = 10^{-1}$. The solid blue curves represent the $\alpha = 0$ scenario, while the dashed orange curves correspond to $\alpha = 5 \times 10^{-2}$. Each trajectory illustrates the evolution of the Universe for distinct initial conditions. In particular, the solid black trajectory is generated from the initial conditions $ x_{1i} = 10^{-12} $, $ x_{2i} = 8.8 \times 10^{-13} $, and $ \Omega_{ri} = 9.99661 \times 10^{-1} $.}. \label{Figura1}
\end{figure}

We present a focused discussion emphasizing how each figure contributes to understanding the dynamics and observational viability of the model under variations in the parameter $ \alpha $:

\textbf{Figure~\ref{Figura1}: Phase Space Trajectories.}  
This figure offers a three-dimensional visualization of the system's dynamical evolution for two cases: the unmodified scenario ($ \alpha = 0 $, solid blue) and the modified regime with a small higher-order contribution ($ \alpha = 5 \times 10^{-2} $, dashed orange). The initial conditions are set to favor radiation dominance, as indicated by the near-unity value of $ \Omega_{ri} $. The trajectories highlight how the inclusion of a nonzero $ \alpha $ alters the curvature and structure of the solution paths, potentially reshaping the stability basin and affecting the approach toward late-time attractors. The solid black curve illustrates a representative trajectory under these initial conditions, helping to contextualize observational predictions, such as the expansion history and effective equation of state evolution.

\begin{figure}[H]
	\centering
		\includegraphics[width=0.8\textwidth]{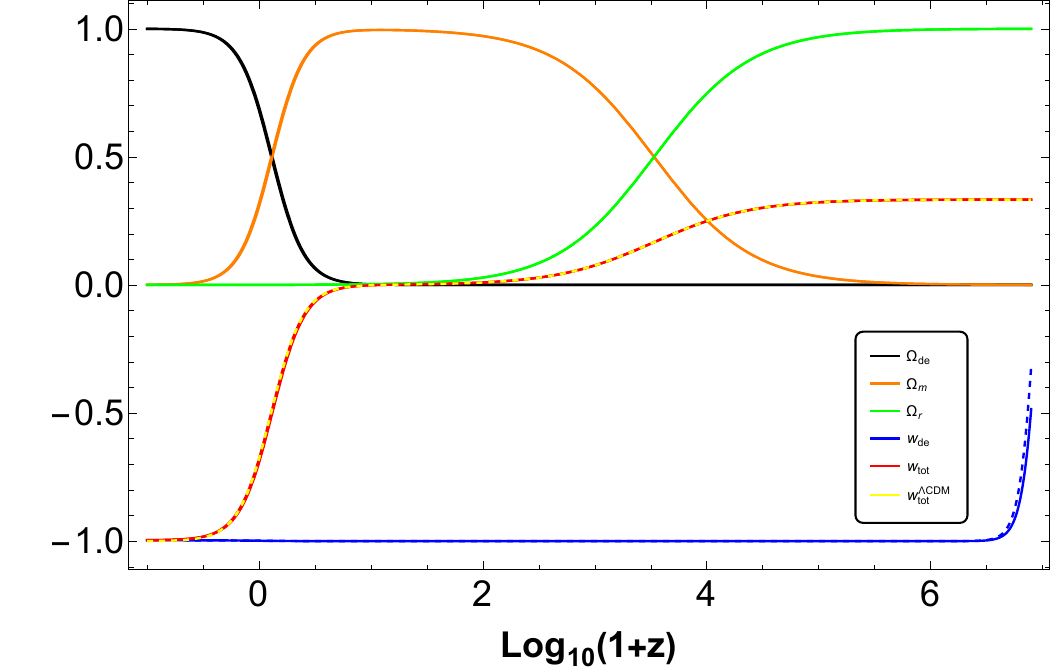}
	\caption{We present the evolution of the density parameters $\Omega_{de}$, $\Omega_{m}$ and $\Omega_{r}$ along with the equation of state (EoS) parameters $ w_{de} $, $ w_{tot} $, and $ w_{tot}^{\Lambda CDM} $, as functions of $\log_{10}(1+z)$. The solid blue lines illustrate the case when $\alpha = 0$, whereas the dashed blue lines depict the scenario where $\alpha = 5 \times 10^{-2}$. The system is initialized with $ x_{1i} = 10^{-12} $, $ x_{2i} = 8.8 \times 10^{-13} $, and $ \Omega_{ri} = 9.99661 \times 10^{-1} $.} 
	\label{Figura2}
\end{figure}

\textbf{Figure~\ref{Figura2}: Energy Densities and EoS Evolution.}  
This figure compares the evolution of density parameters $ \Omega_{de} $, $ \Omega_{m} $, and $ \Omega_{r} $ alongside the equation-of-state parameters for dark energy ($ w_{de} $), the total fluid ($ w_{tot} $), and the standard $\Lambda$CDM prediction $ w_{tot}^{\Lambda\mathrm{CDM}} $. The variable $ \log_{10}(1+z) $ enables targeted analysis of late-time features. Notably, the dashed lines (corresponding to modified $ \alpha $) show slight deviations in $ w_{de} $ near $ z \sim 1 $, suggesting that higher-order terms can mimic quintessence or phantom-like behavior during transitional epochs. The overall evolution remains compatible with observational constraints while enriching the dynamical structure.

\begin{figure}[H]
	\centering
		\includegraphics[width=0.8\textwidth]{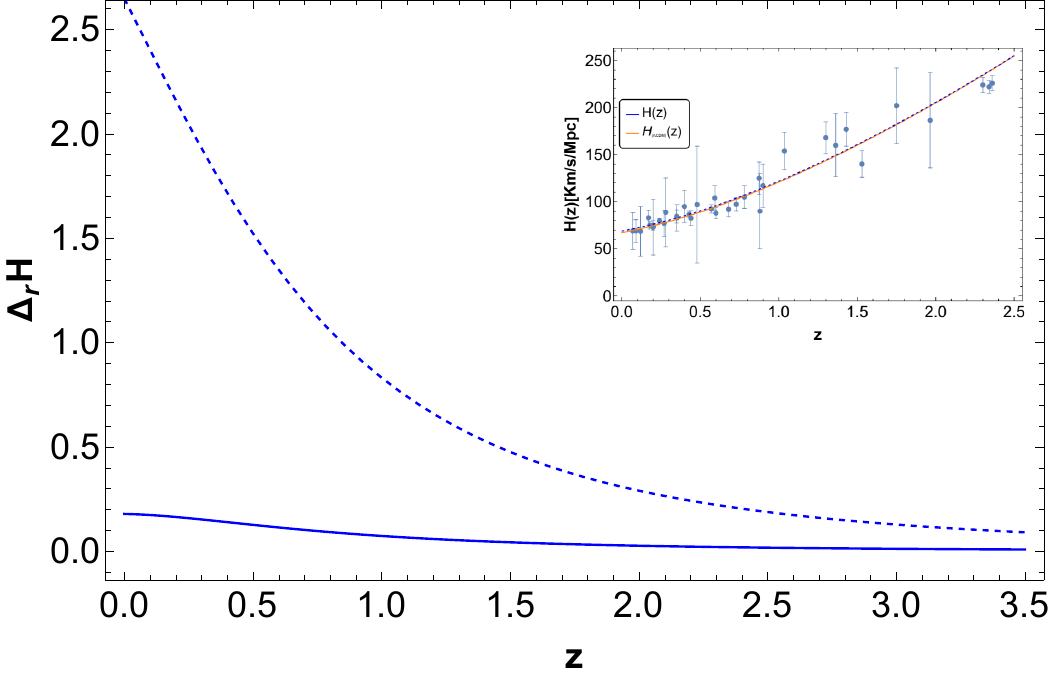}
	\caption{This plot illustrates the evolution of the relative difference $\Delta_r H$ associated with the Hubble parameter as a function of the redshift $z$. Additionally, the inset plot presents the Hubble parameter $H(z)$ alongside the Hubble parameter corresponding to the $\Lambda$CDM model, $H_{(\Lambda CDM)}(z)$, as a function of redshift, including observational data for comparison. The solid blue lines illustrate the case when $\alpha = 0$, whereas the dashed blue lines depict the scenario where $\alpha = 5 \times 10^{-2}$. The system is initialized with $ x_{1i} = 10^{-12} $, $ x_{2i} = 8.8 \times 10^{-13} $, and $ \Omega_{ri} = 9.99661 \times 10^{-1} $.} 
	\label{Figura3}
\end{figure}

\textbf{Figure~\ref{Figura3}: Relative Difference in Hubble Parameter.}  
This plot provides a quantitative measure of how the modified model ($ \alpha \neq 0 $) departs from standard $\Lambda$CDM in terms of expansion rate. The relative error $ \Delta_r H(z) $ captures this deviation. The inset panel enhances interpretability by overlaying observational data with theoretical curves, helping constrain the model's viability. Deviations remain controlled at low redshifts, indicating the model’s capacity to replicate present-day expansion while allowing for altered behavior at earlier epochs. These results support the model's testability in future precision measurements of $ H(z) $.

\begin{figure}[H]
	\centering
		\includegraphics[width=0.8\textwidth]{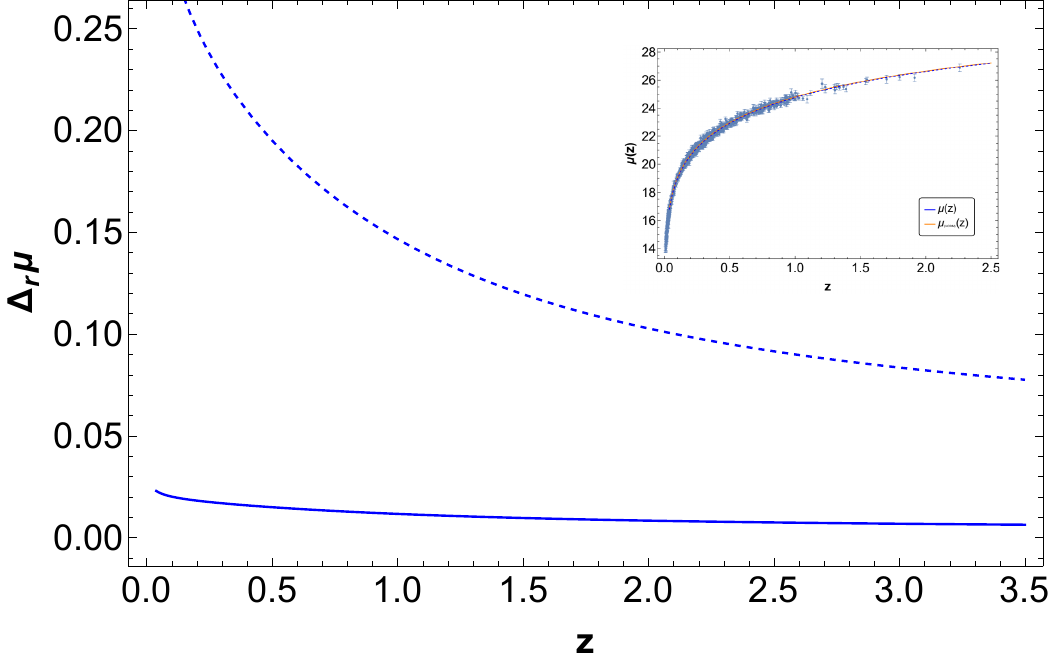}
	\caption{This plot depicts the evolution of the relative difference $\Delta \mu_{r}$ with respect to the $\Lambda$CDM model as a function of the redshift $z$. Additionally, within the same figure, we present the evolution of the distance modulus $\mu(z)$, alongside the distance modulus corresponding to $\Lambda$CDM, denoted as $\mu_{\Lambda CDM}$, as a function of $z$. The solid blue lines illustrate the case when $\alpha = 0$, whereas the dashed blue lines depict the scenario where $\alpha = 5 \times 10^{-2}$. The system is initialized with $ x_{1i} = 10^{-12} $, $ x_{2i} = 8.8 \times 10^{-13} $, and $ \Omega_{ri} = 9.99661 \times 10^{-1} $.} 
	\label{Figura4}
\end{figure}

\textbf{Figure~\ref{Figura4}: Distance Modulus Comparison.}  
This figure presents both the relative deviation $ \Delta \mu_r(z) $ and a direct comparison between $ \mu(z) $ and $ \mu_{\Lambda\mathrm{CDM}}(z) $. In the low-redshift regime, most relevant to supernova observations, the model shows excellent agreement. Small deviations for $ \alpha = 5 \times 10^{-2} $ offer observationally relevant signatures, suggesting that the model remains consistent with current data while potentially improving fits in intermediate redshift ranges. These features may prove useful in resolving tensions between datasets or refining cosmological constraints.

In summary, the figure set collectively demonstrates that a small nonzero $ \alpha $ introduces higher-order corrections capable of subtly reshaping phase space trajectories and the expansion history, without violating observational bounds. These results support the viability of Pais-Uhlenbeck-inspired models as promising alternatives to canonical dark energy formulations, particularly when evaluating transitions between radiation, matter, and accelerated phases.

\subsection{Application 2: Power-Law Potential and Stability of the De Sitter Solution}

We analyze a scalar field model with a power-law potential:
\begin{equation}
    V(\phi) = V_0 \phi^n \quad \implies \quad f(\lambda) = -\frac{\lambda^2}{n}, \label{power-law}
\end{equation}
and explore the stability of the de Sitter solution using both linearization and center manifold theory.

\subsubsection{Critical Points and Cosmological Parameters}

\begin{table*}[h!]
\centering
\caption{Critical points on the slow manifold for $ f(\lambda) = -\lambda^2/n $, including the effective equation of state parameter $ w_{\text{eff}} $.}
\begin{tabular}{c c c c c c c}
\toprule
Name & $ x_{1c} $ & $ x_{2c} $ & $ \Omega_{rc} $ & $ \lambda_c $ & $ u_{5c} $ & $ w_{\text{eff}} $ \\
\midrule
$ a_M $      & $0$  & $0$  & $0$ & arbitrary & $0$ & $0$ \\
$ b_R $      & $0$  & $0$  & $1$ & arbitrary & $0$ & $1/3$ \\
$ c $        & $0$  & $1$  & $0$ & $0$       & arbitrary & $-1$ \\
$ d^{+} $    & $1$  & $0$  & $0$ & $0$       & $0$ & $1$ \\
$ d^{-} $    & $-1$ & $0$  & $0$ & $0$       & $0$ & $1$ \\
\bottomrule
\end{tabular}
\label{table3}
\end{table*}
In table \ref{table3} are presented the critical points on the slow manifold for $ f(\lambda) = -\lambda^2/n $, including the effective equation of state parameter $ w_{\text{eff}} $. They are interpreted as follows. 

\begin{itemize}
    \item \textbf{$ a_M $}: Represents a matter-dominated state with vanishing curvature ($ \Omega_{rc} = 0 $). The scalar field is effectively frozen ($ x_{1c} = x_{2c} = 0 $), but the matter sector satisfies $ \Omega_m = 1 $. The arbitrariness of $ \lambda $ implies degeneracy in the scalar field configuration, typical of critical points with dormant scalar dynamics.

    \item \textbf{$ b_R $}: Corresponds to a curvature-dominated regime ($ \Omega_{rc} = 1 $) where the scalar field and matter contributions vanish. The arbitrariness of $ \lambda $ again reflects that the scalar field is non-dynamical in this configuration. This point signals a geometric phase that can appear transiently in bouncing or loitering scenarios.

    \item \textbf{$ c $}: Characterized by a purely potential-dominated scalar field configuration, with $ x_{2c} = 1 $ and $ x_{1c} = 0 $. The vanishing curvature ($ \Omega_{rc} = 0 $) and slope $ \lambda_c = 0 $ suggest a stationary point of the potential, which might correspond to a cosmological constant-like behavior depending on the model's embedding. The freedom in $u_{5c}$ allows for additional scaling or interaction channels, depending on the field's evolution.

    \item \textbf{$ d^{+} $} and \textbf{$ d^{-} $}: These kinetic-dominated solutions ($ x_{1c} = \pm 1 $, $ x_{2c} = 0 $) represent stiff-fluid-like cosmologies with negligible curvature and scalar potential energy. Fixed slope $ \lambda_c = 0 $ indicates a flat potential, but the dominance of kinetic energy leads to a decelerated expansion. The vanishing $ u_{5c} $ reinforces that there’s no additional coupling or source term active at these points.
\end{itemize}

\subsubsection{Linear Stability Analysis}

The governing dynamical system is:
\begin{equation}
\frac{d}{dN}
\begin{pmatrix}
u_1 \\
u_2 \\
\Omega_r \\
\lambda
\end{pmatrix}
=
\begin{pmatrix}
\frac{1}{2}\left(3u_1^3 + u_1(\Omega_r - 3u_2^2 - 3) + \sqrt{6}\lambda u_2^2 - 6u_1\right) \\
\frac{1}{2}u_2\left(3u_1^2 - \sqrt{6}\lambda u_1 - 3u_2^2 + \Omega_r + 3\right) \\
\Omega_r(3u_1^2 - 3u_2^2 + \Omega_r - 1) \\
\sqrt{6}u_1 \cdot \frac{\lambda^2}{n}
\end{pmatrix}.
\label{systemPL}
\end{equation}

The eigenvalues of the Jacobian at each critical point of system \eqref{systemPL} are presented in Table \ref{eigtable}.

\begin{table}[h!]
\centering
\caption{Eigenvalues of the Jacobian at each critical point of system \eqref{systemPL}}
\label{eigtable}
\renewcommand{\arraystretch}{1.3}
\begin{tabularx}{\textwidth}{@{}cX@{}}
\toprule
\textbf{Point} & \textbf{Eigenvalues} \\
\midrule
$ a_M $ & $ \{-\tfrac{3}{2}, \tfrac{3}{2}, -1, 0\} $: saddle, non-hyperbolic \\
$ b_R $ & $ \{2, -1, 2, 0\} $: saddle, non-hyperbolic \\
$ c $ & $ \{-4, -3, -3, 0\} $: requires nonlinear analysis (center manifold) \\
$ d^{+}, d^{-} $ & $ \{3, 2, 3, 0\} $: repellers, non-hyperbolic \\
\bottomrule
\end{tabularx}
\end{table}

\begin{itemize}
    \item \textbf{$ a_M $}: The eigenvalues $ \{-\tfrac{3}{2}, \tfrac{3}{2}, -1, 0\} $ indicate saddle-type behavior due to the presence of both positive and negative real parts, with one zero eigenvalue.  Physically, it can model a transitional state with partial matter dominance but no attractor behavior.

    \item \textbf{$ b_R $}: With eigenvalues $ \{2, -1, 2, 0\} $, this point is also a saddle, and non-hyperbolic due to the vanishing eigenvalue. The presence of two unstable directions undermines the stability of the curvature-dominated regime, rendering it unsuitable as a late-time attractor. However, it may transiently influence early or intermediate evolution.

    \item \textbf{$ c $}: This point has eigenvalues $ \{-4, -3, -3, 0\} $, suggesting strong contraction in most directions but neutral behavior along one. As the linearized system is non-hyperbolic, a center manifold analysis is crucial for assessing nonlinear stability. This configuration formally corresponds to exact de Sitter expansion but typically fails to act as a dynamical attractor.

    \item \textbf{$ d^{+} $, $ d^{-} $}: These points share eigenvalues $ \{3, 2, 3, 0\} $, marking them as repellers in the linear approximation. The zero eigenvalue again makes them non-hyperbolic, though the dominance of positive eigenvalues ensures strong instability. These solutions model stiff-fluid-like behavior typical of kinetic-dominated regimes near the initial singularity.
\end{itemize}

\subsubsection{Center Manifold Stability of the De Sitter Point $ c $}

The point $ c $ lies at $ (u_1, u_2, \Omega_r, \lambda) = (0,1,0,0) $, where linear analysis yields a zero eigenvalue. According to the Center Manifold Theorem, the qualitative behavior near $ c $ is governed by the reduced gradient-like evolution:
\[
\frac{dy}{dN} = -\nabla U(y), \quad \text{with} \quad U(y) = -\frac{1}{4n} y^4 + \mathcal{O}(y^6).
\]

This implies:
\[
\frac{dy}{dN} = -U'(y) = \frac{1}{n} y^3 + \mathcal{O}(y^5),
\]
where the sign of $ n $ determines the type of extremum at the origin (see Table \ref{table-cm-n}).

\begin{table}[h!]
\centering
\caption{Center manifold classification of de Sitter point $ c $ for power-law potentials.}
\label{table-cm-n}
\renewcommand{\arraystretch}{1.3}
\begin{tabularx}{\textwidth}{@{}Xll@{}}
\toprule
\textbf{Sign of $ n $} & \textbf{Form of $ U(y) $} & \textbf{Stability at $ y=0 $} \\
\midrule
$ n > 0 $ & $ -\frac{1}{4n} y^4 $ (degenerate maximum) &  Unstable \\
$ n < 0 $ & $ +\frac{1}{4|n|} y^4 $ (degenerate minimum) & Stable \\
\bottomrule
\end{tabularx}
\end{table}

The de Sitter solution $ c $ is non-hyperbolic for all $ n $ due to the vanishing eigenvalue associated with the scalar field slope. Applying center manifold theory reveals a bifurcation in its stability:

\begin{itemize}
    \item For $ n > 0 $: the potential exhibits a flat maximum; trajectories generically diverge. The de Sitter point is unstable.
    \item For $ n < 0 $: the potential exhibits a flat minimum; trajectories converge slowly. The de Sitter point becomes locally stable.
\end{itemize}

Hence, inverse power-law potentials ($ n < 0 $) support late-time acceleration via a stable de Sitter attractor, while standard polynomial models ($ n > 0 $) exclude this possibility.

\subsection{Power-Law Potential Dynamics}
\label{sec:powerlawDin}

For the power-law model $ f(\lambda) = -\lambda^2 / n $, we investigate both positive and negative values of the coupling parameter $ n $, using the same initial conditions as in Section~\ref{sec:exp}. This formulation leads to symmetry-breaking vector fields and distinct attractor basins.

\begin{figure}[H]
\centering
\includegraphics[width=0.49\textwidth]{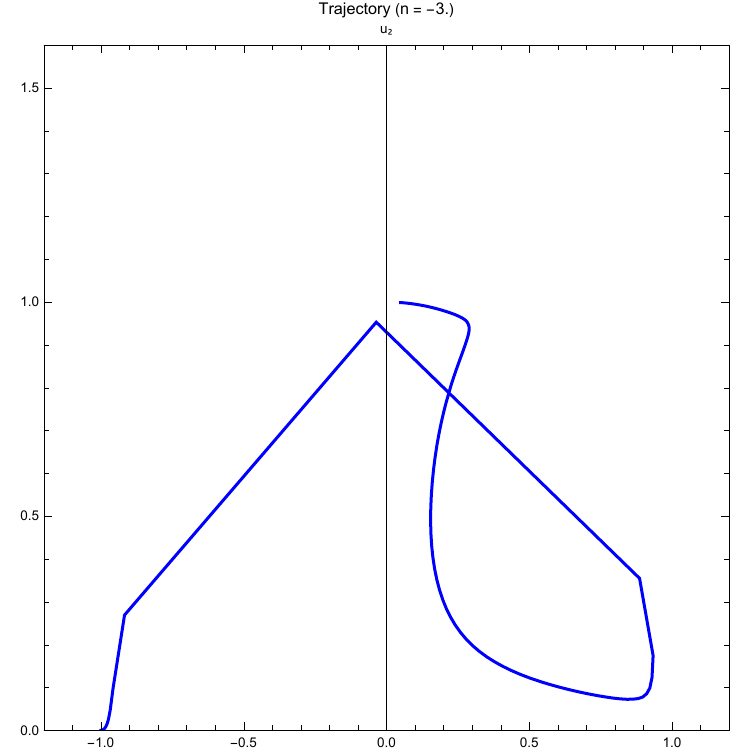}
\includegraphics[width=0.49\textwidth]{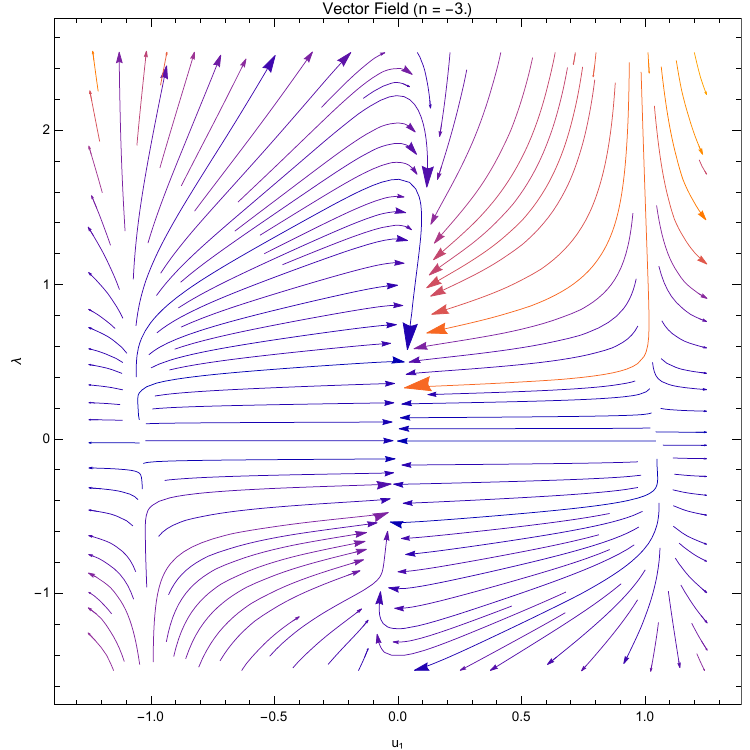}
\includegraphics[width=0.49\textwidth]{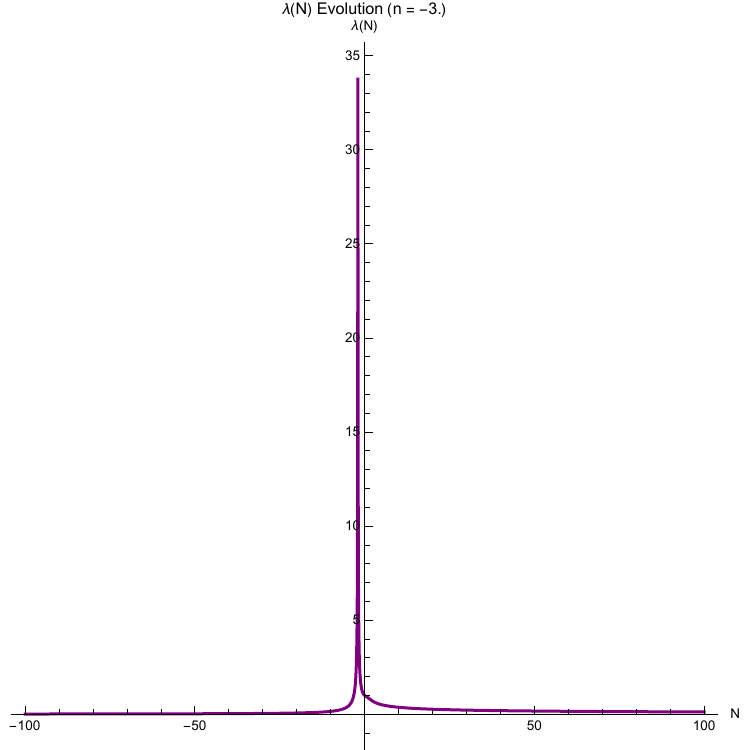}
\includegraphics[width=0.49\textwidth]{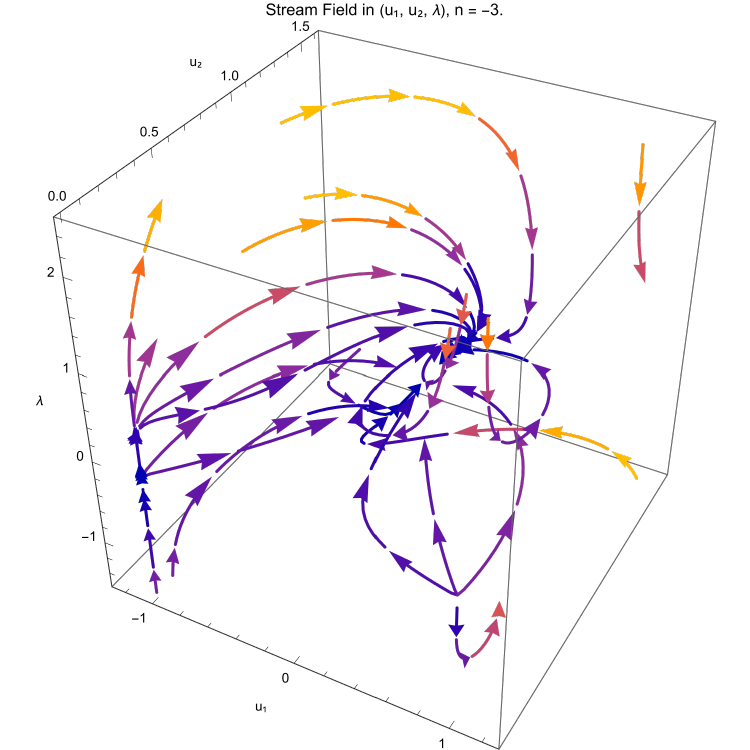}
\caption{
Dynamical behavior for power-law coupling $ f(\lambda) = -\lambda^2/n $ with $ n = -3.0 $, under initial conditions $ u_1(0) = 0.2 $, $ u_2(0) = 0.3 $, $ \Omega_r(0) = 10^{-5} $, and $ \lambda(0) = 1.0 $. Panels show trajectory in $(u_1, u_2)$, stream plot in $(u_1, \lambda)$, evolution of $\lambda(N)$, and full 3D stream plot in $(u_1, u_2, \lambda)$.
}
\label{fig:powerlaw_n-30}
\end{figure}

Figure~\ref{fig:powerlaw_n-30} illustrates the case $ n = -3.0 $, where negative curvature coupling induces rapid growth in $ \lambda(N) $ and sharply inward trajectories in the $ (u_1, u_2) $ plane. Vector reversals occur in the $ (u_1, \lambda) $ projection, and the 3D stream plot reveals folded surfaces with angular divergence. The center manifold at the origin is stable.

\begin{figure}[H]
\centering
\includegraphics[width=0.49\textwidth]{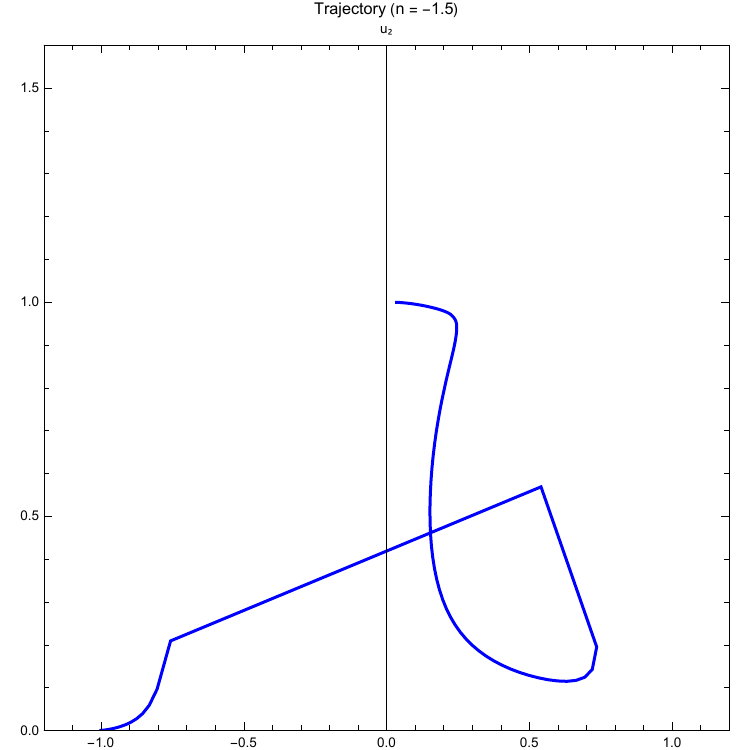}
\includegraphics[width=0.49\textwidth]{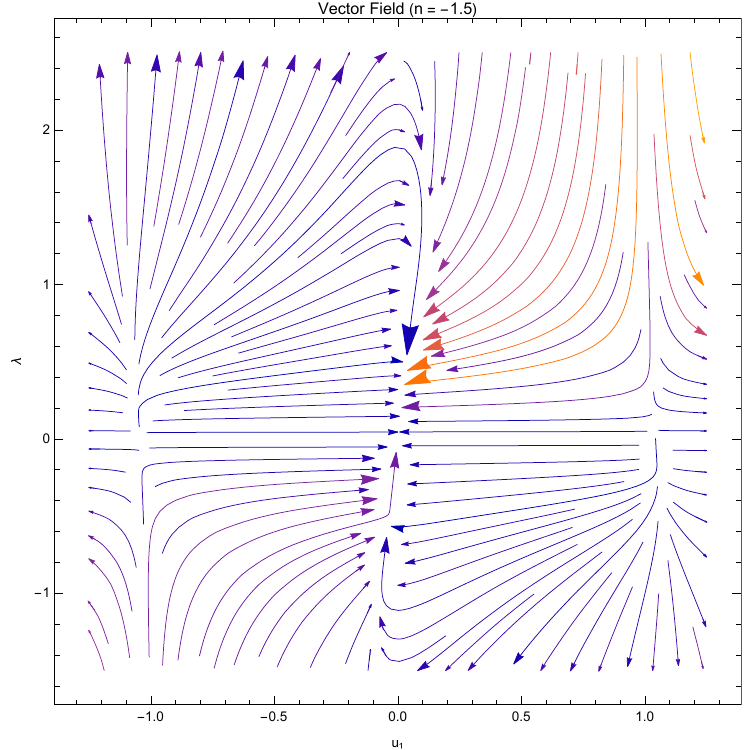}
\includegraphics[width=0.49\textwidth]{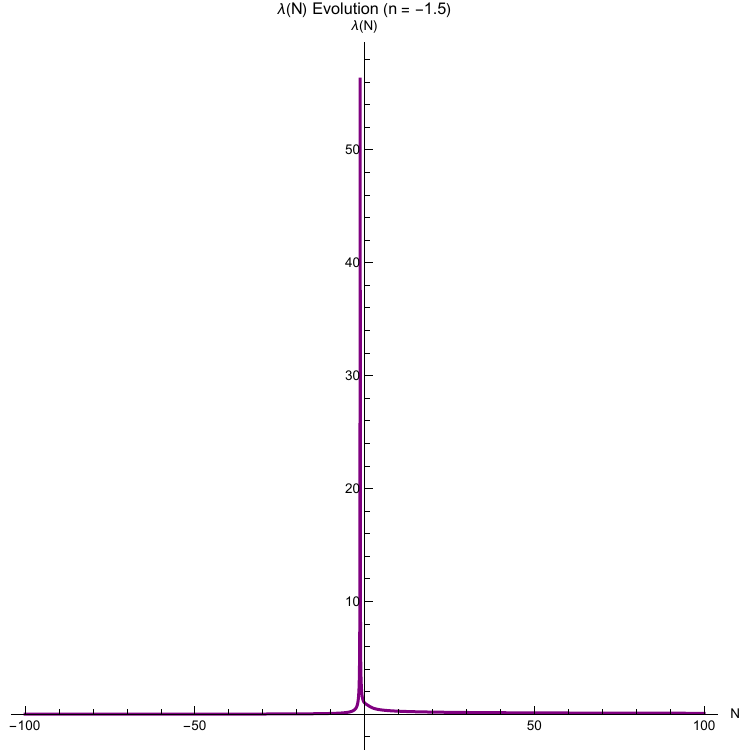}
\includegraphics[width=0.49\textwidth]{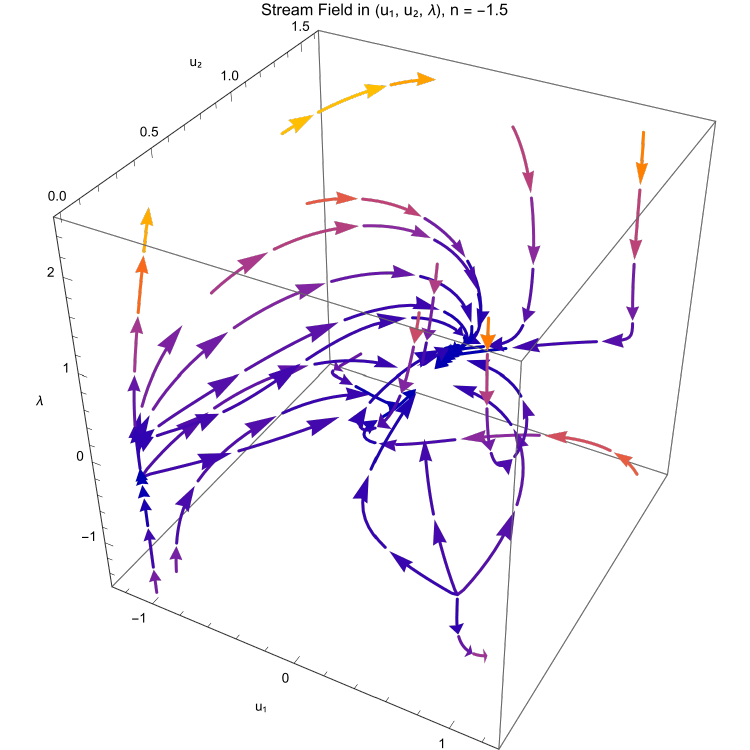}
\caption{
Dynamical behavior for power-law coupling with $ n = -1.5 $. Initial conditions as in Figure~\ref{fig:powerlaw_n-30}.
}
\label{fig:powerlaw_n-15}
\end{figure}

Figure~\ref{fig:powerlaw_n-15}, corresponding to $ n = -1.5 $, displays qualitatively similar behavior with reduced intensity. Trajectories remain inward-bending, though the stream field exhibits partial alignment. The growth of $ \lambda(N) $ slows, indicating a transition toward marginal stability. The center manifold at the origin remains stable.

\begin{figure}[H]
\centering
\includegraphics[width=0.49\textwidth]{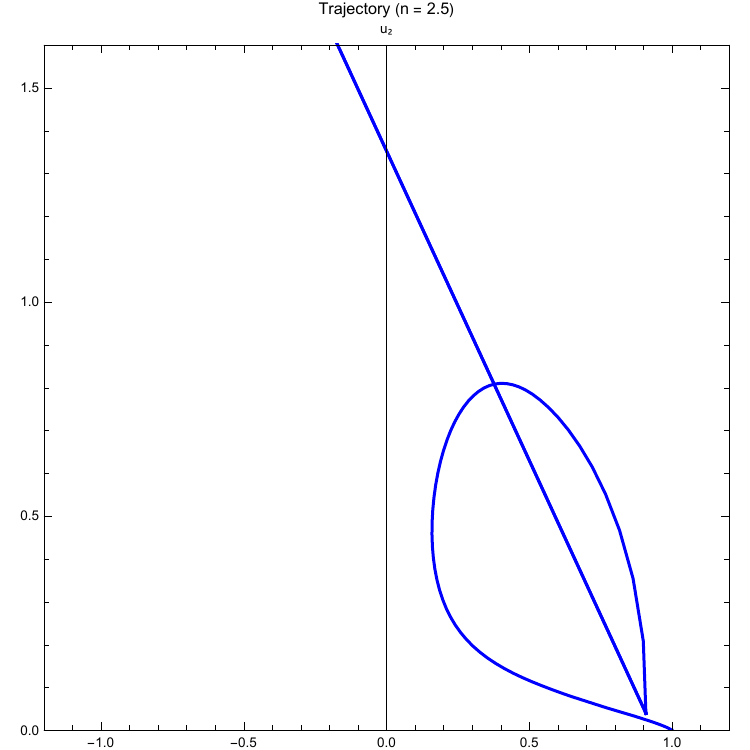}
\includegraphics[width=0.49\textwidth]{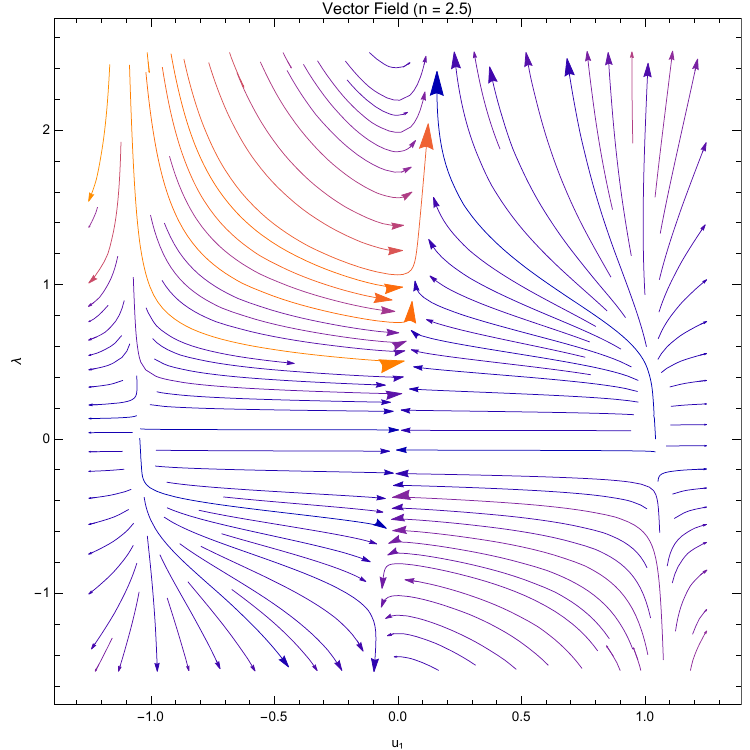}
\includegraphics[width=0.49\textwidth]{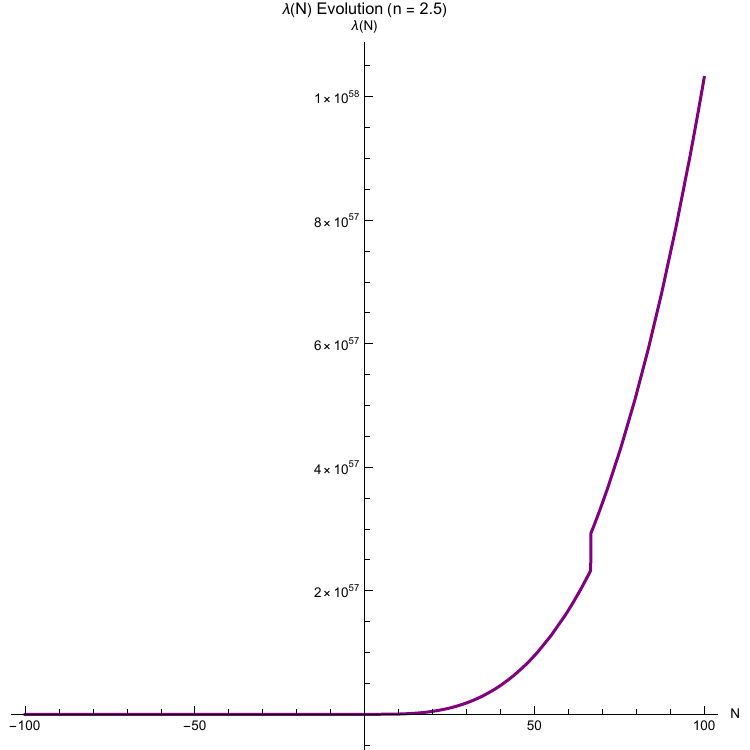}
\includegraphics[width=0.49\textwidth]{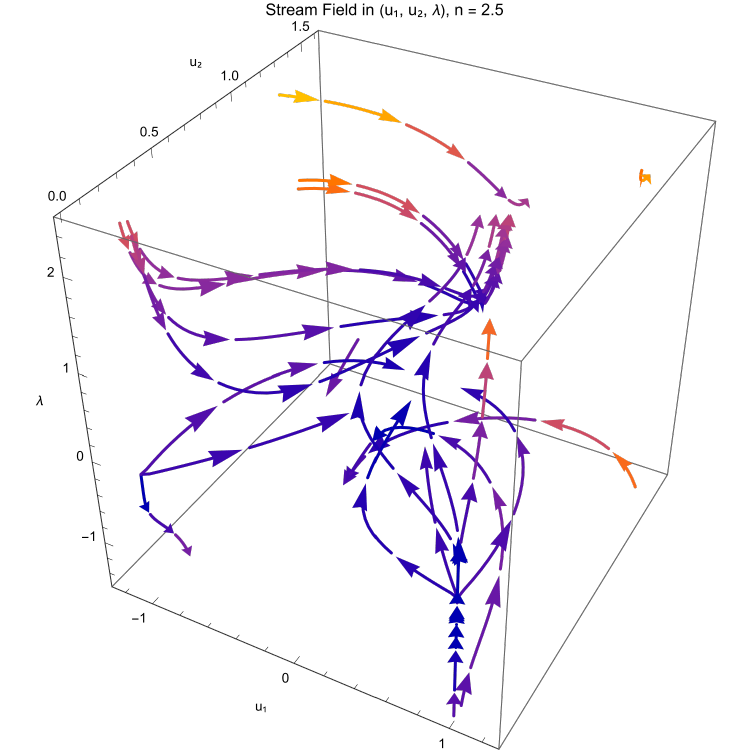}
\caption{
Dynamical behavior for power-law coupling with $ n = 2.5 $. Initial conditions as in Figure~\ref{fig:powerlaw_n-30}.
}
\label{fig:powerlaw_n25}
\end{figure}

In contrast, Figure~\ref{fig:powerlaw_n25} ($ n = 2.5 $) shows late-time divergence from the origin, which now acts as a saddle. Despite the divergence, $ \lambda(N) $ decays smoothly, vector fields straighten, and phase space orbits narrow around fixed points. These results resemble the behavior of the exponential model for small $ \beta $, though the decay profile of $ f(\lambda) $ near $ \lambda = 0 $ introduces distinct partial stabilization features. The center manifold is unstable in this case.
For the power-law model $ f(\lambda) = -\lambda^2 / n $, we investigate both positive and negative values of the coupling parameter $ n $, using the same initial conditions as in Section~\ref{sec:exp}. This formulation leads to symmetry-breaking vector fields and distinct attractor basins.

\subsubsection{Cosmological Interpretation of Critical Points (Power-Law Potential)}

\begin{table}[h!]
\centering
\caption{Cosmological regimes represented by critical points in the power-law potential model.}
\label{table:cosmo-powerlaw-refined}
\renewcommand{\arraystretch}{1.3}
\begin{tabularx}{\textwidth}{@{}c c c c c X@{}}
\toprule
\textbf{Point} & $ \Omega_{de} $ & $ \Omega_m $ & $ \omega_{de} $ & $ \omega_{eff} $ & \textbf{Physical Interpretation} \\
\midrule
$ a_M $ & 0 & 1 & $-1$ & 0 &
Pressureless matter-dominated universe. Scalar field contributes no energy ($ \Omega_{de} = 0 $); expansion rate matches standard dust era. Relevant at intermediate times but dynamically unstable. \\

$ b_R $ & 0 & 0 & $-1$ & $ \tfrac{1}{3} $ &
Radiation-dominated epoch. Scalar field has vacuum-like pressure but zero density; does not affect dynamics. Compatible with early-time expansion. \\

$ c $ & 1 & 0 & $-1$ & $-1$ &
True de Sitter solution: complete scalar field dominance with accelerated expansion, characterized by $ \Omega_{de} = 1 $ and $ \omega_{eff} = -1 $. While this configuration formally reproduces the observed dark-energy–driven background, center manifold analysis reveals that its stability is conditional on the sign of the potential exponent $ n $. For $ n > 0 $, the de Sitter point corresponds to a degenerate maximum of the potential and is structurally unstable; for $ n < 0 $, the potential forms a degenerate minimum, and the de Sitter solution becomes locally stable. Thus, only inverse power-law potentials ($ n < 0 $) can support a dynamically robust late-time attractor. \\

$ d^\pm $ & 1 & 0 & $1$ & $1$ &
Kinetic energy–dominated scalar field phase. Extremely stiff equation of state; causes rapid expansion incompatible with observations. Typically acts as a repeller or early-time transient. \\
\bottomrule
\end{tabularx}
\end{table}

\noindent
 
 \begin{itemize}

\item The only fixed point that reproduces an ideal de Sitter geometry is point $ c $, characterized by $ \Omega_{de} = 1 $ and $ \omega_{\text{eff}} = -1 $. Although its equation of state and energy composition align with cosmological observations of accelerated expansion, nonlinear stability analysis—via the center manifold—reveals that this point is structurally unstable for $ n > 0 $ and therefore cannot serve as a future attractor. In contrast, for $ n < 0 $, the point exhibits asymptotic stability.

The remaining points reflect transitional or early-universe behavior:

\item $ a_M$  and $ b_R$ emulate standard matter and radiation phases, respectively, consistent with early cosmic epochs but lacking acceleration.
\item $ d^\pm$ describes ultra-stiff scalar field states, likely relevant near cosmological singularities but dynamically excluded from the long-term trajectory.
\end{itemize}
Overall, despite admitting a formally correct de Sitter solution, the power-law potential fails to yield a dynamically robust attractor for late-time acceleration, unless modifications (say, taking inverted powers) or alternative mechanisms are introduced.

These results provide insight into cosmological attractors and the viability of dark energy solutions. See the summary in Table \ref{table:cosmo-powerlaw-refined}.

\subsection{Concluding Remarks about Slow manifold dynamics}

This study explored a higher-order scalar field cosmology inspired by the Pais–Uhlenbeck oscillator. The inclusion of fourth-order field equations introduced a slow-fast dynamical structure, described by Eq.~\eqref{singular_ds}, which separates rapid transitions from gradual evolution. Through singular perturbation analysis, we examined the system’s behavior across different regimes. The evolution recovered the Klein–Gordon equation naturally (Eq.~\eqref{EQ.(43)}), while the reduced slow manifold equation (Eq.~\eqref{outer-solution}) constrained the long-term dynamics in phase space.

\subsubsection{Dynamical systems analysis}

Fixed-point analysis revealed cosmological phases corresponding to matter and radiation domination, kinetic scalar field configurations, scaling behavior, and potential de Sitter expansion. The auxiliary function $ f(\lambda) $, derived from the scalar field potential, governs the shape of the gradient-like potential $ U(y) $, which controls nonlinear stability on the center manifold. Truncated expansions near the origin allowed us to classify fixed points using classical criteria for degenerate extrema.

\subsubsection{Typical examples of center manifold application}

Two benchmark potentials were examined: the exponential model defined in Eq.~\eqref{exp}, yielding $ f(\lambda) = -\lambda(\lambda - \beta) $, and the power-law model given in Eq.~\eqref{power-law}, for which $ f(\lambda) = -\lambda^2 / n $. Both admit a de Sitter solution with full scalar field dominance ($ \Omega_{de} = 1 $) and accelerated expansion ($ \omega_{eff} = -1 $). However, their nonlinear stability behaviors differ. In the exponential case, the induced potential near the origin becomes $ U(y) \sim \beta y^3 + \cdots $ when $ \beta \neq 0 $, and $ U(y) \sim -\tfrac{1}{4} y^4 + \cdots $ when $ \beta = 0 $, corresponding to a cubic inflection and a flat maximum, respectively—both unstable.

In contrast, the power-law model yields $ U(y) = -\tfrac{1}{4n} y^4 + \mathcal{O}(y^6) $, and stability depends on the sign of $ n $: the origin is a degenerate maximum and unstable for $ n > 0 $, but becomes a flat minimum and stable for $ n < 0 $. Only inverse power-law potentials support a dynamically consistent late-time attractor.

The cosmological interpretation of the fixed points confirms these findings. While the exponential model presents a broader set of regimes—including matter and radiation scaling, kinetic states, and scalar-matter couplings—none act as stable endpoints of cosmic evolution. The power-law model offers fewer configurations, but successfully produces a viable attractor when $ n < 0 $.

Overall, both models align with observational parameters, but only the inverse power-law potential meets the structural criteria required for long-term stability. These results highlight the importance of analyzing gradient structure and center manifold dynamics, rather than relying solely on background quantities. Future work could explore higher-order corrections, conduct numerical simulations of phase-space trajectories, and extend this framework to alternative scalar field models relevant for inflation and dark energy.

\subsubsection{Slow-Roll Parameters from $ f(\lambda) $ and Potential Geometry}

We begin with the slope parameter for a scalar field potential $ V(\phi) $, defined in equations~\eqref{eq:dV_dphi}--\eqref{eq:d2V_dphi2}. 
The standard slow-roll parameters follow directly from the potential geometry:
\begin{align}
    \epsilon &\equiv \frac{1}{2} \left(\frac{V'}{V}\right)^2 = \frac{1}{2} \lambda^2, \label{eq:epsilon_def} \\
    \eta &\equiv \frac{V''}{V} = f(\lambda) + \lambda^2. \label{eq:eta_def}
\end{align}

Accelerated expansion requires:
\[
\epsilon < 1, \qquad |\eta| \ll 1,
\]
which depend explicitly on the structure of $ f(\lambda) $ and the behavior of $ \lambda $.

\paragraph{Exponential Potential and Slow-Roll Parameters}
Consider the potential 
\[
V(\phi) = V_0 e^{-\beta \phi}
\]
Here:
\[
\lambda = \beta = \text{constant}, \qquad f(\lambda) = 0
\]
Then, using~\eqref{eq:epsilon_def}--\eqref{eq:eta_def}:
\[
\epsilon = \tfrac{1}{2} \beta^2, \qquad \eta = \beta^2
\]
Slow-roll requires $ \beta \ll 1 $, else the field rolls too quickly.

\paragraph{Power-Law Potential and Slow-Roll Parameters}

Consider the potential
\[
V(\phi) = \frac{V_0}{\phi^n},
\]
where $n > 0$. The auxiliary function $\lambda(\phi)$ is defined by
\[
\lambda(\phi) := -\frac{V'}{V} = \frac{n}{\phi}.
\]
Differentiating $\lambda$ with respect to $\phi$, we find
\[
\frac{d\lambda}{d\phi} = -\frac{n}{\phi^2},
\]
and using the definition
\[
f(\lambda) := -\frac{d\lambda}{d\phi},
\]
it follows that
\[
f(\lambda) = \frac{n}{\phi^2} = \frac{\lambda^2}{n},
\]
where we used $\phi = \frac{n}{\lambda}$.

The first and second slow-roll parameters are then given by
\[
\epsilon := \frac{1}{2} \lambda^2 = \frac{n^2}{2\phi^2}, \qquad
\eta := \lambda^2 + f(\lambda) = \lambda^2 \left( 1 + \frac{1}{n} \right).
\]

In the asymptotic regime $\phi \to \infty$, we observe that $\lambda(\phi) \to 0$, implying
\[
\epsilon \to 0, \quad \eta \to 0.
\]
Thus, the slow-roll conditions are satisfied for sufficiently large field values, and inflation becomes viable in the limit $\phi \to \infty$ provided $n > 1$.

\begin{table}[h!]
\centering
\renewcommand{\arraystretch}{1.3}
\caption{Slow-roll behavior across potential types, based on auxiliary functions and derivatives.}
\begin{tabular}{c c c c c c}
\toprule
Potential Type & $ V(\phi) $ & $ \lambda(\phi) $ & $ f(\lambda) $ & $ \epsilon $ & Slow-Roll Viability \\
\midrule
Exponential & $ V_0 e^{-\beta \phi} $ & $ \beta $ (constant) & 0 & $ \tfrac{1}{2} \beta^2 $ & Yes if $ \beta \ll 1 $ \\
Power-law   & $ V_0 \phi^{-n} $       & $ \tfrac{n}{\phi} $ & $ \tfrac{\lambda^2}{n} $ & $ \tfrac{1}{2} \lambda^2 $ & Yes if $ n > 1 $ \\
\bottomrule
\end{tabular}
\label{table:slowroll_compare}
\end{table}

Potentials with vanishing or decaying $ f(\lambda) $ (e.g., power-law $V(\phi) = \frac{V_0}{\phi^n}$ for $ n > 0 $) allow $ \lambda \to 0 $, sustaining slow-roll. In contrast, constant $ \lambda $ potentials require direct tuning. Dynamical systems analysis—especially using center manifold techniques—further clarifies the nonlinear viability of these regimes (see table \ref{table:slowroll_compare}).

\subsubsection{Tracking Solutions and Late-Time Scalar Field Domination}

In contrast to slow-roll inflation, tracking solutions describe scalar field configurations that evolve alongside the dominant cosmological background (e.g., matter or radiation) and eventually overtake it at late times.

\paragraph{General Framework}

Tracking behavior is governed by the slope formalism, and typically requires:
\begin{itemize}
    \item Approximate constancy of $ \lambda(\phi) $,
    \item Small or vanishing $ f(\lambda) := -\frac{d\lambda}{d\phi} $,
    \item Existence of stable attractors in the autonomous system,
    \item An intermediate equation of state: $ w_{\text{eff}} \in (-1, 0) $.
\end{itemize}

From the formal perspective, tracking arises when
\begin{equation}
    \lambda \approx \text{const}, \qquad f(\lambda) \approx 0,
    \label{eq:tracking_condition}
\end{equation}
leading to an effective scalar field equation of state
\[
w_\phi \approx \frac{\lambda^2 - 3}{3}, \qquad \Omega_\phi \to 1 \quad \text{as } N \to \infty.
\]

\subsubsection{Tracking Regimes for Specific Potentials}

\paragraph{Exponential Potential and Scaling Solutions}

Consider the exponential potential
\[
V(\phi) = V_0 e^{-\beta \phi},
\]
with slope parameter $\beta > 0$. This yields a constant auxiliary function
\[
\lambda := -\frac{V'}{V} = \beta, \qquad f(\lambda) := -\frac{d\lambda}{d\phi} = 0.
\]

Since $\lambda$ and $f(\lambda)$ are both constant, the system admits a stable scaling solution. In this regime, the scalar field energy density evolves proportionally to the background fluid:
\[
\frac{\rho_\phi}{\rho_b} = \text{const}, \qquad \Omega_\phi = \text{const}, \qquad w_\phi = w_b.
\]

Tracking behavior emerges robustly across a broad range of initial conditions; however, late-time scalar field domination necessitates a departure from exact scaling. This occurs when
\[
\beta^2 < 3(1 + w_b),
\]
allowing the scalar field to decouple from the background and eventually dominate the total energy density.

\paragraph{Power-Law Potential}

\[
V(\phi) = V_0 \phi^{-n}, \quad \lambda = \frac{n}{\phi}, \quad f(\lambda) = \frac{\lambda^2}{n}.
\]
\begin{itemize}
    \item Initially, $ \lambda \gg 1 \implies $ stiff-like evolution with transient tracking behavior.
    \item For $ n > 0 $, $ \lambda \to 0 $ as $ \phi \to \infty \implies $ late-time scalar field domination.
    \item Transition to acceleration may require crossing saddle-type structures and entering center manifolds.
\end{itemize}

\begin{table}[h!]
\centering
\renewcommand{\arraystretch}{1.2}
\caption{Conditions for tracking behavior and scalar field domination, based on equation~\eqref{eq:tracking_condition}.}
\begin{tabular}{c c c c c c}
\toprule
Potential Type & $ V(\phi) $ & $ \lambda(\phi) $ & $ f(\lambda) $ & Tracking Viable & Late-Time Domination \\
\midrule
Exponential & $ V_0 e^{-\beta \phi} $ & $ \beta $ (constant) & 0 & Yes if $ \beta^2 > 3(1 + w_b) $ & Only if $ \beta^2 < 3(1 + w_b) $ \\
Power-law & $ V_0 \phi^{-n} $ & $ \frac{n}{\phi} $ & $ \frac{\lambda^2}{n} $ & Transient (early-time) & Yes for $ n > 0 $ \\
\bottomrule
\end{tabular}
\label{table:tracking_summary}
\end{table}

Tracking solutions link early-time background-following dynamics with late-time acceleration. While exponential potentials allow for persistent tracking, scalar field domination requires fine-tuning of the slope. Power-law potentials, by contrast, feature evolving $\lambda(\phi)$ that naturally transition toward domination, in agreement with center manifold stability results (see Table~\ref{table:tracking_summary}).

\section{Two-Field Model Reformulation}
\label{sect:two_field_model}

Motivated by the structure of power-law scalar field models, we reformulate higher-derivative scalar theories using a two-field representation, neglecting matter sources. We consider both the quadratic potential
\begin{equation}
V(\phi) = \frac{1}{2} m^2 \phi^2, \label{eq:V_quad}
\end{equation}
and the inverse-square potential
\begin{equation}
V(\phi) = \frac{\mu}{\phi^2}, \label{eq:V_inv_square}
\end{equation}
within the Pais-Uhlenbeck-type action
\begin{equation}
S = \int d^4x\, \sqrt{-g} \left( \frac{R}{2} - \frac{1}{2} \nabla^\nu \phi \nabla_\nu \phi + \frac{\alpha}{2} (\Box\phi)^2 - V(\phi) \right). \label{eq:PU_action}
\end{equation}

The equation of motion becomes
\begin{equation}
\Box\phi + \alpha\, \Box^2\phi + V'(\phi) = 0, \label{eq:PU_eom}
\end{equation}
with $ V'(\phi) = -m^2 \phi $ for the quadratic case, and $ V'(\phi) = \frac{2\mu}{\phi^3} $ for the inverse-square case.

Adopting the Lee–Wick prescription \cite{Li:2005fm}, we introduce auxiliary fields
\begin{equation}
\chi = \alpha\, \Box\phi, \qquad \psi = \phi + \chi, \label{eq:LW_aux_fields}
\end{equation}
which transforms the action into a two-field second-order system:
\begin{equation}
\mathcal{L} = -\frac{1}{2} \nabla^\mu \psi \nabla_\mu \psi + \frac{1}{2} \nabla^\mu \chi \nabla_\mu \chi - U(\psi, \chi), \label{eq:LW_main_lagrangian}
\end{equation}
with effective potential
\begin{equation}
U(\psi, \chi) =
\begin{cases}
\frac{\chi^2}{2\alpha} + \frac{1}{2} m^2 (\psi - \chi)^2 & \text{(quadratic)}, \\
\frac{\chi^2}{2\alpha} + \frac{\mu}{(\psi - \chi)^2}      & \text{(inverse-square)}.
\end{cases} \label{eq:LW_effective_potential}
\end{equation}

In a spatially flat FLRW background, the system evolves according to:
\begin{align}
\ddot{\chi} + 3H \dot{\chi} &= \frac{\chi}{\alpha} + \frac{\partial U}{\partial \chi}, \label{eq:eom_chi} \\
\ddot{\psi} + 3H \dot{\psi} &= -\frac{\partial U}{\partial \psi}, \label{eq:eom_psi} \\
3H^2 &= \frac{1}{2} \dot{\psi}^2 - \frac{1}{2} \dot{\chi}^2 + U(\psi,\chi). \label{eq:Friedmann_equation}
\end{align}

\subsection{Quadratic Potential}
\label{subsec:quad_potential}

The ghost–quintessence Lagrangian is
\begin{equation}
\mathcal{L} = -\frac{1}{2} \nabla^\mu \psi \nabla_\mu \psi + \frac{1}{2} \nabla^\mu \chi \nabla_\mu \chi - U(\psi, \chi), \label{eq:lagrangian_quad}
\end{equation}
with
\begin{equation}
U(\psi, \chi) = \frac{\chi^2}{2\alpha} + \frac{m^2}{2} (\psi - \chi)^2. \label{eq:potential_quad}
\end{equation}

Energy density and pressure are given by:
\begin{align}
\rho_{DE}^{(q)} &= \frac{1}{2} \left( m^2 (\chi - \psi)^2 + \frac{\chi^2}{\alpha} - \dot{\chi}^2 + \dot{\psi}^2 \right), \label{eq:rho_quad} \\
p_{DE}^{(q)}   &= \frac{1}{2} \left( -m^2 (\chi - \psi)^2 - \frac{\chi^2}{\alpha} - \dot{\chi}^2 + \dot{\psi}^2 \right). \label{eq:p_quad}
\end{align}

Diagonalizing into canonical fields:
\begin{equation}
\phi_1 = \frac{a_2 \chi - a_1 \psi}{a_1^2 - a_2^2}, \qquad
\phi_2 = \frac{a_1 \chi - a_2 \psi}{a_1^2 - a_2^2}, \label{eq:canonical_fields_quad}
\end{equation}
with coefficients:
\begin{equation}
a_1 = \frac{\sqrt{4\alpha m^2 + 1} - 1}{2 \sqrt[4]{4\alpha m^2 + 1}}, \qquad
a_2 = \frac{\sqrt{4\alpha m^2 + 1} + 1}{2 \sqrt[4]{4\alpha m^2 + 1}}. \label{eq:diagonal_coeff_quad}
\end{equation}

The equations of motion are
\begin{align}
\ddot{\phi}_1 + 3H\dot{\phi}_1 - m_1^2 \phi_1 &= 0, \label{eq:eom_phi1_quad} \\
\ddot{\phi}_2 + 3H\dot{\phi}_2 + m_2^2 \phi_2 &= 0, \label{eq:eom_phi2_quad}
\end{align}
with masses
\begin{equation}
m_1^2 = \frac{\sqrt{4\alpha m^2 + 1} + 1}{2\alpha}, \qquad
m_2^2 = \frac{\sqrt{4\alpha m^2 + 1} - 1}{2\alpha}. \label{eq:masses_quad}
\end{equation}

The diagonal energy density is:
\begin{equation}
\rho_{DE}^{(q)} = -\frac{1}{2} \dot{\phi}_1^2 + \frac{1}{2} \dot{\phi}_2^2 + \frac{1}{2} m_1^2 \phi_1^2 + \frac{1}{2} m_2^2 \phi_2^2. \label{eq:rho_diag_quad}
\end{equation}

Classical stability requires $ \alpha < 0 $ and $ 4|\alpha|m^2 \leq 1 $. In particular, for $ |\alpha| m^2 \ll 1 $, the ghost is strongly suppressed due to $ m_1^2 \approx \frac{1}{|\alpha|} \gg m^2 $.

\subsection{Inverse-Square Potential}
\label{subsec:inv_square_potential}

We consider a Lee–Wick-type ghost–quintessence system with a Coulomb-like interaction:
\begin{equation}
L = -\frac{1}{2} \nabla^\mu \psi \nabla_\mu \psi + \frac{1}{2} \nabla^\mu \chi \nabla_\mu \chi - U(\psi, \chi), \label{eq:LW_lagrangian_invS}
\end{equation}
where the effective potential is given by
\begin{equation}
U(\psi, \chi) = \frac{\chi^2}{2\alpha} + \frac{\mu}{(\psi - \chi)^2}. \label{eq:eff_potential_invS}
\end{equation}
This form retains the field-space translational symmetry $ \psi \to \psi + \lambda, \chi \to \chi + \lambda $, but introduces a singularity as $ \psi \to \chi $.

The energy density and pressure of the dark energy sector become
\begin{align}
\rho_{DE} &= \frac{1}{2} \left( \frac{\chi^2}{\alpha} + \frac{2\mu}{(\psi - \chi)^2} - \dot{\chi}^2 + \dot{\psi}^2 \right), \label{eq:rho_de_invS} \\
p_{DE}   &= \frac{1}{2} \left( -\frac{\chi^2}{\alpha} - \frac{2\mu}{(\psi - \chi)^2} - \dot{\chi}^2 + \dot{\psi}^2 \right). \label{eq:p_de_invS}
\end{align}

\subsubsection{Local stability}

To study local stability, we expand around a background configuration where $ \psi - \chi = \phi_0 \neq 0 $, and define:
\begin{equation}
\phi_1 = \psi - \chi - \phi_0, \qquad \phi_2 = \chi. \label{eq:local_fields_invS}
\end{equation}

Expanding the potential to quadratic order yields:
\begin{equation}
U(\phi_1, \phi_2) \approx \frac{\mu}{\phi_0^2} - \frac{2\mu}{\phi_0^3} \phi_1 + \frac{3\mu}{\phi_0^4} \phi_1^2 + \frac{\phi_2^2}{2\alpha} + \cdots. \label{eq:U_expanded_invS}
\end{equation}

We identify the effective masses:
\begin{equation}
m_1^2 = \frac{6\mu}{\phi_0^4}, \qquad m_2^2 = \frac{1}{\alpha}. \label{eq:masses_invS}
\end{equation}

For $ \alpha < 0 $ and $ |\alpha| \mu \ll \phi_0^4 $, the ghost field $ \phi_2 $ becomes heavy and energetically suppressed. Classical-level stability is maintained in this regime, as the singularity at $ \psi = \chi $ is avoided by fixing $ \phi_0 \neq 0 $, and the ghost excitation is disfavored by a large mass hierarchy.

\subsubsection{Scaling Behavior for Inverse-Square Potential}

To explore attractor solutions, we consider the ansatz:
\begin{equation}
a(t) = a_0 t^p, \quad \phi(t) = \phi_0 t^\kappa, \quad H(t) = \frac{p}{t}, \label{eq:scaling_ansatz_kappa}
\end{equation}
and substitute into the inverse-square PU equation:
\begin{equation}
\Box \phi + \alpha\, \Box^2 \phi + \frac{2\mu}{\phi^3} = 0. \label{eq:PU_eom_inverse_scaled}
\end{equation}

Each term scales as:
\begin{align}
\Box \phi &\sim t^{\kappa - 2}, \label{eq:Box_scaling_kappa} \\
\Box^2 \phi &\sim t^{\kappa - 4}, \label{eq:Box2_scaling_kappa} \\
\frac{2\mu}{\phi^3} &\sim t^{-3\kappa}. \label{eq:V_scaling_kappa}
\end{align}

Matching all exponents is impossible: equating $ \kappa - 2 = -3\kappa $ yields $ \kappa = \frac{1}{2} $; equating $ \kappa - 4 = -3\kappa $ yields $ \kappa = 1 $. This contradiction implies that no exact scaling symmetry exists across all terms.

Although a full scaling solution to Eq.~\eqref{eq:PU_eom_inverse_scaled} cannot be realized with all terms contributing equally, we identify a regime in which the higher-derivative correction becomes subdominant. Selecting $ \kappa = \frac{2}{3} $, the time dependence of each term is:
\begin{equation}
\Box \phi \sim t^{-4/3}, \qquad
\Box^2 \phi \sim t^{-10/3}, \qquad
\frac{2\mu}{\phi^3} \sim t^{-2}. \label{eq:scaling_estimates}
\end{equation}
Here, $ \Box^2 \phi $ decays more rapidly than the other contributions and can be neglected at leading order. The reduced equation
\begin{equation}
\Box \phi + \frac{2\mu}{\phi^3} \approx 0
\end{equation}
admits the self-consistent solution
\begin{equation}
\phi(t) = \phi_0 t^{2/3}, \qquad V(\phi) = \frac{\mu}{\phi^2} = \frac{\mu}{\phi_0^2} t^{-4/3}, \label{eq:scaling_solution_kappa}
\end{equation}
valid when the background evolves as $ a(t) \propto t^1 $ (i.e., $ p = 1 $).

This trajectory corresponds to a coasting universe, characterized by linear expansion and a Hubble rate $ H(t) = 1/t $. In this regime, the deceleration parameter vanishes:
\begin{equation}
q \equiv -\frac{\ddot{a} a}{\dot{a}^2} = 0,
\end{equation}
signifying an expansion that is neither accelerating nor decelerating. The effective equation-of-state parameter is
\begin{equation}
w_{\text{eff}} = -\frac{1}{3},
\end{equation}
marking the threshold between matter-dominated deceleration ($ w > -1/3 $) and dark energy–driven acceleration ($ w < -1/3 $).

The attractor solution $ \phi(t) = \phi_0 t^{2/3} $ ensures that the field’s energy density dilutes more slowly than radiation or matter but does not dominate prematurely. This behavior is typical of tracker scalar field models and emerges naturally from inverse-square potentials. Within the PU framework, higher-derivative instabilities are suppressed in this regime, enabling the system to asymptotically approach a stable coasting phase that interpolates between decelerating expansion and potential late-time acceleration.

\subsection{Concluding Remarks on Local Stability around the background configuration}

To assess local stability, we expanded $ U(\psi, \chi) $ around the background configuration $ \psi - \chi = \phi_0 $, and introduced fluctuations as
\[
\phi_1 = \psi - \chi - \phi_0, \qquad \phi_2 = \chi.
\]

This yielded the expanded potential
\[
U(\phi_1, \phi_2) \approx
\begin{cases}
\frac{1}{2} m_1^2 \phi_1^2 + \frac{1}{2} m_2^2 \phi_2^2 & \text{(quadratic)}, \\[6pt]
\frac{\mu}{\phi_0^2} - \frac{2\mu}{\phi_0^3} \phi_1 + \frac{3\mu}{\phi_0^4} \phi_1^2 + \frac{\phi_2^2}{2\alpha} + \cdots & \text{(inverse-square)},
\end{cases}
\]

with corresponding mass terms given by
\begin{equation}
m_1^2 =
\begin{cases}
\frac{\sqrt{4\alpha m^2 + 1} + 1}{2\alpha} & \text{(quadratic)}, \\[6pt]
\frac{6\mu}{\phi_0^4} & \text{(inverse-square)},
\end{cases}
\qquad
m_2^2 =
\begin{cases}
\frac{\sqrt{4\alpha m^2 + 1} - 1}{2\alpha} & \text{(quadratic)}, \\[6pt]
\frac{1}{\alpha} & \text{(inverse-square)}.
\end{cases}\label{eq:masses}
\end{equation}

For $ \alpha < 0 $, the ghost field $ \phi_2 $ became heavy and energetically inaccessible. As long as $ |\alpha| m^2 \ll 1 $ or $ |\alpha| \mu \ll 1 $, classical stability was maintained, with ghost suppression ensured via mass hierarchy.

The two-field Lagrangian in Eq.~\eqref{eq:LW_main_lagrangian} can be extended to analytic potentials $ V(\phi) $, as long as the curvature $ V''(\phi) $ stays bounded and the ghost regulator term $ \chi^2 / 2\alpha $ dominates. The interaction $ V(\psi - \chi) $ preserves translational symmetry in field space, and the mass hierarchy in Eq.~\eqref{eq:masses} introduces a natural suppression.

This setup allows stable higher-derivative scalar field models with flexible dynamics. If $ V(\phi) \in \mathcal{C}^2(\mathbb{R}) $, the theory keeps its decoupling and admits attractor solutions relevant for cosmology. Examples include:

\begin{itemize}
    \item Exponential potentials $ V(\phi) = V_0 e^{-\lambda \phi} $: enable scaling and late-time acceleration.
    \item Plateau-like potentials: allow inflationary or slow-roll phases with low curvature.
    \item Tracker and thawing potentials: suit evolving dark energy models with controlled behavior.
\end{itemize}

In all cases, the regulator $ \chi^2 / 2\alpha $ keeps the ghost mode heavy and suppressed, especially when $ \alpha < 0 $ and $ |\alpha| V''(\phi) \ll 1 $. The symmetry $ (\psi, \chi) \rightarrow (\psi + \delta, \chi + \delta) $ keeps the interaction $ V(\psi - \chi) $ unchanged, offering flexibility for inflation or dark energy model building.

Overall, the reformulated theory remains simple and adaptable, making it suitable for embedding higher-derivative dynamics into cosmological models.

\section{Inflationary dynamics within the Pais–Uhlenbeck model}
\label{sect:V} 
We explore inflationary dynamics within the Pais–Uhlenbeck model under the constraint
\begin{equation}
1 - 3\alpha \dot{\phi}^2 = 0, \qquad \alpha > 0,
\end{equation}
which corresponds to an extremal configuration of the effective kinetic coupling. Solving yields a constant scalar velocity:
\begin{equation}
\dot{\phi} = \pm \sqrt{\frac{1}{3\alpha}} \quad \implies \quad \phi(t) = \pm \sqrt{\frac{1}{3\alpha}}\, t + \phi_0.
\end{equation}

Substituting into the background equations, we obtain the reduced dynamical system:
\begin{align}
\dot{H} + \frac{3}{2} H^2 - V - \frac{1}{6\alpha} - \rho &= 0, \label{eq:inflationH1} \\
\dot{H} + \frac{3}{2} H^2 - V + \frac{1}{6\alpha} + p &= 0, \label{eq:inflationH2} \\
\frac{V_{,\phi}}{\dot{\phi}} + 3H - 9\alpha H \dot{H} - 3\alpha \ddot{H} &= 0. \label{eq:inflationKG}
\end{align}

Differentiating Eq.~\eqref{eq:inflationH1} with respect to time, solving for $ \ddot{H} $, and substituting into Eq.~\eqref{eq:inflationKG} yields a consistency condition:
\begin{equation}
V_{,\phi}(1 - 3\alpha \dot{\phi}^2) + 3\dot{\phi}(H - \alpha \dot{\rho}) = 0.
\end{equation}
Using the constraint $ 1 - 3\alpha \dot{\phi}^2 = 0 $, the first term vanishes, and the equation simplifies to
\begin{equation}
\dot{\rho} = \frac{H}{\alpha} \quad \implies \quad \rho(t) = \rho_0 + \ln\left(a^{\frac{1}{\alpha}}\right).
\end{equation}

Subtracting Eqs.~\eqref{eq:inflationH1} and \eqref{eq:inflationH2}, we find the effective equation of state:
\begin{equation}
-3(\rho + p) = \frac{1}{\alpha} \quad \implies \quad p = -\frac{1}{3\alpha} - \rho.
\end{equation}
Substituting back, the evolution equation becomes
\begin{equation}
\dot{H} + \frac{3}{2} H^2 - V - \frac{1}{6\alpha} - \rho_0 - \ln\left(a^{\frac{1}{\alpha}}\right) = 0, \qquad \dot{a} = aH.
\end{equation}

To reconstruct viable inflationary potentials, we prescribe functional forms for $ a(t) $ and deduce $ V(t) $, which can be reparametrized as $ V(\phi) $ using the linear solution $ \phi(t) = \pm \sqrt{1/3\alpha}\, t + \phi_0 $.

\subsection{Standard Evolution (Power-Law Expansion)}
\begin{equation}
a\left(  t\right)  =a_{0}t^{\frac{2}{3\left(
1+w\right)  }}.
\end{equation}
This model describes the standard evolution of a universe dominated by pressureless matter (e.g., dust and dark matter). The exponent depends on the equation of state parameter $ w $, given by:
\begin{equation}
    w = \frac{P}{\rho}.
\end{equation}
where $ P $ is pressure and $ \rho $ is energy density. For $ w = 0 $, the scale factor follows:
\begin{equation}
    a(t) \propto t^{2/3}
\end{equation}
\begin{equation}
    V(t)= -\ln \left(\left(a_0 t^{\frac{2}{3 (\omega
   +1)}}\right)^{\frac{1}{\alpha }}\right)-\frac{4 \alpha  \omega +t^2
   (\omega +1)^2 (6 \alpha   \rho_{0}+1)}{6 \alpha  t^2 (\omega
   +1)^2}.
\end{equation}
The potential that accommodates this behavior is 
\begin{equation}
    V(\phi)=  -\frac{6 \alpha  \rho_0+1}{6 \alpha }-\frac{2 \omega }{9
   \alpha  (\omega +1)^2 (\phi -\phi_0)^2}-\ln
   \left(3^{\frac{1}{\alpha  (3 \omega +3)}} \alpha ^{\frac{1}{3 \alpha 
   (\omega +1)}} a_0^{\frac{1}{\alpha }} (\pm (\phi
   -\phi_0))^{\frac{2}{3 \alpha  (\omega +1)}}\right).
\end{equation}

\subsection{De Sitter Inflation (Exponential Growth)}
\begin{equation}
  a\left(  t\right)  =a_{0}e^{H_0 t}.
\end{equation}
Characterizes a universe dominated by vacuum energy ($ \Lambda $), leading to accelerated expansion. It addresses major cosmological problems:
\begin{itemize}
    \item Horizon problem (ensuring causal connectivity)
    \item Flatness problem (smoothing out curvature anomalies)
    \item Monopole problem (avoiding excess relics from high-energy physics)
\end{itemize}
We deduce 
\begin{equation}
   V(t)=  -\frac{1}{6 \alpha }-\frac{\ln
   (a_0)+H_0 t}{\alpha }+\frac{3 H_0^2}{2}- \rho_0.
\end{equation}
The potential that accommodates this behavior is 
\begin{equation}
   V(\phi)= -\frac{1}{6 \alpha }-\frac{\ln (a_0)}{\alpha }+\frac{3
   H_0^2}{2} \mp \frac{\sqrt{3} H_0 (\phi - \phi_{0})}{\sqrt{\alpha } }-\rho_0.
\end{equation}

\subsection{Gaussian Expansion (Modified Early Universe Growth)}
\begin{equation}
  a\left(  t\right)  =a_{0} \exp \left(-a_{1}t^{2}\right). 
\end{equation}
Unlike other inflationary models, this suggests initial acceleration followed by significant slowdown, possibly linked to Gaussian-shaped potentials in high-energy physics.

We deduce
\begin{equation}
    V(t)= -\frac{1}{6 \alpha }-\frac{\ln (a_0)-a_1 t^2}{\alpha }+6 a_1^2 t^2-2 a_1-\rho_0.
\end{equation}
The potential that accommodates this behavior is 
\begin{equation}
    V(\phi)= 3 a_1 (6 \alpha  a_1+1) (\phi -\phi_0)^2-\frac{6 \ln (a_0)+6 \alpha  (2 a_1+\rho_0)+1}{6 \alpha }. 
\end{equation}

\subsection{Hybrid Expansion (Power $\times$ Exponential Growth)}
\begin{equation}
  a\left(  t\right)  =a_{0}t^{\alpha_{1}}e^{\alpha_{2}t}.
\end{equation}
Represents an intermediate scenario, transitioning from power-law to inflationary expansion. Possible origins include:
\begin{itemize}
    \item Evolving scalar field potential (e.g., slow-roll inflation)
    \item Modified gravity models incorporating dynamical dark energy effects.
\end{itemize}

This scale factor lead to $H= \frac{\alpha_1}{t} + \alpha_2$ 
\begin{equation}
    V(t)= -\frac{-9 \alpha  \alpha_2^2+6 \alpha  \rho_0+6 \ln (a_0)+6 \alpha_1 \ln (t)+1}{6 \alpha }+\frac{\alpha_1 (3 \alpha_1-2)}{2 t^2}-\frac{\alpha_2
   t}{\alpha }+\frac{3 \alpha_1 \alpha_2}{t}.
\end{equation}
The potential that accommodates this behavior is 
\begin{align}
 V(\phi) &= \pm  \frac{\sqrt{3} \sqrt{\frac{1}{\alpha }} \alpha_1 \alpha_2 }{\phi -\phi_0}-\frac{\alpha_1 \ln \left(\pm \frac{ (\phi -\phi_0)}{\sqrt{\frac{1}{\alpha }}}\right)}{\alpha
   }+\frac{\alpha_1 (3 \alpha_1-2)}{6 \alpha  (\phi -\phi_0)^2}\mp \sqrt{3} \sqrt{\frac{1}{\alpha }} \alpha_2   (\phi -\phi_0) \nonumber \\
   & -\frac{-9 \alpha  \alpha_2^2+6 \alpha 
   \rho_{0}+6 \ln (a_0)+\alpha_1 \ln (27)+1}{6 \alpha }.
\end{align}

\subsection{Extended Inflation (Generalized Exponential)}
\begin{equation}
  a\left(  t\right)  =a_{0} e^{A t^f}
\end{equation}
where $A>0$ and $ f $ is fractional ($0<f<1$). This model refines traditional inflation by allowing nonlinear accelerations, capturing alternative early-universe scenarios.

This scale factor lead to $H= A f t^{f-1}$ and
\begin{equation}
    V(t)=  -\frac{1}{6 \alpha }-\frac{A t^f+\ln (a_0)}{\alpha }+\frac{1}{2} A f t^{f-2} \left(f \left(3 A t^f+2\right)-2\right)- \rho_{0}
\end{equation}
The potential that accommodates this behavior is 
\begin{align}
  V(\phi)&=\frac{3^{\frac{f}{2}-1} A f \left(\frac{\pm  (\phi -\phi_0)}{\sqrt{\frac{1}{\alpha }}}\right)^f \left(f \left(3^{\frac{f}{2}+1} A \left(\frac{\pm  (\phi -\phi_0)}{\sqrt{\frac{1}{\alpha
   }}}\right)^f+2\right)-2\right)}{2 \alpha  (\phi -\phi_0)^2} \nonumber \\
   & -\frac{6 \alpha  \rho_0+2 \times 3^{\frac{f}{2}+1} A \left(\frac{\pm  (\phi -\phi_0)}{\sqrt{\frac{1}{\alpha }}}\right)^f+6 \ln
   (a_0)+1}{6 \alpha }.
\end{align}

\subsection{Logarithmic Inflation (Log-based Growth)}
\begin{equation}
    a(t) = e^{A (\ln t)^\lambda}
\end{equation}
Introduces logarithmic modifications to standard expansion laws, leading to slower inflationary growth. It can arise from modified gravity models or non-minimally coupled scalar fields.

This scale factor $a(t) =\exp[A(\ln t)^\lambda]$ leads to $H=\frac{A \lambda  \ln ^{\lambda -1}(t)}{t}$ and
\begin{equation}
    V(t)= -\frac{1}{6 \alpha }+\frac{A \lambda  \ln ^{\lambda -2}(t) \left(3 A \lambda  \ln ^{\lambda }(t)+2 \lambda -2 \ln (t)-2\right)}{2 t^2}-\frac{A \ln ^{\lambda }(t)}{\alpha }-\rho_0
\end{equation}
The potential supporting this behavior is 
\begin{align}
    V(\phi)& = \frac{A \lambda  \ln ^{\lambda -2}\left(\frac{\sqrt{3} \epsilon  (\phi -\phi_0)}{\sqrt{\frac{1}{\alpha }}}\right) \left(-2 \ln \left(\frac{\sqrt{3} \epsilon  (\phi -\phi_0)}{\sqrt{\frac{1}{\alpha
   }}}\right)+3 A \lambda  \ln ^{\lambda }\left(\frac{\sqrt{3} \epsilon  (\phi -\phi_0)}{\sqrt{\frac{1}{\alpha }}}\right)+2 \lambda -2\right)}{6 \alpha  (\phi -\phi_0)^2} \nonumber \\
   & -\frac{6 \alpha  \rho_0+6 A
   \ln ^{\lambda }\left(\frac{\sqrt{3} \epsilon  (\phi -\phi_0)}{\sqrt{\frac{1}{\alpha }}}\right)+1}{6 \alpha }.
\end{align}
These examples are summarized in Table~\ref{tab:my_label}, illustrating various expansion scenarios including power-law, exponential, and generalized log-growth inflation.
\begin{table}[H]
    \centering
       \caption{Reconstruction of inflationary potentials from the PU model and prescribed scale factor.}
    \label{tab:my_label}
    \resizebox{\columnwidth}{!}{%
\begin{tabular}{c c}\hline
    Scale factor &  Potential reconstruction \\ \hline
   $a_1\left(  t\right)  =a_{0}t^{\frac{2}{3\left(
1+w\right)  }}$ & $  V_1(\phi)=  -\frac{6 \alpha  \rho_0+1}{6 \alpha }-\frac{2 \omega }{9
   \alpha  (\omega +1)^2 (\phi -\phi_0)^2}-\ln
   \left(3^{\frac{1}{\alpha  (3 \omega +3)}} \alpha ^{\frac{1}{3 \alpha 
   (\omega +1)}} a_0^{\frac{1}{\alpha }} (\pm (\phi
   -\phi_0))^{\frac{2}{3 \alpha  (\omega +1)}}\right).$\\
  $a_2\left(  t\right)  =a_{0}%
e^{H_0 t}$  & $V_2(\phi)= -\frac{1}{6 \alpha }-\frac{\ln (a_0)}{\alpha }+\frac{3
   H_0^2}{2} \mp \frac{\sqrt{3} H_0 (\phi - \phi_{0})}{\sqrt{\alpha } }-\rho_0.$ \\
   $a_3\left(  t\right)  =a_{0} e^{-a_{1}t^{2}}$ & $V_3(\phi)= 3 a_1 (6 \alpha  a_1+1) (\phi -\phi_0)^2-\frac{6 \ln (a_0)+6 \alpha  (2 a_1+\rho_0)+1}{6 \alpha }.$ \\
 $a_4\left(  t\right)  =a_{0}t^{\alpha_{1}}e^{\alpha
_{2}t}$ &   $ V_4(\phi) = \pm  \frac{\sqrt{3} \sqrt{\frac{1}{\alpha }} \alpha_1 \alpha_2 }{\phi -\phi_0}-\frac{\alpha_1 \ln \left(\pm \frac{ (\phi -\phi_0)}{\sqrt{\frac{1}{\alpha }}}\right)}{\alpha
   }+\frac{\alpha_1 (3 \alpha_1-2)}{6 \alpha  (\phi -\phi_0)^2}\mp \sqrt{3} \sqrt{\frac{1}{\alpha }} \alpha_2   (\phi -\phi_0)  -\frac{-9 \alpha  \alpha_2^2+6 \alpha 
   \rho_{0}+6 \ln (a_0)+\alpha_1 \ln (27)+1}{6 \alpha }.$\\
   $a_5\left(  t\right)  =a_{0} e^{A t^f}$ & $  V_5(\phi) = \frac{3^{\frac{f}{2}-1} A f \left(\frac{\pm  (\phi -\phi_0)}{\sqrt{\frac{1}{\alpha }}}\right)^f \left(f \left(3^{\frac{f}{2}+1} A \left(\frac{\pm  (\phi -\phi_0)}{\sqrt{\frac{1}{\alpha
   }}}\right)^f+2\right)-2\right)}{2 \alpha  (\phi -\phi_0)^2}  -\frac{6 \alpha  \rho_0+2 \times 3^{\frac{f}{2}+1} A \left(\frac{\pm  (\phi -\phi_0)}{\sqrt{\frac{1}{\alpha }}}\right)^f+6 \ln
   (a_0)+1}{6 \alpha }.$\\
   $a_5(t) =\exp[A(\ln t)^\lambda]$ & $   V_6(\phi) = \frac{A \lambda  \ln ^{\lambda -2}\left(\frac{\sqrt{3} \epsilon  (\phi -\phi_0)}{\sqrt{\frac{1}{\alpha }}}\right) \left(-2 \ln \left(\frac{\sqrt{3} \epsilon  (\phi -\phi_0)}{\sqrt{\frac{1}{\alpha
   }}}\right)+3 A \lambda  \ln ^{\lambda }\left(\frac{\sqrt{3} \epsilon  (\phi -\phi_0)}{\sqrt{\frac{1}{\alpha }}}\right)+2 \lambda -2\right)}{6 \alpha  (\phi -\phi_0)^2} \nonumber  -\frac{6 \alpha  \rho_0+6 A
   \ln ^{\lambda }\left(\frac{\sqrt{3} \epsilon  (\phi -\phi_0)}{\sqrt{\frac{1}{\alpha }}}\right)+1}{6 \alpha }.$\\\hline
   \end{tabular}}
\end{table}

\begin{figure}[H]
    \centering
    \includegraphics[width=0.75\textwidth]{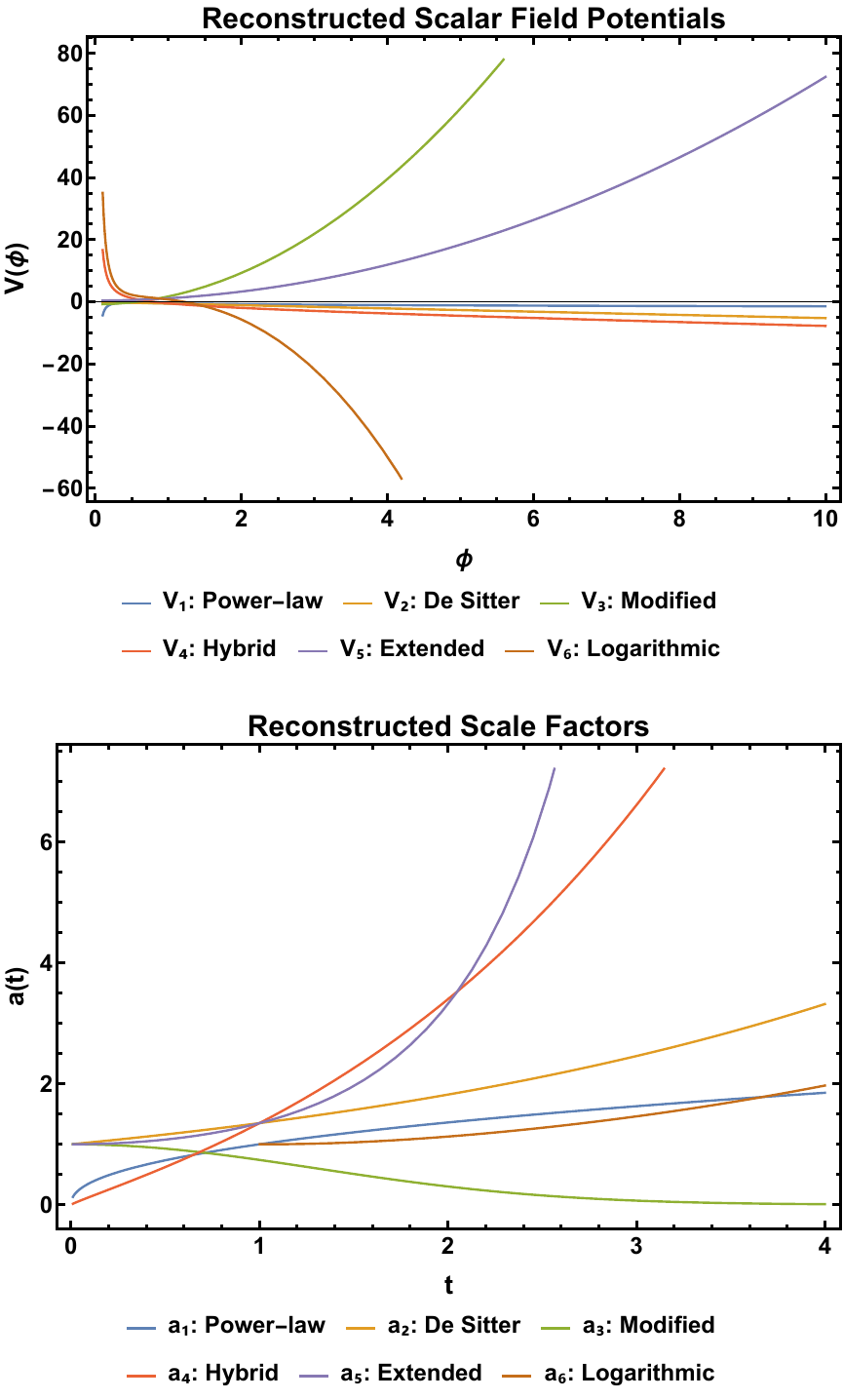}
\caption{
Reconstructed scalar field potentials $V_i(\phi)$ (top panel) and their associated scale factors $a_i(t)$ (bottom panel) across six inflationary scenarios derived from the PU model. The potentials $V_1$ through $V_6$ correspond to: (1) power-law, (2) de Sitter, (3) modified Gaussian, (4) hybrid, (5) extended inflation with $a(t) = a_0 e^{A t^f}$, and (6) logarithmic inflation with $a(t) = a_0 e^{A (\log t)^\lambda}$. Potentials are plotted over $\phi \in [0.1, 10]$, and scale factors over $t \in [0.01, 4]$. Parameter values are set to $\alpha = 1$, $\omega = 0.5$, $\rho_0 = 0.01$, $H_0 = 0.3$, $a_0 = 1$, $A = 0.3$, $f = 2.0$, $\lambda = 2.5$, $\epsilon = 1.2$, and $\phi_0 = 0$. All curves are shown with boxed frames and uniform styling to facilitate direct visual comparison.
} \label{scale-factor-V(phi)}
\end{figure}

Figure~\ref{scale-factor-V(phi)} presents a comparative illustration of reconstructed scalar field potentials $V_i(\phi)$ and their corresponding background expansions $a_i(t)$, derived from six inflationary scenarios within the PU framework. The top panel displays the potentials over the domain $\phi \in [0.1, 10]$, revealing distinct analytic structures and asymptotic behaviors.

For instance, the power-law expansion yields
\[
V_1(\phi) = -0.0100002 - \frac{1.6}{\phi^2} - \ln \left(1.73205\, \phi^{1.33333}\right),
\]
characterized by an inverse-square decay and logarithmic suppression. The de Sitter case corresponds to a linear slope:
\[
V_2(\phi) = 0.134167 - 0.519615\, \phi,
\]
while the modified Gaussian expansion leads to a quadratic form:
\[
V_3(\phi) = 1.929\, \phi^2 - 0.016667.
\]

The hybrid scenario combines inverse, logarithmic, and linear terms:
\[
V_4(\phi) = \frac{0.4677}{\phi} - 0.9\, \ln(\phi) + \frac{2.07}{\phi^2} - 0.5196\, \phi - 0.034168.
\]

Although the extended and logarithmic inflation potentials share structural features, they are analytically and visually distinct. The extended inflation model simplifies to:
\[
V_5(\phi) \approx 0.1194\, \phi^2 + 0.239999,
\]
dominated by polynomial growth. By contrast, the logarithmic inflation potential exhibits nonlinear modulation:
\[
\begin{aligned}
V_6(\phi) &= \frac{0.75\, \ln^{0.5}(2.078\, \phi)
\left[-2 \ln(2.078\, \phi) + 2.25\, \ln^{2.5}(2.078\, \phi) + 3\right]}{6\, \phi^2} \\
&\quad - \left[0.45\, \ln^{2.5}(2.078\, \phi) + 0.060001\right],
\end{aligned}
\]
highlighting deformation introduced by logarithmic time dependence.

The bottom panel displays the scale factors $a_i(t)$ over cosmic time $t \in [0.01, 4]$, corresponding to the functions:
\[
a(t) = \{t^{2/3},\, e^{0.3t},\, t^{0.9} e^{0.3t},\, e^{0.3 t^{0.5}},\, e^{0.2 (\log t)^2},\, e^{-0.3 t^2}\}.
\]
These capture a range of expansion behaviors, from polynomial and exponential growth to hybrid and nonstandard logarithmic forms.

All curves are plotted with boxed frames, consistent labeling, and uniform domain ranges to support direct visual comparison. The parameter values used throughout are: $\alpha = 1$, $\omega = 0.5$, $\rho_0 = 0.01$, $H_0 = 0.3$, $a_0 = 1$, $A = 0.3$, $f = 2.0$, $\lambda = 2.5$, $\epsilon = 1.2$, and $\phi_0 = 0$.

\subsection{Concluding Remarks on Inflationary Scenarios}

The Pais–Uhlenbeck scalar field framework encompasses a diverse landscape of inflationary solutions, each corresponding to a specific cosmic scale factor and its associated reconstructed potential. These scenarios highlight the flexibility of higher-derivative theories in capturing both standard and non-standard phases of cosmic expansion.

Power-law expansions such as $ a(t) \propto t^{2/3(1 + w)} $ reproduce the expected behavior for matter-dominated and radiation-dominated eras. While not inflationary in the strict sense, they serve as transitional benchmarks. The associated potentials typically include inverse-square terms and logarithmic corrections, consistent with non-accelerating expansion.

The exponential scale factor $ a(t) = a_0 e^{H_0 t} $ models de Sitter inflation, driven by near-constant vacuum energy. Its corresponding potential is linear in $ \phi $, superposed with constant and logarithmic contributions. This scenario satisfies the classical inflationary requirements: causal horizon stretching, curvature smoothing, and relic suppression.

Modified growth models, including Gaussian ($ a(t) = a_0 e^{-a_1 t^2} $) and hybrid ($ a(t) = a_0 t^{\alpha_1} e^{\alpha_2 t} $), represent more nuanced expansion mechanisms. These can interpolate between decelerated and accelerated phases or accommodate asymmetric potentials. The reconstructed potentials typically involve quadratic terms and a mixed polynomial–logarithmic structure, offering fertile ground for fine-tuning the behavior of the early universe.

Extended and logarithmic inflationary forms—such as $ a(t) = a_0 e^{A t^f} $ with $ 0 < f < 1 $, or $ a(t) = \exp[A (\ln t)^\lambda] $—allow for nontrivial acceleration profiles. These generalizations produce potentials with logarithmic powers and fractional exponents in $ \phi $, reflecting their nonlinear origin. They are naturally suited to model departures from strict de Sitter behavior and may be relevant for pre-inflationary or reheating regimes.

Overall, the reconstructed potentials demonstrate that inflationary behavior can emerge from a broad class of scale factors within the PU framework, each yielding a distinct effective field theory. The analytic structure of $ V(\phi) $ is sensitive to both the growth rate of $ a(t) $ and the underlying higher-derivative coupling $ \alpha $, which controls field stiffness and energy descent. As such, the formalism supports inflation with tunable features and offers a systematic method for potential reconstruction from desired expansion histories.

\section{System Reduction and Analytical Perspectives}
\label{sect:VI} 
The singular perturbation analysis of this slow-fast dynamical system offers a structured approach to understanding cosmological evolution. Identifying fixed points helps reveal potential attractor solutions, while stability analysis determines whether these configurations remain viable over time.

Future studies could focus on numerical simulations of phase-space trajectories, refining the behavior of attractors, and extending models to include higher-order corrections for a more comprehensive description of cosmic dynamics.

The full dynamical system \eqref{syst1} exhibits a rational structure in which the denominator $ s(\mathbf{u}; \alpha) = u_5^2 - 9 \alpha u_1^2 $ introduces potential singular behavior in specific regions of the phase space. Despite this, the vector field $ \mathbf{G}(\mathbf{u}, \lambda; \alpha) $ remains smooth away from the singular surface. The symbolic representation of the system,
\[
\frac{d \mathbf{u}}{d N} = \frac{\mathbf{G} (\mathbf{u}, \lambda; \alpha)}{s(\mathbf{u}; \alpha)}, \quad
 \frac{d \lambda}{d N}= -\sqrt{6} f(\lambda) u_1,
\]
emphasizes this underlying smoothness and rational dependence.

A key structural simplification arises from enforcing the slow manifold constraint:
\[
\sqrt{6} \lambda u_2^2 - 6 u_1 + 2 u_3 = 0, \quad v = \text{constant},
\]
which permits algebraic elimination of $ u_3 $. Specifically, the relation
\[
u_3 = 3u_1 - \sqrt{\frac{3}{2}} \lambda u_2^2,
\]
decouples the evolution of $ u_3 $ from the remaining variables, and, due to restricted dependence, isolates $ u_5 $ and $ v $ to specific subblocks of the vector field (namely Eqs.~\eqref{eq51c} and \eqref{eq51f}). These decouplings yield a reduced dynamical core consisting of $ (u_1, u_2, \Omega_r, \lambda) $, governed by the regular system \eqref{reduced}.

\noindent\textbf{Phase Space Geometry.} The reduced system evolves in the invariant region:
\[
\left\{(u_1,u_2, \Omega_r, \lambda)\in\mathbb{R}^4:\quad u_1^2+u_2^2+\Omega_r\leq 1,\quad u_2\geq 0,\quad \lambda \in \mathbb{R} \right\},
\]
where the constraint follows directly from the normalized Friedmann equation. This bounded domain supports global existence results and compactness-based stability methods.

\medskip

\noindent\textbf{Singular Limit: $ \alpha \rightarrow 0 $.} In this regime, the fast subsystem becomes dominant, and the dynamics concentrate near the slow manifold. Singular perturbation techniques—such as matched asymptotic expansions or Fenichel theory—could be used to construct invariant manifolds near the singular surface $ s(\mathbf{u}; \alpha) = 0 $ and study regularized flows.

The reduced system \eqref{reduced} is precisely the $ \alpha \rightarrow 0 $ leading-order truncation. Stability of critical points can be assessed via center manifold expansions, using the fact that $ f(\lambda) $ controls the residual curvature of the scalar field potential through
\[
f(\lambda) = \frac{V_{,\phi\phi}}{V} - \left( \frac{V_{,\phi}}{V} \right)^2.
\]

\noindent\textbf{Large Coupling: $ \alpha \rightarrow \infty $.} For large $ \alpha $, the denominator $ s(\mathbf{u}; \alpha) $ becomes dominated by the term $ -9\alpha u_1^2 $, potentially leading to divergence if $ u_1 \neq 0 $. However, this regime can be regularized by redefining a rescaled time variable $ d\tau = \alpha^{-1} dN $ and applying scaling arguments to analyze asymptotic behavior. In this limit, higher-order terms become negligible, and the system may exhibit slow-roll–like trajectories with enhanced frictional damping.

Further investigation of this regime could involve:

\begin{itemize}
    \item Asymptotic expansions in $ 1/\alpha $ for fixed initial conditions.
    \item Identification of emergent slow attractors.
    \item Robustness of the inflationary potential reconstruction when $ \alpha $ is large but finite.
\end{itemize}

\noindent\textbf{Future Directions.} We classified the center manifold profiles for generic potentials $V(\phi)$ by examining the sign-definiteness of $f(\lambda)$ and evaluating higher-order derivatives at the zeros of $f$. However, several analytical and numerical avenues remain open:
\begin{enumerate}
    \item Lyapunov analysis of reduced system \eqref{reduced}, with focus on stability domains under perturbations.
    \item Rigorous treatment of the fast–slow transition layer across $ s(\mathbf{u}; \alpha) = 0 $ using geometric singular perturbation theory.
    \item Extension to multi-field generalizations by promoting $ \phi $ to a vector-valued field with multiple interacting directions.
\end{enumerate}

These approaches aim to consolidate the structural understanding of singular systems in cosmology and clarify the role of higher-order dynamics in inflationary and dark energy models.

\section{Final Conclusions}
\label{sect:VII}
This study examined a cosmological model based on a higher-order scalar field governed by the Pais–Uhlenbeck (PU) oscillator. The inclusion of fourth-order differential equations introduced a slow–fast dynamical hierarchy, enabling the separation of rapid transients from long-term evolution. This structure
provides a consistent framework for analyzing nonlinear scalar field dynamics (see equations \eqref{syst-symbolic}). A constraint on the slow manifold,
\begin{equation}
\sqrt{6} \lambda u_2^2 - 6 u_1 + 2 u_3 = 0, \quad v = \text{constant},
\end{equation}
naturally recovers the Klein–Gordon equation:
\begin{equation}
\ddot{\phi} + 3H\dot{\phi} + V'(\phi) = 0.
\end{equation}

Fixed-point analysis revealed various cosmological regimes, including matter and radiation domination, kinetic configurations, scaling behavior, and exact de Sitter expansion. Stability was evaluated using center manifold theory, where the auxiliary function $ f(\lambda) $ determines the form of the mechanical potential $ U(y) $. Truncated expansions near the origin allowed classification of degenerate extrema and attractor behavior.

Two benchmark potentials were analyzed: the exponential model, defined by $ f(\lambda) = -\lambda(\lambda - \beta) $, and the power-law model, with $ f(\lambda) = -\lambda^2 / n $. Both admit a de Sitter solution characterized by full scalar field dominance ($ \Omega_{de} = 1 $) and accelerated expansion ($ \omega_{eff} = -1 $). However, their stability properties differ significantly. For the exponential potential, the induced mechanical potential near the origin becomes $ U(y) \sim \beta y^3 + \cdots $ when $ \beta \neq 0 $, and $ U(y) \sim -\tfrac{1}{4} y^4 + \cdots $ when $ \beta = 0 $, corresponding to a cubic inflection and a flat maximum—both structurally unstable. In contrast, the power-law potential yields $ U(y) = -\tfrac{1}{4n} y^4 + \mathcal{O}(y^6) $, which is unstable for $ n > 0 $ but becomes stable for $ n < 0 $, where the origin corresponds to a flat minimum. Only inverse power-law potentials support a dynamically viable late-time attractor.

Cosmological interpretations confirm these conclusions. Although the exponential model offers a broader spectrum of regimes—including scaling, kinetic, and mixed scalar–matter configurations— de Sitter does not act as a stable endpoint of evolution. The power-law model offers fewer configurations, but admits a viable attractor when $n < 0$. These results underscore the importance of analyzing gradient structure and center manifold stability, beyond background-level behavior.

To clarify the role of higher-order derivatives, the theory was reformulated as a two-field system under the quadratic potential via the transformation
\[
\chi = \alpha \Box \phi, \qquad \psi = \phi + \chi.
\]
The resulting Lagrangian,
\[
L = -\frac{1}{2} \nabla^\mu \psi \nabla_\mu \psi + \frac{1}{2} \nabla^\mu \chi \nabla_\mu \chi - V(\psi - \chi) - \frac{\chi^2}{2\alpha},
\]
describes a ghost–quintessence pair with translational field-space symmetry. Despite the phantom-like nature of $ \psi $, classical stability is maintained when $ \alpha < 0 $ and $ |\alpha| m^2 \ll 1 $, since the ghost field acquires a large mass that suppresses low-energy excitations. This reformulation applies to arbitrary analytic potentials $V(\phi)$, provided they are bounded from below and have a well-defined minimum.

The Lee–Wick-type model with an inverse-square (Coulomb-like) interaction is governed by the Lagrangian
\begin{equation}
L = -\frac{1}{2} \nabla^\mu \psi \nabla_\mu \psi + \frac{1}{2} \nabla^\mu \chi \nabla_\mu \chi - \left( \frac{\chi^2}{2\alpha} + \frac{\mu}{(\psi - \chi)^2} \right). \label{eq:summary_lagrangian_invS}
\end{equation}
This setup retains translational symmetry in field space but introduces a singularity at $ \psi = \chi $. By expanding around a finite background separation $ \phi_0 = \psi - \chi $, the theory admits a regular quadratic approximation with fluctuation masses
\begin{equation}
m_1^2 = \frac{6\mu}{\phi_0^4}, \qquad m_2^2 = \frac{1}{\alpha}. \label{eq:summary_masses_invS}
\end{equation}
Provided $ \alpha < 0 $ and $ |\alpha| \mu \ll \phi_0^4 $, classical-level stability is maintained: the ghost field $ \phi_2 \equiv \chi $ becomes heavy and energetically suppressed, and the singularity is avoided by enforcing $ \phi_0 \neq 0 $. The model thus remains viable under controlled field separation and appropriate parameter/masses hierarchy.

The framework also enables exact inflationary reconstruction. Imposing the constraint $ 1 - 3\alpha \dot{\phi}^2 = 0 $ leads to linear evolution of the scalar field and allows analytic reconstruction of $ V(\phi) $ from chosen scale factors. A broad class of inflationary scenarios was considered, including power-law, de Sitter, Gaussian, hybrid, extended, and logarithmic inflation, each producing distinct potential structures involving inverse powers, logarithmic corrections, and mixed terms. These cases demonstrate the PU model’s ability to emulate both conventional and generalized inflationary behavior, featuring tunable analytic properties.

Power-law expansions such as $ a(t) \propto t^{2/3(1 + w)} $ describe matter- and radiation-dominated eras. While not strictly inflationary, they serve as transitional benchmarks and yield potentials with inverse-square and logarithmic characteristics. De Sitter inflation ($ a(t) = a_0 e^{H_0 t} $) arises from constant vacuum energy and corresponds to linear potentials with additive logarithmic terms—solving the horizon, flatness, and relic problems. Modified growth models, including Gaussian and hybrid forms, enable smoother transitions between decelerated and accelerated phases and are associated with polynomial–logarithmic potentials. Extended and logarithmic expansions enable nontrivial acceleration profiles, resulting in potentials with fractional powers and nested logarithms, which are suitable for pre-inflationary or reheating regimes.

Altogether, the reconstructed potentials demonstrate how inflationary dynamics can arise from a broad class of scale factors within the PU framework. The analytic structure of $ V(\phi) $ is highly sensitive to the expansion profile and the higher-order coupling $ \alpha $, which governs field stiffness and descent behavior.

In conclusion, the Pais–Uhlenbeck scalar field model offers a versatile and consistent framework for investigating nonlinear cosmological dynamics, late-time acceleration, and inflation. It integrates slow–fast decomposition, center manifold analysis, multi-field reformulation, and potential reconstruction into a unified structure. These results clarify the role of higher-order derivatives in cosmology and provide new pathways for constructing stable and viable models of dark energy and inflation. Future studies could extend this approach to include cosmological perturbations, mechanisms for ghost suppression, and tests of consistency with alternative theories of gravity, as well as observational constraints.

\section*{Acknowledgments}
M. Gonzalez-Espinoza acknowledges the financial support of FONDECYT de Postdoctorado, N° 3230801. GL would like to express his gratitude towards faculty member Alan Coley and staff members Anna Maria Davis, Nora Amaro, Jeanne Clyburne, and Mark Monk for their warm hospitality during the implementation of the final details of the research in the Department of Mathematics and Statistics at Dalhousie University.  He dedicates this work to the memory of his father. GL and AP are grateful for the support of Vicerrector\'{\i}a de Investigaci\'{o}n y Desarrollo Tecnol\'{o}gico (Vridt) at Universidad Cat\'{o}lica del Norte through N\'{u}cleo de Investigaci\'{o}n Geometr\'{\i}a Diferencial y Aplicaciones, Resoluci\'{o}n Vridt No - 096/2022 and Resoluci\'{o}n Vridt No - 098/2022. GL and AP were economically supported by the Proyecto Fondecyt Regular 2024, Folio 1240514, Etapa 2025. 
Aleksander Kozak is supported by Project No. 3250036, Concurso Fondecyt de Postdoctorado 2025. 

\bibliographystyle{elsarticle-num}

\bibliography{bio.bib}

\end{document}